\documentclass[twocolumn,tighten, twocolappendix]{aastex631}

\usepackage{amsmath,amstext,eqnarray}
\usepackage[T1]{fontenc}
\usepackage{apjfonts} 
\usepackage[figure,figure*]{hypcap}
\usepackage{commath}
\usepackage{CJKutf8}
\usepackage{tabularx}
\usepackage{graphicx,array,bm,booktabs}
\usepackage{float}
\usepackage{longtable}
\usepackage{rotating}
\usepackage{color} 
\usepackage{xcolor}
\usepackage[flushleft]{threeparttable}
\usepackage{CJKutf8}
\usepackage{multirow}

\usepackage[none]{hyphenat}
\usepackage{array}
\newcolumntype{+}{>{\global\let\currentrowstyle\relax}}
\newcolumntype{^}{>{\currentrowstyle}}

\newcommand\clearrow{\global\let\rowmac\relax}
\clearrow

\shorttitle{}
\shortauthors{Chung et al.}
\graphicspath{{./}{figures/}}

\begin{document}
\begin{CJK}{UTF8}{bsmi}

\title{The 4--400 GHz Survey for the 32 Class II Disks in the Taurus Molecular Cloud}

\author[0009-0007-3677-8040]{Chia-Ying Chung}
\affiliation{Department of Physics, National Sun Yat-Sen University, No. 70, Lien-Hai Road, Kaohsiung City 80424, Taiwan, R.O.C.}

\author[0000-0002-3211-4219]{An-Li Tsai}
\affiliation{Department of Physics, National Sun Yat-Sen University, No. 70, Lien-Hai Road, Kaohsiung City 80424, Taiwan, R.O.C.}

\author[0000-0002-9154-2440]{Melvyn Wright}
\affiliation{Department of Astronomy, University of California, Berkeley, 501 Campbell Hall, Berkeley, CA 94720-3441, USA}


\author[0000-0002-9408-2857]{Wenrui Xu}
\affiliation{Center for Computational Astrophysics, Flatiron Institute, 162 5th Ave, New York, NY10010, USA}

\author[0000-0002-7607-719X]{Feng Long}
\affiliation{Lunar and Planetary Laboratory, University of Arizona, Tucson, AZ 85721, USA}
\altaffiliation{NASA Hubble Fellowship Program Sagan Fellow}

\author[0000-0003-0685-3621]{Mark A. Gurwell}
\affiliation{Center for Astrophysics \textbar\, Harvard \& Smithsonian, 60 Garden St., Cambridge, MA 02138, USA}

\author[0000-0003-2300-2626]{Hauyu Baobab Liu}
\affiliation{Department of Physics, National Sun Yat-Sen University, No. 70, Lien-Hai Road, Kaohsiung City 80424, Taiwan, R.O.C.}
\affiliation{Center of Astronomy and Gravitation, National Taiwan Normal University, Taipei 116, Taiwan}



\begin{abstract}
We have compiled the $\sim$4--400 GHz broad spectra of 32 Class II protoplanetary disks in the Taurus-Auriga region, which represents the brightest one-third of sources detected in the submillimeter band in this region.
The spectra at $>$20 GHz frequency can be described with a piecewise function: (1) a power law with a spectral index $\sim$2 at $>$200 GHz, (2) a power law  with spectral index in the range  0.3--4.2 at 20-50 GHz, and (3) a transition region in between these two power laws which can be characterized by a sigmoid function. 
This suggests that the flux densities at $>$200 GHz and $<$50 GHz are dominated by distinct emission components. 
At $>$200 GHz, the emission is likely dominated by the optically thick dust thermal emission in the bulk of the disks. 
In some sources that were not detected at 6.8 GHz or 10 GHz, embedded high-density dust substructures may contribute to a significant fraction of the flux densities at 30–50 GHz, and the spectral indices are mostly consistent with 2.0. 
Although, at 30–50 GHz, free-free and/or synchrotron emission may be significant, and some sources in our sample have spectral indices $<$ 2.0.
Based on these results, we hypothesize that high-density dust substructures (e.g., vortices) are often found in resolved Class II protoplanetary disks, and are a precursor to the formation of kilometer-sized planetesimals and rocky planets.
They may not present high contrast at $>$200 GHz frequencies owing to the high optical depth. 
To probe these dust substructures, high angular resolution observations at $<$100 GHz are necessary to distinguish them from free-free and synchrotron emission sources.
Otherwise, in the analyses of the spatially unresolved spectra, one needs to simultaneously constrain the flux densities of free-free, synchrotron, and dust emission with the observations at $\sim$5--50 GHz.
\end{abstract}



\section{Introduction} \label{sec:intro}

Dust in protoplanetary disks (PPDs) must be processed to form the ubiquitously detected rocky planets.
Initially, the dust grains in PPDs are very small (e.g., in sub-$\mu$m sizes) and are well coupled with the gas in the disk. 
The first step in forming planets is the coagulation of these tiny dust grains into 10--100 $\mu$m sized aggregates (for a recent review see \citealt{Birnstiel2023arXiv231213287B}).
Once formed, these aggregates become dynamically loosely coupled with the bulk of the parent gaseous PPDs due to their small geometric cross-section per unit mass.
They tend to migrate inward until they are trapped at a pressure maximum, which may be manifested as the dusty rings that were ubiquitously resolved in the ALMA observations (e.g., \citealt{Andrews2018ApJ...869L..41A,Long2018ApJ...869...17L,Francis2020ApJ...892..111F,Cieza2021MNRAS.501.2934C,Yamaguchi2024PASJ...76..437Y}, etc).
Owing to the accumulation of a large amount of dust, the recent theoretical and observational studies showed that the dust rings are likely very optically thick (i.e., optical depth $\tau\gtrsim$3) at $\gtrsim$100 GHz frequencies (e.g., \citealt{Liu2016ApJ...816L..29L,Liu2019ApJ...877L..22L,Carrasco2019ApJ...883...71C,Chung2024ApJS..273...29C,Delussu2024A&A...688A..81D}).
We note that when maximum dust grain size $a_{\rm max}$ is comparable or larger than $\sim$100 $\mu$m, the effective dust scattering opacity may outweigh the dust absorption opacity (\citealt{Birnstiel2018ApJ...869L..45B}).
In this case, brightness temperature of dust thermal emission will be lower than dust temperature.
The analyses that do not self-consistently consider the effect of dust scattering may misinterpret the observations as optically thin dust emission, potentially leading to underestimates of dust optical depths by an order of magnitude (\citealt{Liu2019ApJ...877L..22L,Zhu2019ApJ...877L..18Z}).

\begin{deluxetable*}{clrclccl}
\tabletypesize{\footnotesize}
\tablecolumns{9}
\tablewidth{0pt}
\tablecaption{New JVLA observations (Project code: 22B-033; PI: C.-Y. Chung)  \label{tab:obs_summary}}
\tablehead{ 
\colhead{Date} & \colhead{Band} & \colhead{Frequency} & \colhead{uv--range} & \colhead{Observed sources} & \colhead{Flux Calib.} & \colhead{Passband Calib.} & \colhead{Gain Calib.} \\
\colhead{(UTC)} & \colhead{} & \colhead{(GHz)}& \colhead{ (k$\lambda$)} & \colhead{} & \colhead{} & \colhead{} & \colhead{} }
\startdata 
2022 09 30 & X & 8.0--12.0 & 1.1--134 & DE Tau, CW Tau, Haro 6-39 & 3C147 & 3C84 & J0403+2600, J0443+3411 \\
2022 10 15 & Ka & 29.0--37.0 &  5.2--350 & DE Tau, BP Tau, CW Tau,  Haro 6-39 & 3C147 & 3C84 & J0403+2600, J0438+3004 \\
2022 11 02 & Ka & 29.0--37.0 &  3.9--370 & DE Tau, BP Tau, CW Tau, Haro 6-39 & 3C147 & 3C84 & J0403+2600, J0438+3004 \\
2022 10 12 & Q & 40.0--48.0 &  4.0--450 & DE Tau, BP Tau, CW Tau, Haro 6-39 & 3C147 & 3C84 & J0403+2600, J0438+3004 \\
\hline
2022 10 03 & X & 8.0--12.0 & 0.7--108 & IP Tau & 3C147 & 3C84 & J0403+2600 \\
2022 10 01 & Ka & 29.2--36.9 & 2.8--384 & IP Tau & 3C147 & 3C84 & J0403+2600\\
2022 10 11 & Ka & 29.0--37.0 &  6.7--350 & IP Tau & 3C147 & 3C84 & J0403+2600 \\
2022 11 13 & Q & 40.0--48.0 &  5.9--525 & IP Tau & 3C147 & 3C84 & J0403+2600 \\
2022 11 18 & Q & 40.1--48.0 &  5.3--490 & IP Tau & 3C147 & 3C84 & J0403+2600 \\
\hline
2022 10 02 & X & 8.0--12.0 & 0.8--119 & BP Tau, DS Tau, V836 Tau & 3C147 & 3C84 & J0403+2600, J0443+3441\\
2022 11 08 & Ka & 29.0--37.0 &  4.1--390 & DS Tau, V836 Tau & 3C147 & 3C84 & J0438+3004, J0431+2037 \\
2022 11 13 & Ka & 29.0--37.0 &  3.5--390 & DS Tau, V836 Tau & 3C147 & 3C84 & J0438+3004, J0431+2037 \\
2022 10 02 & Q & 40.2--47.4 & 0.5--507 & DS Tau, V836 Tau & 3C147 & 3C84 & J0438+3004, J0431+2037 \\
\hline
 2022 10 03 & X & 8.0--12.0 & 1.1--134 & SU Aur & 3C147 & 3C84 & J0443+3441\\
 2022 10 03 & Ka & 29.2--36.9 & 3.1--377 & SU Aur & 3C147 & 3C84 & J0438+3004\\
 2022 10 30 & Ka & 29.2--36.8 &  3.7--380 & SU Aur & 3C147 & 3C84 & J0438+3004 \\
 2022 10 13 & Q & 40.0--48.0 &  5.5--500 & SU Aur & 3C147 & 3C84 & J0438+3004 \\
 2022 11 07 & Q & 40.0--48.0 &  4.3--480 & SU Aur & 3C147 & 3C84 & J0438+3004 \\
\hline
 2022 09 30 & X & 8.0--12.0 & 0.9--126 & (23 targets)* & 3C147 & 3C84 & J0403+2600, J0431+2037, J0443+3441 \\
\hline
\enddata
    \begin{tablenotes}
        \item[a] All observations were taken in the C array configuration.
        \item[b]  * These 23 targets are: CY Tau, V892 Tau, RY Tau, FT Tau, IQ Tau, UZ Tau, DL Tau, AA Tau, DN Tau, CI Tau, T Tau, UX Tau, V710 Tau, DM Tau, LkCa 15, DO Tau, GO Tau, DQ Tau, Haro 6-37, DR Tau, UY Aur, GM Aur, AB Aur.
    \end{tablenotes}
\end{deluxetable*}

In the rings where there are azimuthally asymmetric pressure maxima, such as the vortices induced by Rossby wave instability or planet-disk interactions, the aggregates will migrate towards the vortices and form dusty crescents or clumps (e.g., the numerical simulations presented in \citealt{Li2020ApJ...892L..19L,Wu2024ApJ...970...25W}, and references therein).
When such localized dust substructures reach high dust-to-gas ratios ($>$0.1), a dynamic instability, namely streaming instability (\citealt{Youdin2005ApJ...620..459Y}), can be activated to efficiently collect aggregates into high-density concentrations.
It is likely that pebbles or kilometer-sized planetesimals form in these high-density concentrations. 
Therefore, detecting these high-density dust substructures and determining their masses ($M_{\rm dust}$) and maximum dust grain sizes ($a_{\rm max}$) are crucial to understand the formation of rocky planets or solid cores of gas giants. 
In particular, $M_{\rm dust}$ and $a_{\rm max}$ can only be reliably estimated by observations in the optically thin regime, where flux density is proportional to optical depth, and the frequency variation of dust absorption opacity can be inferred from the measurements of spectral indices. 
Observations at optically thinner wavelengths (e.g. $\lesssim$100 GHz in frequency) are therefore an outstanding science case for the Karl G. Jansky Very Large Array (JVLA), ALMA Band 1 (e.g., \citealt{Di2013arXiv1310.1604D}), the next generation Very Large Array (e.g., \citealt{vanderMarel2018ASPC..517..199V}) and Square Kilometer Array (e.g., \citealt{Wu2024ApJ...965..110W}) in the future. 

Observationally, a spatially resolved example of this scenario is the large transitional disk DM~Tau: the previous ALMA 225 GHz observations revealed a bright and optically thick ring at the inner rim of the bulk of the transitional disk (\citealt{Hashimoto2021ApJ...911....5H}); with lower optical depths, the JVLA Q band (44 GHz; $\sim$6.8 mm) observations further detected multiple localized dust clumps in the ring (\citealt{Liu2024A&A...685A..18L,Wu2024ApJ...970...25W}).
Fitting the spectrum indicated that $a_{\mbox{\scriptsize max}}$ is $\sim$50 $\mu$m in the bulk of the DM~Tau disk, while in the localized dust clumps the lower limit of $a_{\mbox{\scriptsize max}}$ is 300 $\mu$m (\citealt{Liu2024A&A...685A..18L}). 
This is consistent with promoted grain growth in the localized dust substructures. 
Another similar example is the transitional disk LkCa~15 (\citealt{Isella2014ApJ...788..129I}).
Recent observations found lopsided distributions of (grown) dust embedded in a few more disks (e.g., PDS~70: \citealt{Liu2024ApJ...972..163L,Doi2024arXiv240809216D}; LkCa~15: \citealt{Long2022ApJ...937L...1L}; MWC~758: \citealt{Casassus2019MNRAS.483.3278C}; CIDA1: \citealt{Hashimoto2023AJ....166..186H}; WL~17: \citealt{Han2023ApJ...956....9H}, etc), which may signify planet-disk interactions or the presence of vortices (e.g., \citealt{Li2020ApJ...892L..19L,Wu2024ApJ...970...25W}).

\begin{longrotatetable}
\begin{deluxetable*}{ccccccccccc}
\tabletypesize{\footnotesize}
\tablecolumns{9}
\tablewidth{0pt}
\tablecaption{Archival Disk$@$EVLA observations (Project code: AC982; PI: C. Chandler) 
\label{tab:EVLA_obs_summary}}
\tablehead{ 
 \colhead{Date} & \colhead{Band} & \colhead{Frequency} & \colhead{Array Config.} & \colhead{uv--range} & \colhead{Observed sources} & \colhead{Flux Calib.} & \colhead{Passband Calib.} & \colhead{Gain Calib.} \\
 \colhead{(UTC)} & \colhead{} & \colhead{(GHz)} & \colhead{} & \colhead{ (k$\lambda$)} & \colhead{} & \colhead{} & \colhead{} & \colhead{} }
\startdata 
 \multirow{4}{*}{2010 07 11} & \multirow{4}{*}{C} & \multirow{4}{*}{4.8$-$6.8} & \multirow{4}{*}{D} & \multirow{4}{*}{0.5--21.5} & MHO 1/2, CY Tau, V892 Tau, RY Tau, IQ Tau & \multirow{4}{*}{3C147} & \multirow{4}{*}{3C84} & J0403+2600\\
 & & & & & GO Tau, DO Tau, DL Tau, UZ Tau, FT Tau, AA Tau, DN Tau, CI Tau, LkCa 15 & & & \multirow{2}{*}{J0431+2037} \\
 & & & & &  T Tau, UX Tau, V710 Tau, Haro 6-37, DR Tau, DM Tau, DQ Tau & & &  \\
 & & & & & GM Aur, AB Aur, UY Aur & & & J0443+3441 \\
  \hline
 \multirow{3}{*}{2010 08 23} & \multirow{3}{*}{K} & \multirow{3}{*}{20.0$-$22.0} & \multirow{3}{*}{D} & \multirow{3}{*}{1.8--73} & MHO 1/2, CY Tau, V892 Tau, RY Tau & \multirow{3}{*}{3C147} & \multirow{3}{*}{3C84} & J0403+2600 \\
 & & & & & GO Tau, DO Tau, DL Tau, UZ Tau, IQ Tau, FT Tau& & & \multirow{2}{*}{J0431+2037} \\
 & & & & & AA Tau, DN Tau, CI Tau, LkCa 15, T Tau & & &  \\
  \hline
 \multirow{2}{*}{2010 08 24} & \multirow{2}{*}{K} & \multirow{2}{*}{20.0$-$22.0} & \multirow{2}{*}{D} & \multirow{2}{*}{1.8--72} & UX Tau, V710 Tau, DM Tau, Haro 6-37, DQ Tau, DR Tau & \multirow{2}{*}{3C147} & \multirow{2}{*}{3C84} & J0431+2037 \\
 & & & & & GM Aur, AB Aur, UY Aur & & & J0443+3441 \\
 \hline
  \multirow{3}{*}{2010 08 20} & \multirow{3}{*}{Ka} & \multirow{3}{*}{31.7$-$33.6} & \multirow{3}{*}{D} & \multirow{3}{*}{3.6--107} & MHO 1/2, CY Tau, V892 Tau, RY Tau & \multirow{3}{*}{3C147} & \multirow{3}{*}{3C84} & J0403+2600 \\
 & & & & & GO Tau, DO Tau, DL Tau, UZ Tau, FT Tau, IQ Tau & & & \multirow{2}{*}{J0431+2037} \\
 & & & & & DN Tau, AA Tau, CI Tau, LkCa 15, T Tau & & &  \\
   \hline
 \multirow{2}{*}{2010 08 21} & \multirow{2}{*}{Ka} & \multirow{2}{*}{31.7$-$33.6} & \multirow{2}{*}{D} & \multirow{2}{*}{3.3--98} & UX Tau, V710 Tau, DM Tau, Haro 6-37, DQ Tau, DR Tau & \multirow{2}{*}{3C147} & \multirow{2}{*}{3C84} & J0431+2037 \\
 & & & & & GM Aur, AB Aur, UY Aur & & & J0443+3441 \\
 \hline
  \multirow{3}{*}{2010 09 12} & \multirow{3}{*}{Ka} & \multirow{3}{*}{31.7$-$33.6} & \multirow{3}{*}{D} & \multirow{3}{*}{4.1--114} & CY Tau & \multirow{3}{*}{3C147} & \multirow{3}{*}{3C84} & J0403+2600 \\
 & & & & & DO Tau, DL Tau, UZ Tau, IQ Tau, AA Tau & & & \multirow{2}{*}{J0431+2037} \\
 & & & & & FT Tau, DN Tau, CI Tau, DM Tau & & &  \\
 \hline
 \multirow{3}{*}{\tablenotemark{a}2010 08 10} & \multirow{3}{*}{Q} & \multirow{3}{*}{41.0$-$43.0} & \multirow{3}{*}{D} & \multirow{3}{*}{$-$} & MHO 1/2, CY Tau, V892 Tau, RY Tau & \multirow{3}{*}{3C147} & \multirow{3}{*}{3C84} & J0403+2600 \\
 & & & & & GO Tau, DO Tau, DL Tau, UZ Tau, IQ Tau, FT Tau & & & \multirow{2}{*}{J0431+2037} \\
 & & & & & AA Tau, DN Tau, CI Tau, LkCa 15 & & &  \\
  \hline
 \multirow{2}{*}{2010 08 10} & \multirow{2}{*}{Q} & \multirow{2}{*}{41.0$-$43.0} & \multirow{2}{*}{D} & \multirow{2}{*}{5.5--145} & GM Aur, AB Aur, UY Aur & \multirow{2}{*}{3C147} & \multirow{2}{*}{3C84} & J0443+3441 \\
 & & & & & T Tau, UX Tau, V710 Tau, DM Tau, Haro 6-37, DQ Tau, DR Tau & & & J0431+2037 \\
 \hline
 \multirow{3}{*}{2010 08 11} & \multirow{3}{*}{Q} & \multirow{3}{*}{41.0$-$43.0} & \multirow{3}{*}{D} & \multirow{3}{*}{4.6--137} & MHO 1/2, CY Tau, V892 Tau, RY Tau & \multirow{3}{*}{3C147} & \multirow{3}{*}{3C84} & J0403+2600 \\
 & & & & & GO Tau, DO Tau, DL Tau, UZ Tau, IQ Tau, FT Tau& & & \multirow{2}{*}{J0431+2037} \\
 & & & & & AA Tau, DN Tau, CI Tau, LkCa 15 & & &  \\
  \hline
   \multirow{2}{*}{2010 08 12} & \multirow{2}{*}{Q} & \multirow{2}{*}{41.0$-$43.0} & \multirow{2}{*}{D} & \multirow{2}{*}{4.2--134} & T Tau, UX Tau, V710 Tau, DM Tau, Haro 6-37, DQ Tau, DR Tau & \multirow{2}{*}{3C147} & \multirow{2}{*}{3C84} & J0431+2037 \\
 & & & & & GM Aur, AB Aur, UY Aur & & & J0443+3441 \\
 \hline
 \multirow{3}{*}{2010 10 28} & \multirow{3}{*}{Q} & \multirow{3}{*}{41.0$-$43.0} & \multirow{3}{*}{C} & \multirow{3}{*}{11.3--450} & MHO 1/2, CY Tau, V892 Tau, RY Tau & \multirow{3}{*}{3C147} & \multirow{3}{*}{3C84} & J0403+2600 \\
 & & & & & GO Tau, DO Tau, DL Tau, UZ Tau, IQ Tau, FT Tau & & & \multirow{2}{*}{J0431+2037} \\
 & & & & & AA Tau, DN Tau, CI Tau, LkCa 15 & & &  \\
 \hline
 \multirow{3}{*}{2010 08 30} & \multirow{3}{*}{Q} & \multirow{3}{*}{41.0$-$43.0} & \multirow{3}{*}{D} & \multirow{3}{*}{4.1--114} & CY Tau & \multirow{3}{*}{3C147} & \multirow{3}{*}{3C84} & J0403+2600 \\
 & & & & & GO Tau, DO Tau, DL Tau, UZ Tau, IQ Tau, FT Tau & & & \multirow{2}{*}{J0431+2037} \\
 & & & & & AA Tau, DN Tau, CI Tau, LkCa 15, V710 Tau, DM Tau, Haro 6-37, DR Tau & & &  \\
\hline
 \multirow{2}{*}{2010 09 11} & \multirow{2}{*}{K} & \multirow{2}{*}{20.0$-$22.0} & \multirow{2}{*}{D} & \multirow{2}{*}{2.2--74} & V710 Tau, DM Tau, Haro 6-37, DR Tau & \multirow{2}{*}{3C147} & \multirow{2}{*}{3C84} & J0431+2037 \\
 & & & & & GM Aur, AB Aur & & & J0443+3441 \\
\tablenotemark{a}2010 09 10 & Ka & 31.7$-$33.6 & D & $-$ & GM Aur, AB Aur & 3C147 & 3C84 & J0443+3441 \\
\tablenotemark{a}2010 09 13 & Ka & 31.7$-$33.6 & D & $-$ & GM Aur & 3C147 & 3C84 & J0443+3441 \\
 \multirow{2}{*}{2010 11 01} & \multirow{2}{*}{Ka} & \multirow{2}{*}{30.0$-$38.0} & \multirow{2}{*}{C} & \multirow{2}{*}{7.6--422} & T Tau, DQ Tau & \multirow{2}{*}{3C147} & \multirow{2}{*}{3C84} & J0431+2037 \\
 & & & & & UY Aur & & & J0443+3441 \\
2010 11 13 & Ka & 30.0$-$38.0 & C & 8.0--427 & DL Tau & 3C147 & 3C84 & J0431+2037 \\
 2010 11 14 & Ka & 30.0$-$38.0 & C & 8.0--349 & GM Aur & 3C147 & 3C84 & J0443+3441 \\
 2012 03 06 & Ka & 30.0$-$38.0 & C & 7.9--415 & GM Aur & 3C147 & 3C84 & J0439+3045 \\
 2010 09 02 & Q & 41.0$-$43.0 & D & 4.5--127 & GM Aur, AB Aur & 3C147 & 3C84 & J0443+3441 \\
 2010 11 13 & Q & 41.0$-$43.0 & C & 8.0--441 & CY Tau & 3C147 & 3C84 & J0403+2600 \\
 2010 11 15 & Q & 41.0$-$43.0 & C & 8.8--457 & DN Tau & 3C147 & 3C84 & J0431+2037 \\
 2010 11 28 & Q & 41.0$-$43.0 & C & 4.3--454 & DR Tau & 3C147 & 3C84 & J0431+2037 \\
 2010 11 23 & Q & 41.0$-$43.0 & C & 6.1--457 & UX Tau & 3C147 & 3C84 & J0431+2037 \\
 2010 11 19 & Q & 41.0$-$43.0 & C & 5.6--445 & AB Aur & 3C147 & 3C84 & J0443+3441 \\
 2010 12 09 & Q & 41.0$-$43.0 & C & 4.3--456 & Haro 6-37 & 3C147 & 3C84 & J0431+2037 \\
 2010 12 13 & Q & 41.0$-$43.0 & C & 4.9--456 & DM Tau & 3C147 & 3C84 & J0431+2037 \\
\enddata
\tablenotetext{a}{Calibration failed.}
\end{deluxetable*}
\end{longrotatetable}

\begin{figure}
    \hspace{-0.5cm}
    \includegraphics[width=9.2cm]{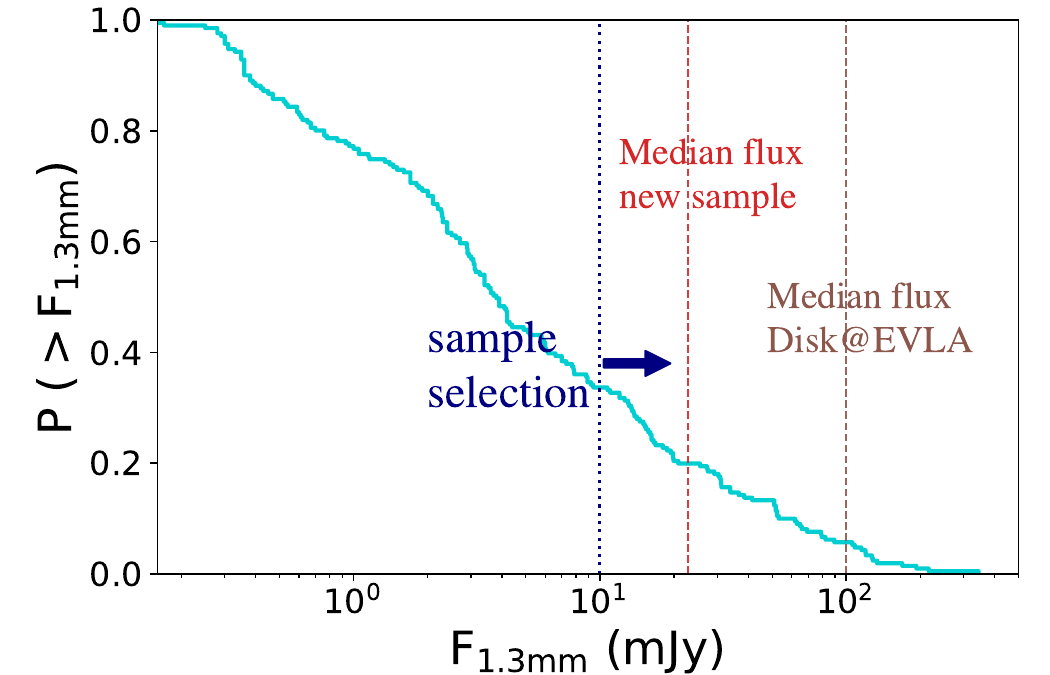}
    \caption{ The cumulative distribution of $F_{\scriptsize \mbox{1.3 mm}}$ of 211 Class II disks in Taurus-Auriga region (quoted from \citealt{Akeson_2019}). The brown and red dashed lines label the median $F_{\scriptsize \mbox{1.3 mm}}$ of the Disk@EVLA sample and the median $F_{\scriptsize \mbox{1.3 mm}}$ the new 8 sources sample, respectively. 
    The blue dotted line label the sample selection criterion on flux density. 
    }
    \label{fig:total_ClassII_230_flux}
\end{figure}

However, the majority of the Class II PPDs have $<$50 au dust disk radii at 1.3 mm wavelength (c.f., \citealt{Long2019ApJ...882...49L,Ciezza2019MNRAS.482..698C,Cazzoletti2019A&A...626A..11C}).
Due to the sensitivity and angular resolution limitations of current observing facilities, we cannot resolve the high-density substructures as clearly as cases like DM~Tau, PDS~70, MWC~758, LkCa~15, and CIDA1.
To test whether or not the aforementioned physical picture is common for Class II disks,  a more practical approach is examining the spectra over a broad frequency range (more below).
Observations at shorter wavelengths 
are sensitive to the emission of small dust particles in the bulk of the PPDs; 
larger (e.g., mm--cm sized) dust particles emit more efficiently at longer wavelengths.
Observations at longer wavelengths (e.g., $\nu<$50 GHz) preferentially traces the grown dust (e.g., $a_{\rm max} >$1 mm) that is concentrated in the high column density substructures (\citealt{Hildebrand1983QJRAS..24..267H,Draine2006ApJ...636.1114D}).

In a recent study, \citet{Chung2024ApJS..273...29C} presented a survey of 47 Class II protoplanetary disks (PPDs) in the Taurus-Auriga region, focusing on sources with high 850 $\mu$m fluxes (top 32\% percentile). 
They found that the spectral indices in the 200--400 GHz frequency range (hereafter $\alpha_{\rm 200-400GHz}$) for the selected sources were concentrated within an extremely narrow range of 2.0$\pm$0.2, which is consistent with the dust thermal emission in the bulk of the optically thick dust rings. 
To probe the embedded, high-density (grown) dust substructures in these disks, longer wavelengths observations are demanded to penetrate into the disk mid-planes. 
However, considering the relatively long observing time required for achieving adequate signal-to-noise ratios at frequencies below 50 GHz, in this work, we focus on the bright sub-sample selected from  \citet{Chung2024ApJS..273...29C} and present their $\sim$4--200 GHz spectra. 

The target source selections and observations are introduced in Section \ref{sec:observation}.
Details of our data processings are described in Section \ref{sec:reduction}.
Specifically, we included the JVLA observations at 4--50 GHz frequencies which were taken through our own project (22B-033; PI: C.-Y. Chung) and through the archival Disk@EVLA observations (AC982; PI: C. Chandler). 
We also quoted various previous observations in between 4 GHz and 400 GHz.
The results are summarized in Section \ref{sec:result}.
The physical implications of our observations are briefly discussed in Section \ref{sec:discussions}.
Our conclusion is given in Section \ref{sec:conclusion}.
In Appendix \ref{appendix:SED_fitting}, we characterize the observed spectra with piecewise functions that are independent of specific physical models.

\section{Observations}\label{sec:observation}
\subsection{Sample selection}\label{sub:source}

We chose to analyze a subset of bright sources from the SMA Taurus-Auriga survey (\citealt{Chung2024ApJS..273...29C}; Figure \ref{fig:total_ClassII_230_flux}). 
A total of 32 Class II objects were selected.

\begin{deluxetable*}{ccccccccccccc}[h]
\tabletypesize{\footnotesize}
\tablecolumns{13}
\tablewidth{0pt}
\tablecaption{Imaging profiles \label{tab:img_parameter}}
\tablehead{ 
 \colhead{Source} & \colhead{RA} & \colhead{Dec} & \colhead{$\theta_{\mbox{\tiny 6.8 GHz}}$} & \colhead{$\theta_{\mbox{\tiny 10.0 GHz}}$} &  \colhead{$\theta_{\mbox{\tiny 21.0 GHz}}$} & \colhead{$\theta_{\mbox{\tiny 33.0 GHz}}$} & 
 \colhead{$\theta_{\mbox{\tiny 44.0 GHz}}$} & \colhead{$\sigma_{\mbox{\tiny 6.8 GHz}}$} & \colhead{$\sigma_{\mbox{\tiny 10.0 GHz}}$} & \colhead{$\sigma_{\mbox{\tiny 21.0 GHz}}$} & \colhead{$\sigma_{\mbox{\tiny 33.0 GHz}}$} & \colhead{$\sigma_{\mbox{\tiny 44.0 GHz}}$} \\
 \colhead{} & \colhead{(hh:mm:ss)} & \colhead{(dd$^{\circ}$mm$'$ss$''$)} & \colhead{($''$)} & \colhead{($''$)} & \colhead{($''$)} & \colhead{($''$)} & \colhead{($''$)} & \colhead{(mJy)} & \colhead{(mJy)} & \colhead{(mJy)} & \colhead{(mJy)} & \colhead{(mJy)} }
\startdata
AA Tau & 04:34:55.43 & $24^{\circ}28'52\farcs70$ & $15.4\times13.8$ & $2.9\times2.4$ & $4.5\times3.5$ & $2.9\times2.2$ & $2.7\times1.7$ & 0.012 & 0.034 & 0.06 & 0.03 & 0.10\\
AB Aur & 04:55:45.85 & $30^{\circ}33'03\farcs91$ & $15.9\times14.1$ & $3.0\times2.4$ & $3.7\times3.4$ & $2.8\times2.4$ & $1.0\times0.6$ & 0.030 & 0.033 & 0.03 & 0.05 & 0.03\\
BP Tau & 04:19:15.84 & $29^{\circ}06'26\farcs51$ & $-$ & $3.2\times2.5$ & $-$ & $1.0\times0.7$ & $1.0\times0.6$ & $-$ & 0.025 & $-$ & 0.02 & 0.11\\
CI Tau & 04:33:52.02 & $22^{\circ}50'29\farcs82$ & $15.3\times13.2$ & $3.2\times2.6$ & $4.3\times3.4$ & $3.1\times2.2$ & $1.9\times1.4$ & 0.012 & 0.036 & 0.05 & 0.06 & 0.16\\
CW Tau & 04:14:17.01 & $28^{\circ}10'57\farcs38$ & $-$ & $2.4\times2.3$ & $-$ & $1.2\times0.7$ & $1.2\times0.6$ & $-$ & 0.029 & $-$ & 0.04 & 0.13\\
CY Tau & 04:17:33.74 & $28^{\circ}20'46\farcs40$ & $18.8\times14.7$ & $3.0\times2.4$ & $5.7\times3.5$ & $2.9\times2.6$ & $0.9\times0.6$ & 0.055 & 0.048 & 0.10 & 0.08 & 0.03\\
DE Tau & 04:21:55.65 & $27^{\circ}55'05\farcs75$ & $-$ & $2.5\times2.3$ & $-$ & $1.1\times0.7$ & $1.0\times0.6$ & $-$ & 0.011 & $-$ & 0.02 & 0.08\\
DL Tau & 04:33:39.09 & $25^{\circ}20'37\farcs81$ & $16.1\times14.4$ & $3.1\times2.5$ & $5.5\times3.5$ & $0.7\times0.7$ & $2.6\times1.8$ & 0.027 & 0.040 & 0.11 & 0.02 & 0.19\\
DM Tau & 04:33:48.75 & $18^{\circ}10'09\farcs68$ & $15.4\times13.9$ & $3.3\times2.4$ & $4.2\times3.6$ & $2.9\times2.1$ & $0.9\times0.7$ & 0.013 & 0.033 & 0.01 & 0.03 & 0.02\\
DN Tau & 04:35:27.38 & $24^{\circ}14'58\farcs58$ & $15.6\times13.6$ & $3.1\times2.6$ & $4.3\times3.4$ & $2.9\times2.4$ & $0.7\times0.6$ & 0.015 & 0.034 & 0.07 & 0.07 & 0.03\\
DO Tau & 04:38:28.60 & $26^{\circ}10'49\farcs13$ & $17.9\times14.4$ & $3.0\times2.4$ & $5.6\times3.5$ & $3.1\times2.7$ & $2.3\times1.5$ & 0.043 & 0.035 & 0.15 & 0.10 & 0.16\\
DQ Tau & 04:46:53.06 & $16^{\circ}59'59\farcs92$ & $15.8\times14.0$ & $3.2\times2.4$ & $4.1\times3.4$ & $1.6\times0.9$ & $2.3\times1.8$ & 0.037 & 0.032 & 0.04 & 0.55 & 0.11\\
DR Tau & 04:47:06.22 & $16^{\circ}58'42\farcs59$ & $15.3\times13.2$ & $3.2\times2.4$ & $3.9\times3.5$ & $3.1\times2.4$ & $1.0\times0.6$ & 0.016 & 0.034 & 0.01 & 0.04 & 0.03\\
DS Tau & 04:47:48.60 & $29^{\circ}25'10\farcs80$ & $-$ & $3.0\times2.5$ & $-$ & $1.1\times0.7$ & $0.6\times0.5$ & $-$ & 0.015 & $-$ & 0.01 & 0.05\\
FT Tau & 04:23:39.20 & $24^{\circ}56'13\farcs90$ & $15.9\times14.1$ & $3.1\times2.5$ & $5.7\times3.3$ & $2.8\times2.1$ & $2.2\times1.7$ & 0.025 & 0.035 & 0.14 & 0.11 & 0.12\\
GM Aur & 04:55:10.99 & $30^{\circ}21'58\farcs99$ & $15.4\times14.1$ & $3.0\times2.4$ & $3.8\times3.4$ & $0.9\times0.8$ & $2.6\times1.9$ & 0.021 & 0.035 & 0.01 & 0.01 & 0.11\\
GO Tau & 04:43:03.08 & $25^{\circ}20'18\farcs39$ & $16.6\times14.4$ & $3.0\times2.4$ & $6.0\times3.5$ & $3.4\times2.8$ & $2.6\times1.7$ & 0.018 & 0.035 & 0.06 & 0.08 & 0.11\\
Haro 6-37 & 04:46:59.09 & $17^{\circ}02'39\farcs73$ & $15.6\times13.0$ & $3.2\times2.4$ & $4.2\times3.5$ & $3.1\times2.5$ & $1.2\times0.9$ & 0.074 & 0.024 & 0.02 & 0.04 & 0.03\\
Haro 6-39 & 04:52:09.71 & $30^{\circ}37'45\farcs02$ & $-$ & $2.4\times2.2$ & $-$ & $1.2\times0.7$ & $1.0\times0.6$ & $-$ & 0.008 & $-$ & 0.02 & 0.09\\
IP Tau & 04:24:57.09 & $27^{\circ}11'56\farcs12$ & $-$ & $3.9\times2.6$ & $-$ & $1.3\times0.8$ & $0.7\times0.6$ & $-$ & 0.010 & $-$ & 0.01 & 0.04\\
IQ Tau & 04:29:51.56 & $26^{\circ}06'44\farcs52$ & $19.7\times15.1$ & $3.0\times2.5$ & $5.0\times3.5$ & $3.1\times2.3$ & $2.5\times1.7$ & 0.022 & 0.033 & 0.05 & 0.03 & 0.09\\
LkCa 15 & 04:39:17.80 & $22^{\circ}21'03\farcs11$ & $15.0\times12.0$ & $3.1\times2.4$ & $3.9\times3.4$ & $3.0\times2.3$ & $2.4\times1.8$ & 0.008 & 0.034 & 0.03 & 0.05 & 0.13\\
MHO 1/2 & 04:14:26.28 & $28^{\circ}06'02\farcs85$ & $18.9\times14.8$ & $-$ & $5.7\times3.5$ & $2.9\times2.7$ & $2.1\times1.4$ & 0.026 & $-$ & 0.14 & 0.15 & 0.18\\
RY Tau & 04:21:57.42 & $28^{\circ}26'35\farcs12$ & $19.4\times15.2$ & $3.0\times2.4$ & $5.5\times3.5$ & $2.9\times2.7$ & $2.1\times1.5$ & 0.041 & 0.035 & 0.15 & 0.15 & 0.17\\
SU Aur & 04:55:59.39 & $30^{\circ}34'01\farcs11$ & $-$ & $2.5\times2.2$ & $-$ & $0.9\times0.7$ & $0.7\times0.6$ & $-$ & 0.012 & $-$ & 0.01 & 0.03\\
T Tau & 04:21:59.45 & $19^{\circ}32'06\farcs20$ & $16.5\times11.8$ & $3.2\times2.4$ & $3.8\times3.3$ & $1.8\times1.1$ & $2.3\times1.8$ & 0.026 & 0.039 & 0.22 & 0.14 & 0.29\\
UX Tau & 04:30:04.01 & $18^{\circ}13'49\farcs19$ & $15.5\times12.0$ & $3.2\times2.4$ & $6.7\times3.4$ & $3.3\times2.8$ & $1.2\times0.8$ & 0.015 & 0.026 & 0.04 & 0.04 & 0.03\\
UY Aur & 04:51:47.40 & $30^{\circ}47'13\farcs14$ & $17.2\times14.2$ & $3.0\times2.4$ & $3.6\times3.4$ & $1.3\times0.9$ & $2.3\times1.8$ & 0.028 & 0.011 & 0.05 & 0.02 & 0.12\\
UZ Tau & 04:32:43.08 & $25^{\circ}52'30\farcs66$ & $17.4\times14.2$ & $2.9\times2.4$ & $5.4\times3.5$ & $3.2\times2.7$ & $1.9\times1.3$ & 0.029 & 0.034 & 0.13 & 0.11 & 0.20\\
V710 Tau & 04:31:57.81 & $18^{\circ}21'37\farcs67$ & $15.4\times12.4$ & $3.2\times2.4$ & $4.8\times3.5$ & $3.1\times2.7$ & $2.3\times1.8$ & 0.009 & 0.023 & 0.02 & 0.03 & 0.08\\
V836 Tau & 05:03:06.60 & $25^{\circ}23'19\farcs33$ & $-$ & $3.1\times2.5$ & $-$ & $1.0\times0.7$ & $0.6\times0.5$ & $-$ & 0.082 & $-$ & 0.02 & 0.11\\
V892 Tau & 04:18:40.62 & $28^{\circ}19'15\farcs16$ & $18.6\times14.8$ & $3.1\times2.5$ & $5.7\times3.5$ & $2.9\times2.7$ & $2.1\times1.5$ & 0.032 & 0.039 & 0.18 & 0.15 & 0.18\\
\enddata
\tablecomments{ Column 1: Source name. Column 2, 3: J2000 coordinates. Column 4, 5, 6, 7, 8: FWHM of the synthesized beam of the C band (6.8 GHz), X band (10.0 GHz), K band (21.0 GHz), Ka band (33.0 GHz) and Q band (44.0 GHz) observations. Column 9, 10, 11, 12, 13: The achived rms noise of the C band, X band, K band, Ka band and Q band observations.}
\end{deluxetable*}

Out of these, 24 sources that are relatively bright at 1.2 mm wavelength were observed by the Disk@EVLA project (VLA project code: AC982) at C (4.8--6.8 GHz), K (20.0--22.0 GHz), Ka (31.7--33.6 GHz), and Q (41.0--43.0 GHz) bands (more in Section \ref{sub:obsDiskatEVLA}).
We utilized the Disk@EVLA data taken in the D array configuration and C array configuration, which are not subject to missing short-spacing. 
We also carried out only new X band (8--12 GHz) observations for 23\footnote{We missed to select MHO~1/2 due to the different naming convention in SMA (\citealt{Chung2024ApJS..273...29C}) and Disk$@$EVLA surveys. Nevertheless, MHO~1/2 was detected in the C band observations of the Disk@EVLA project, and in the $<$10 GHz observations of \citet{Dzib2015ApJ...801...91D} (Figure \ref{fig:SED_1}).}
of these 24 sources (see Section \ref{sub:obs}). 

According to recent ALMA surveys (e.g., \citealt{Akeson_2019}), the Disk@EVLA samples are considerably brighter at 1.3 mm band compared to typical Class II disks (see Figure \ref{fig:total_ClassII_230_flux}).
Such a target source selection strategy was likely due to the limited, 2 GHz available continuum bandwidth during the Disk@EVLA observations.
We took advantage of the presently available, 8 GHz continuum bandwidth at Ka and Q bands to observe another 8 objects that are fainter at 1.2 mm wavelength (Figure \ref{fig:total_ClassII_230_flux}). 
When selecting these 8 sources, we (i) required their flux densities at 1.3 mm wavelength ($F_{\scriptsize \mbox{1.3 mm}}$) to be larger than 10 mJy to ensure a $\gtrsim 5$-$\sigma$ significance in every 8 GHz bandwidth in our new Ka and Q band observations (see Section \ref{sub:obs}), and (ii) avoided the known binary or multiple system.

\subsection{New JVLA observations (JVLA/22B-033)}\label{sub:obs}

We carried out new JVLA X band (8.0--12.0 GHz) observations for 31 of the 32 selected sources (Section \ref{sub:source}).
In addition, we carried out Ka (29.0--37.0 GHz) and Q (40.0--48.0 GHz) observations for the 8 objects that were not previously observed by the Disk@EVLA project. 
Our JVLA observations took place between September and November 2022, with observing details summarized in Table \ref{tab:obs_summary}. 

We employed the standard 3-bit continuum setups for the Q, Ka, and X bands observations. 
In Q and Ka band observations, we employed the WIDAR correlator that was configured with four baseband pairs observing dual polarization; each baseband covers a 2 GHz intermediate frequency (IF) range, and the four basebands were tuned to separated frequencies to provide 8 GHz aggregate bandwidth.
In X band observations, the WIDAR correlator was configured with two baseband pairs with the IF coverages to provide 4 GHz aggregate bandwidth for dual polarization.

The observations were conducted in the C array configuration. 
In these new JVLA observations, the achieved synthesized beam full width at half maxima (FWHM) are 0\farcs71 ($\sim$99.4 AU), 0\farcs95 ($\sim$133 AU), 3\farcs14 ($\sim$440 AU) at Q, Ka, and X bands, respectively. 
Our observations are not affected by missing short-spacing as the known disk sizes fall well within the maximum recoverable scale provided by the {\it $uv$} coverage (Table \ref{tab:obs_summary}). 

\begin{figure*}
    \hspace{-1.8cm} 
        \begin{tabular}{ lll }
        \includegraphics[width=6.5cm]{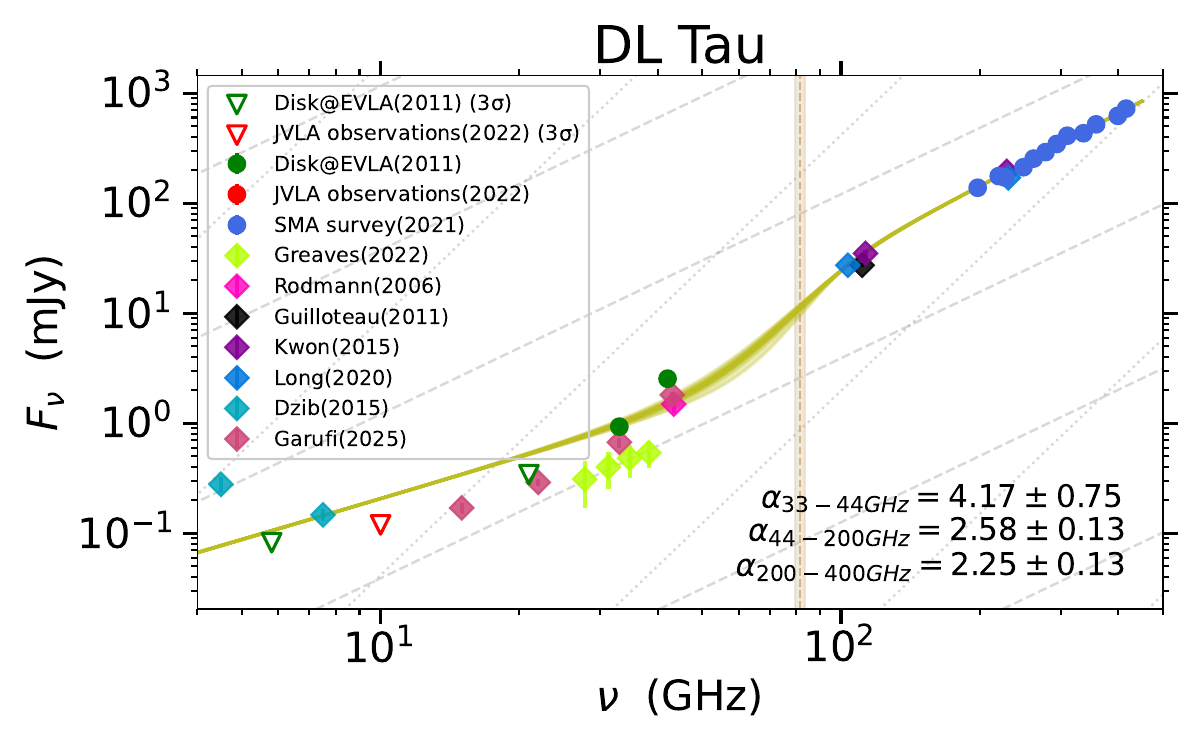}&
        \hspace{-0.5cm}
        \includegraphics[width=6.5cm]{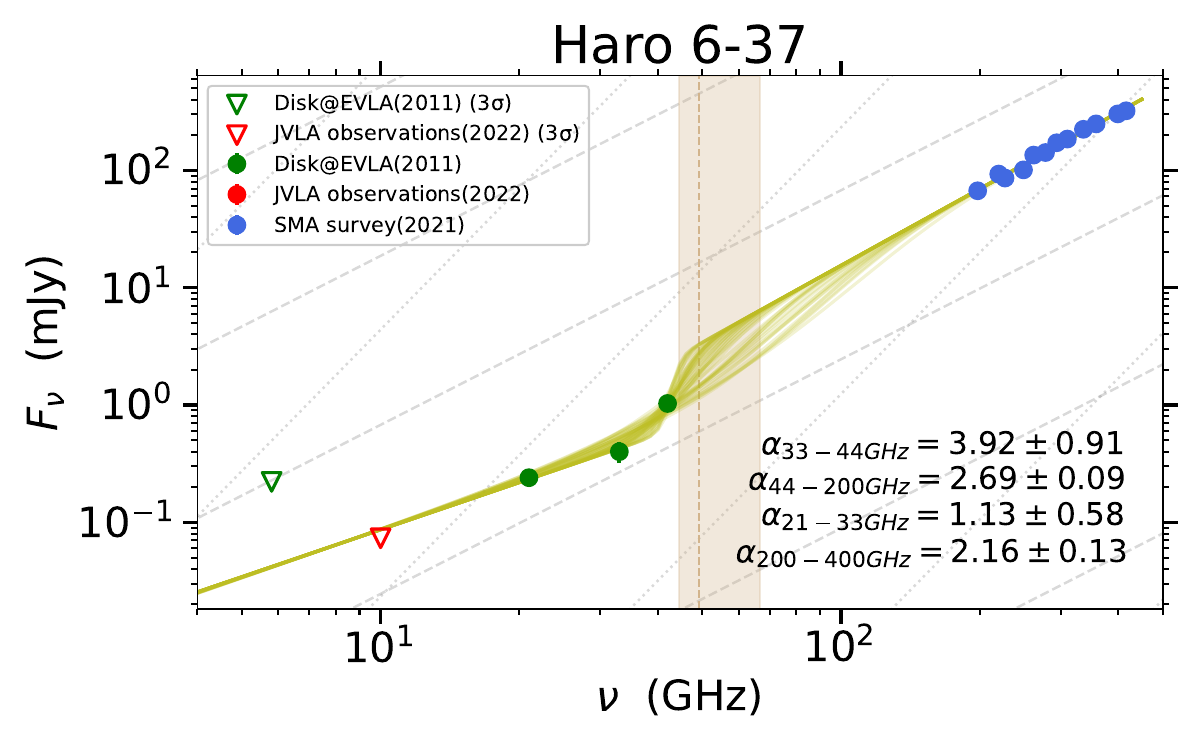}&
        \hspace{-0.5cm}
        \includegraphics[width=6.5cm]{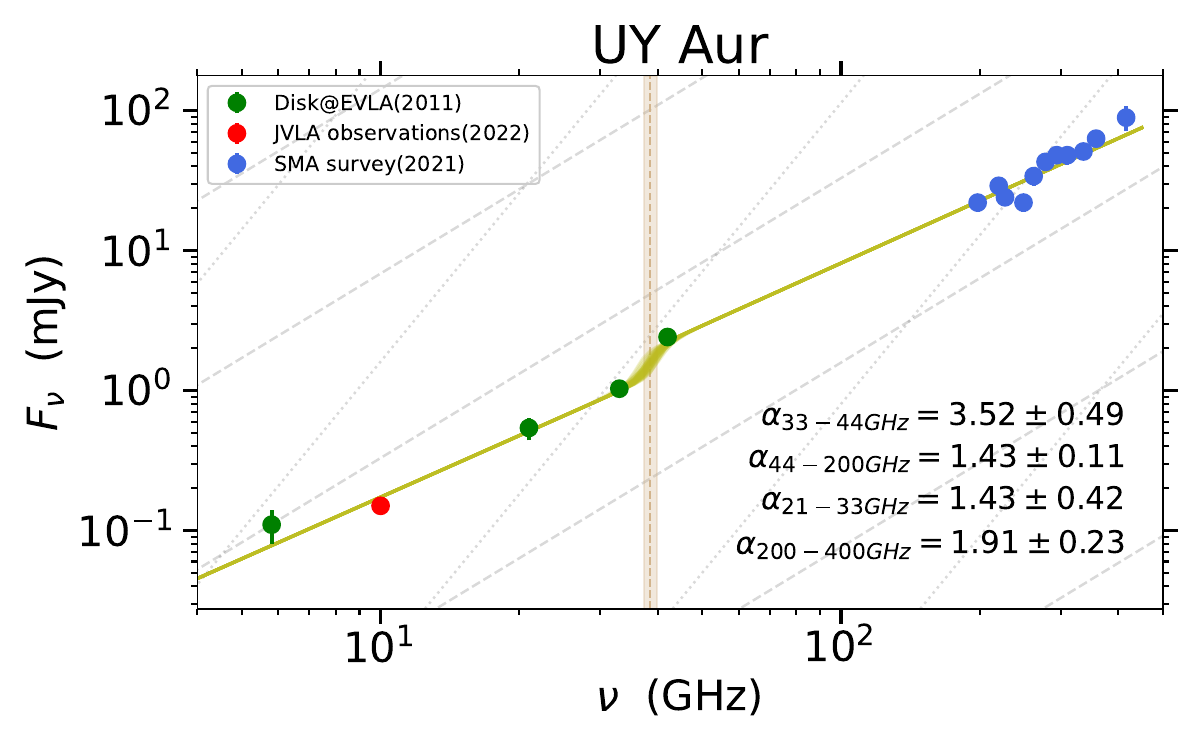}\\
        \includegraphics[width=6.5cm]{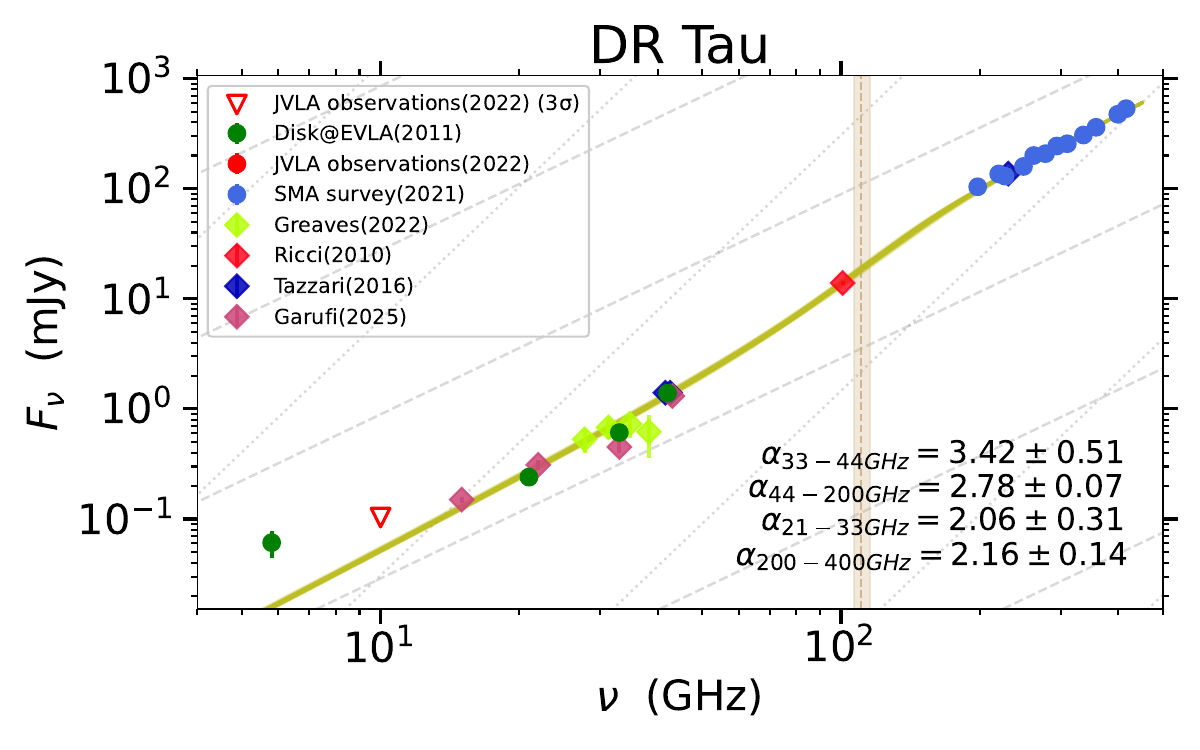}& 
        \hspace{-0.5cm}
        \includegraphics[width=6.5cm]{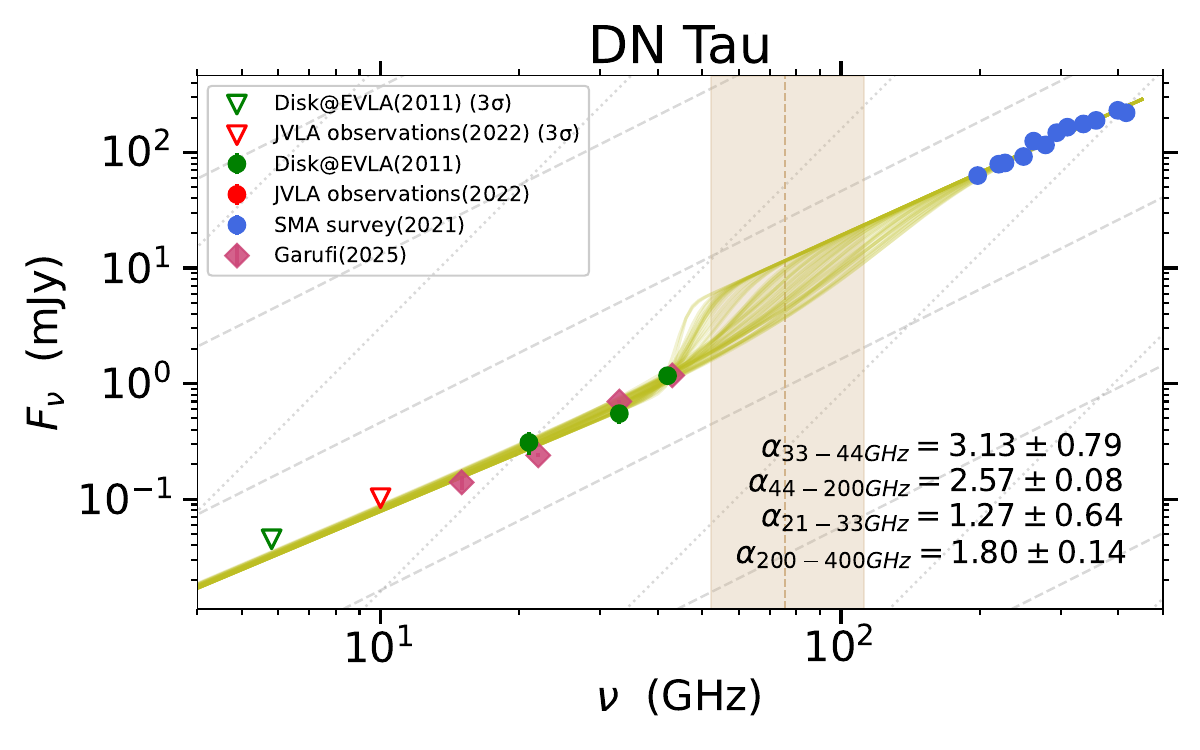}&
        \hspace{-0.5cm}
        \includegraphics[width=6.5cm]{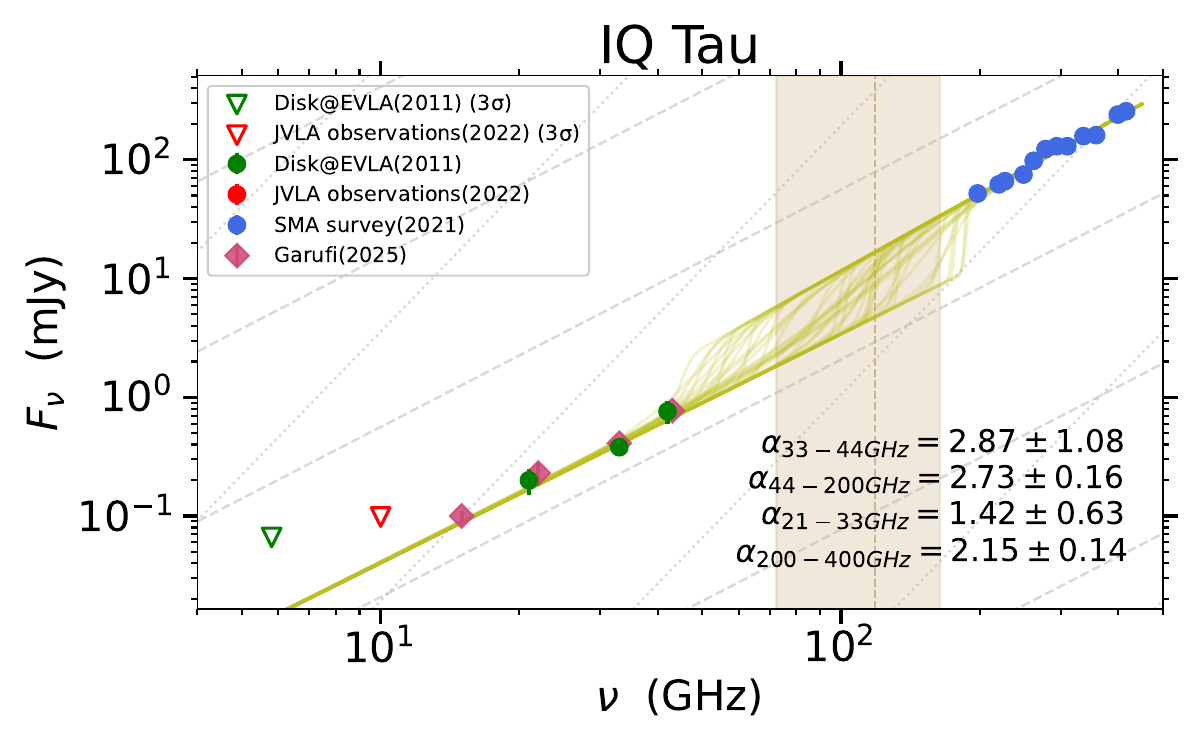}\\
        \includegraphics[width=6.5cm]{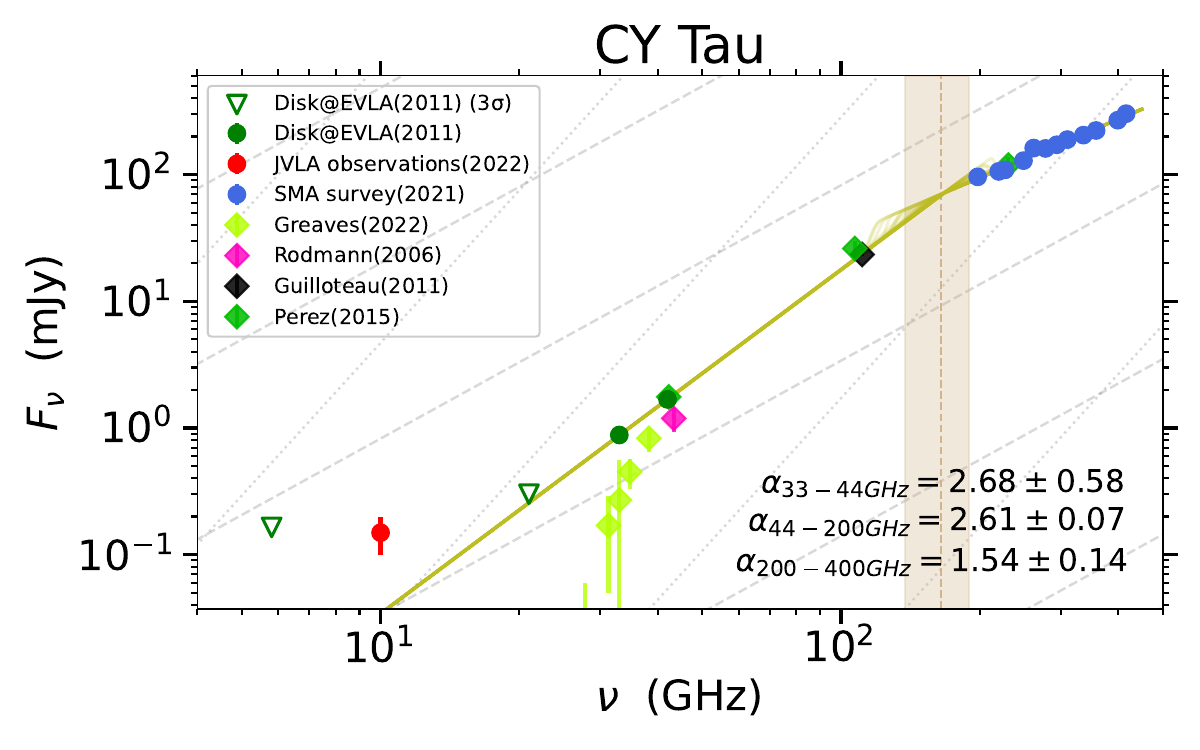}&
        \hspace{-0.5cm}
        \includegraphics[width=6.5cm]{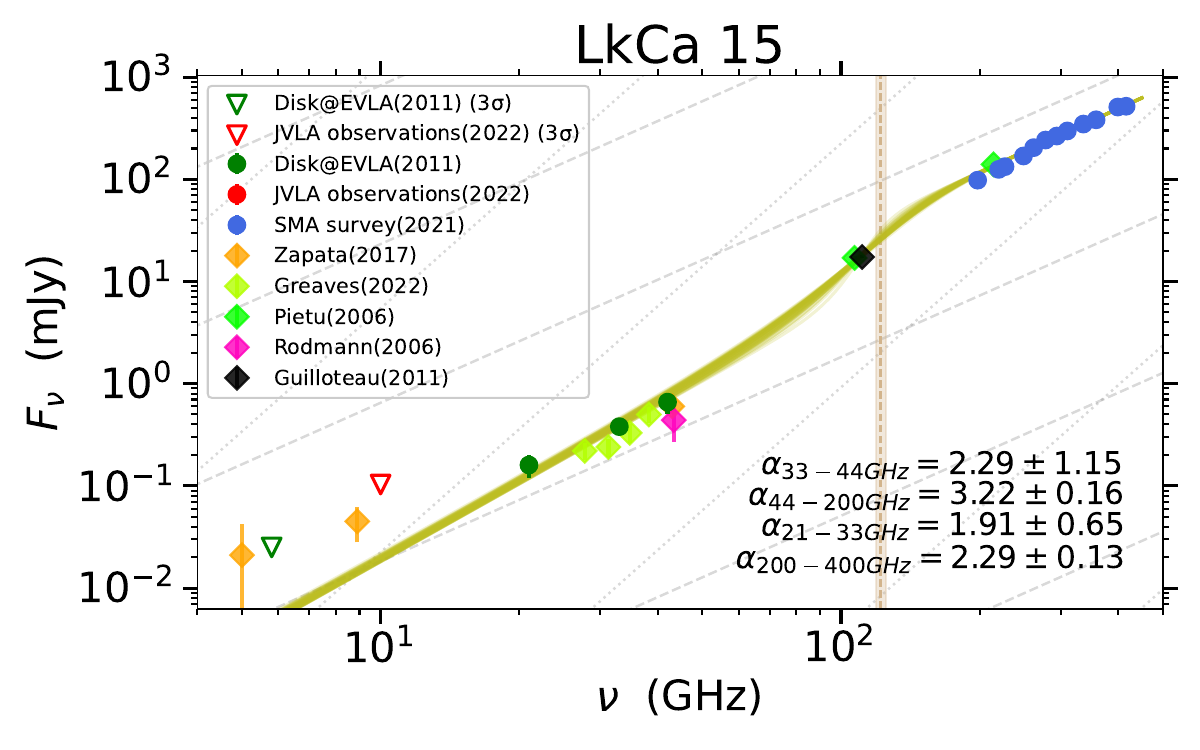}&
        \hspace{-0.5cm}
        \includegraphics[width=6.5cm]{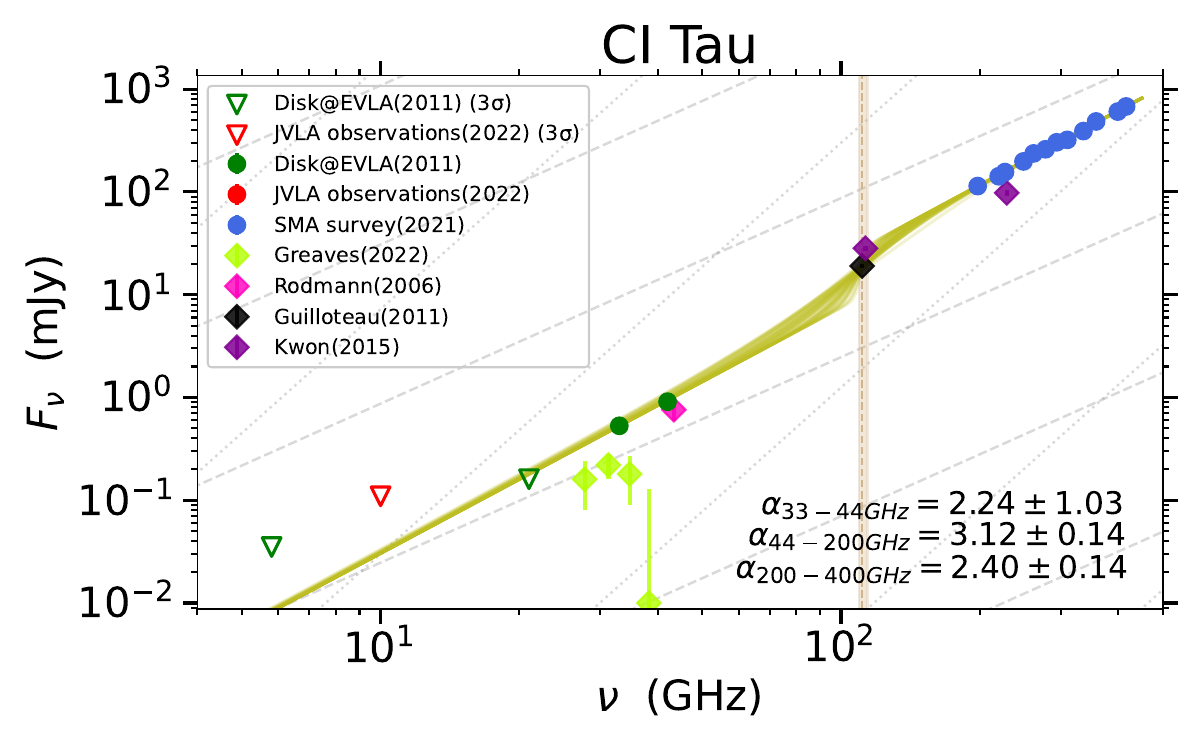}\\
        \includegraphics[width=6.5cm]{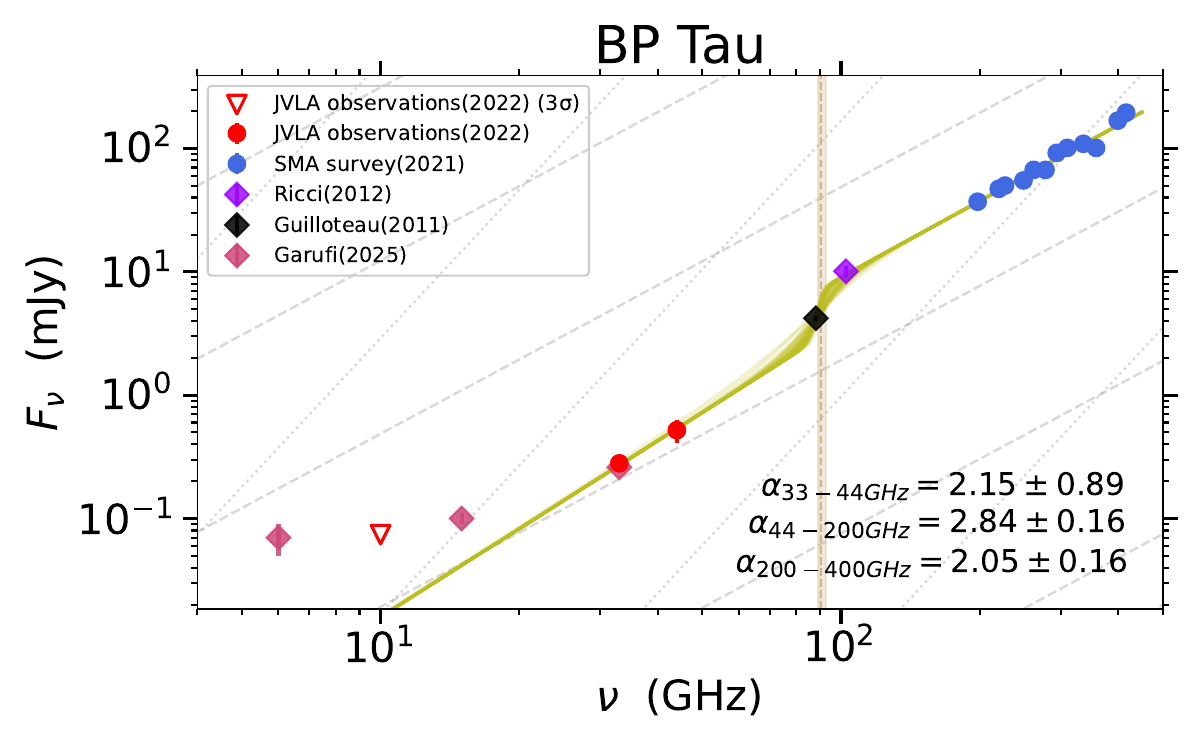}&
        \hspace{-0.5cm}
        \includegraphics[width=6.5cm]{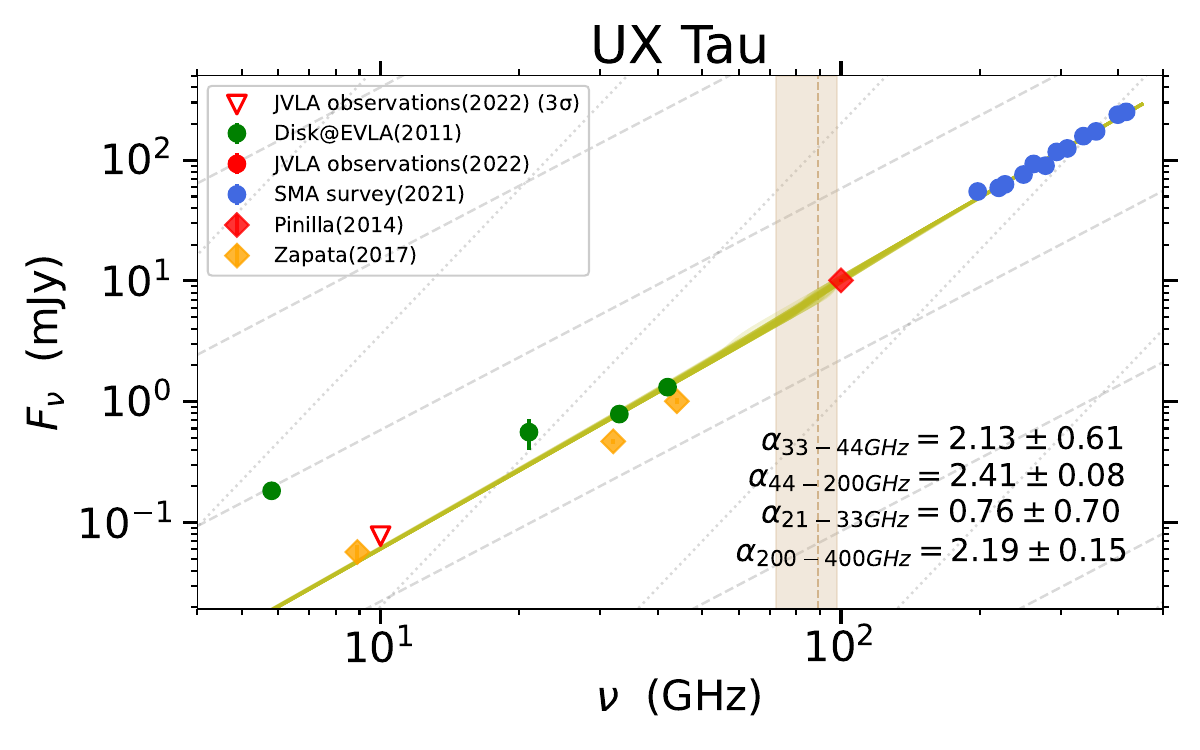}&
        \hspace{-0.5cm}
        \includegraphics[width=6.5cm]{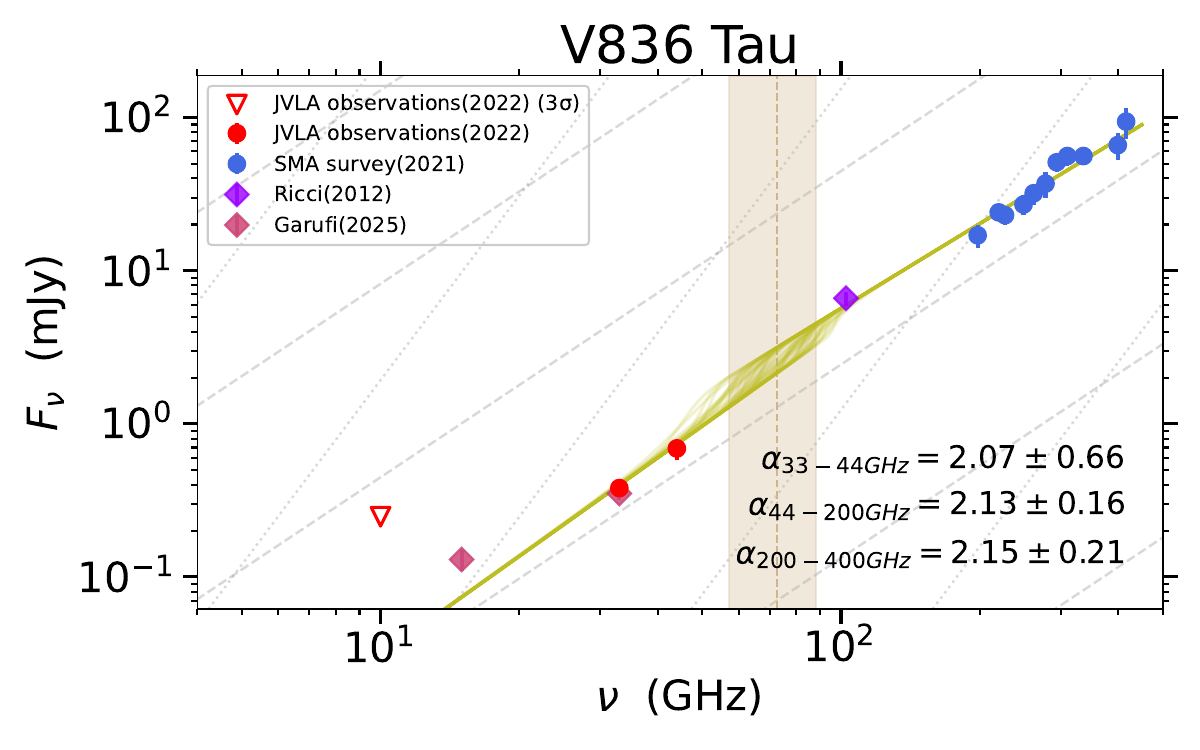}\\
        \end{tabular}
    \caption{
    The 4--400 GHz spectra of our sample of 32 Class II protoplanetary disks in the Taurus-Auriga region.  
    The flux densities in the Disk$@$EVLA observations (Section \ref{sub:obsDiskatEVLA}), the SMA survey (\citealt{Chung2024ApJS..273...29C}), and the new JVLA observations (Section \ref{sub:obs}) are shown by the green, blue and red filled points, respectively. 
    The 3-$\sigma$ upper limit of flux densities in the Disk$@$EVLA and the new JVLA observations are shown by green and red inverted triangles. 
    We compiled and quoted some flux densities measurements published over 2004--2025 (\citealt{Acke2004A&A...422..621A,Pietu2006A&A...460L..43P,Rodmann2006A&A...446..211R,Ricci2010A&A...512A..15R,Guilloteau2011A&A...529A.105G,Ricci2012A&A...540A...6R,Pinilla2014A&A...564A..51P,Rodriguez2014ApJ...793L..21R,Dzib2015ApJ...801...91D,Perez2015ApJ...813...41P,Tazzari2016A&A...588A..53T,Zapata2017ApJ...834..138Z,Tripathi2018ApJ...861...64T,Huang2020ApJ...891...48H,Long2020ApJ...898...36L,Greaves2022MNRAS.513.3180G,Ueda2022ApJ...930...56U,Garufi2025arXiv250111686G}). 
    The 32 sources are arranged in the descending order of $\alpha_{\mbox{\scriptsize 33-44 GHz}}$. 
    The $\alpha_{\mbox{\scriptsize 33-44 GHz}}$, $\alpha_{\mbox{\scriptsize 44-200 GHz}}$ and $\alpha_{\mbox{\scriptsize 21-33 GHz}}$ were calculated between two adjacent bands and shown in the lower right of each panel; the $\alpha_{\mbox{\scriptsize 200-400GHz}}$ was quoted from \citet{Chung2024ApJS..273...29C}. 
    The olive semitransparent lines are 50 random draws of the fitting parameters extracted from the MCMC samplings.    
    The best fit value of the parameter ${b/\omega}$ (Appendix \ref{appendix:SED_fitting}), with its uncertainty, are shown by the brown vertical line and shaded area. 
    The spectra are overlaid with the gray dashed and dotted lines, which indicate the spectral indices of $\alpha = 2.0$ and $\alpha = 4.0$, respectively. 
    }
    \label{fig:SED_0}
\end{figure*}

\begin{figure*}
    \hspace{-1.8cm} 
        \begin{tabular}{ lll }
        \includegraphics[width=6.5cm]{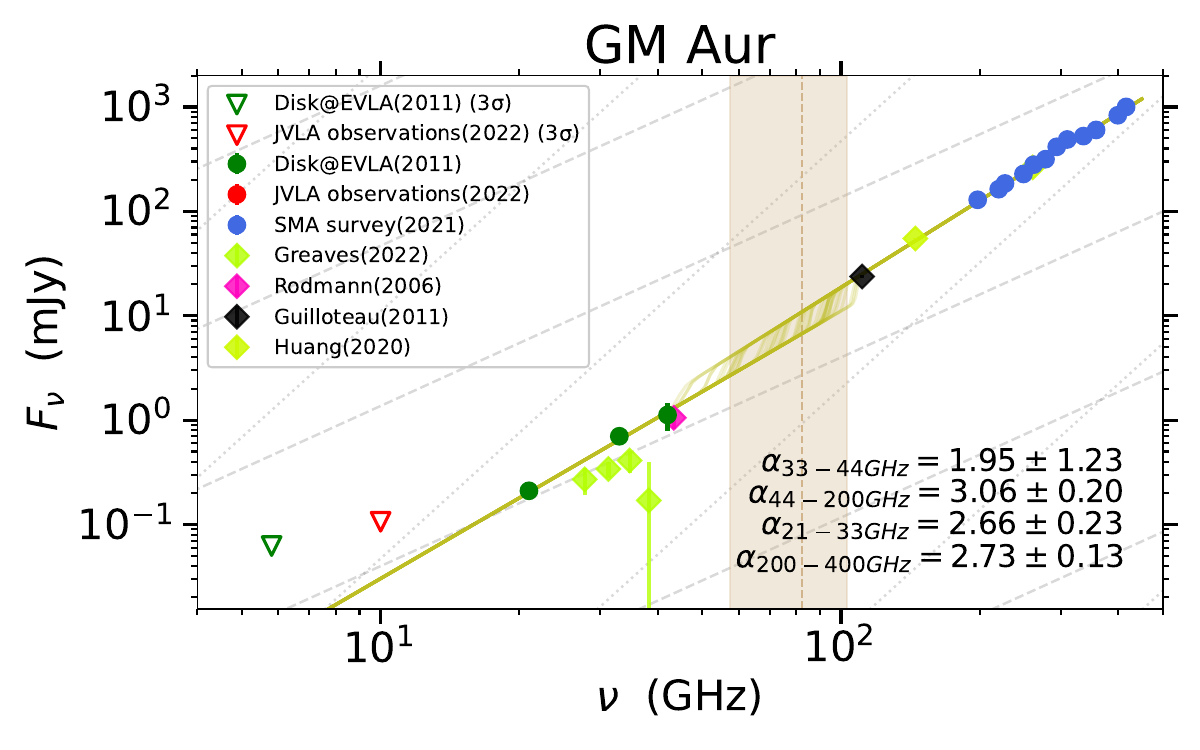}&
        \hspace{-0.5cm}
        \includegraphics[width=6.5cm]{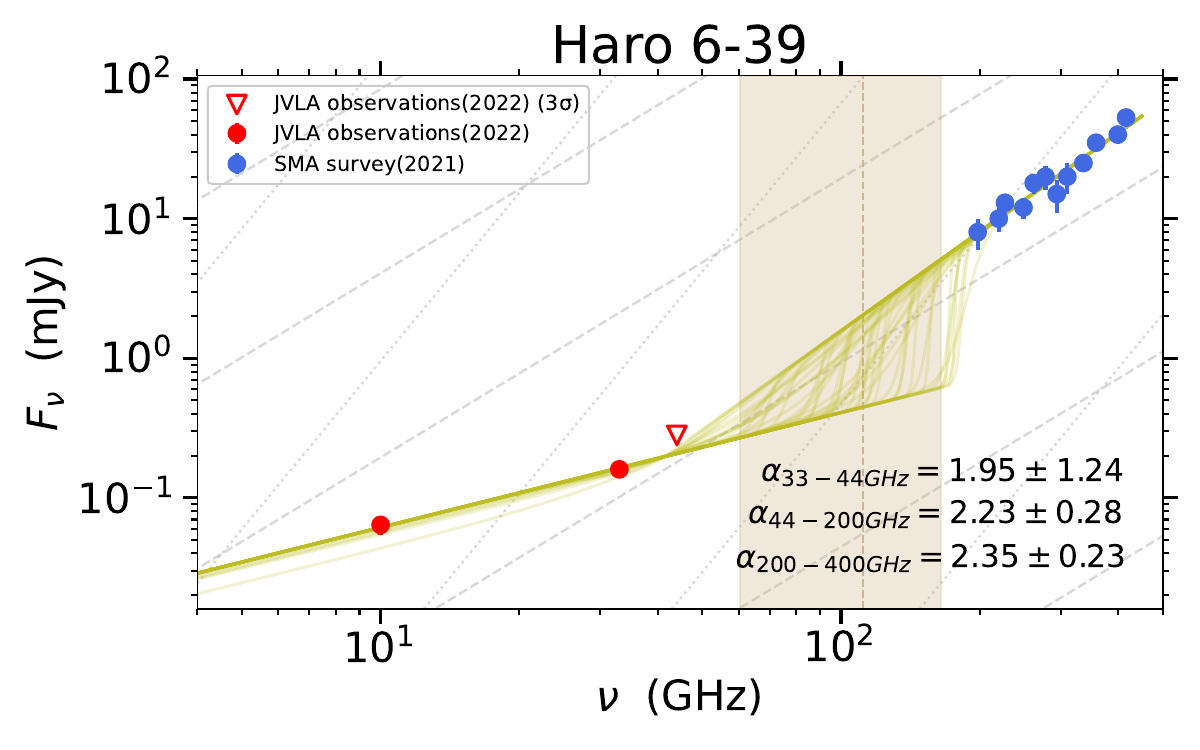}&
        \hspace{-0.5cm}
        \includegraphics[width=6.5cm]{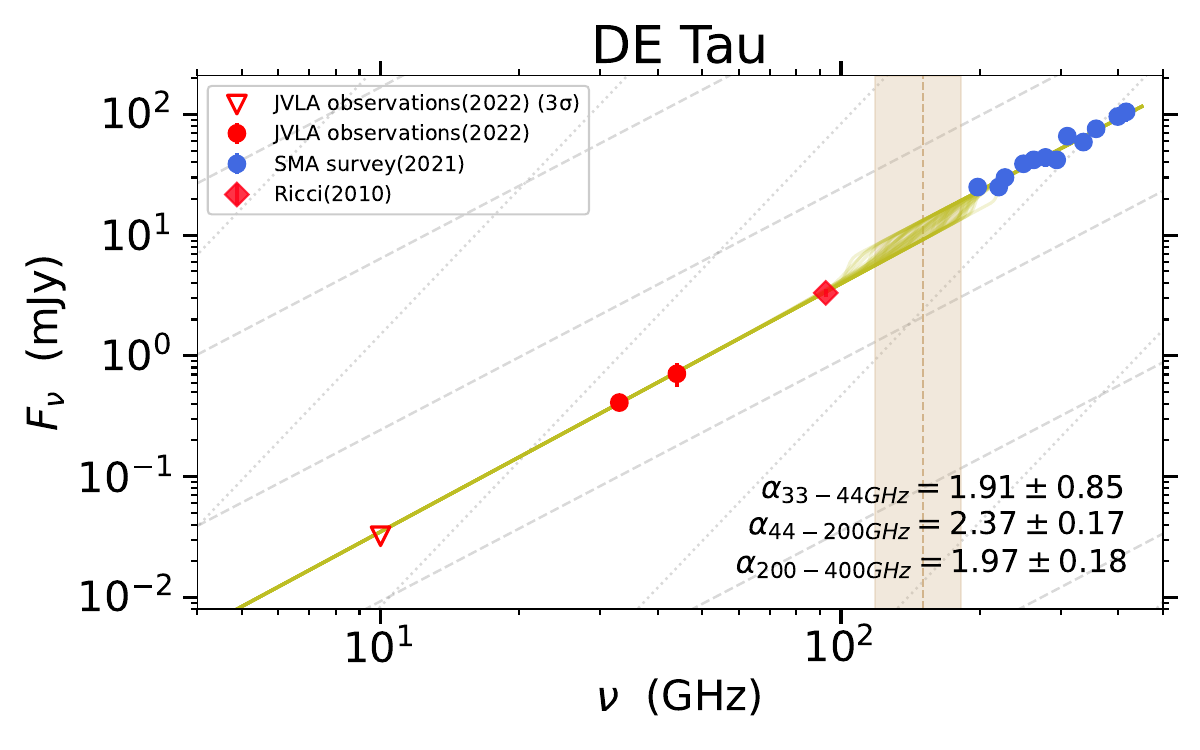}\\
        \includegraphics[width=6.5cm]{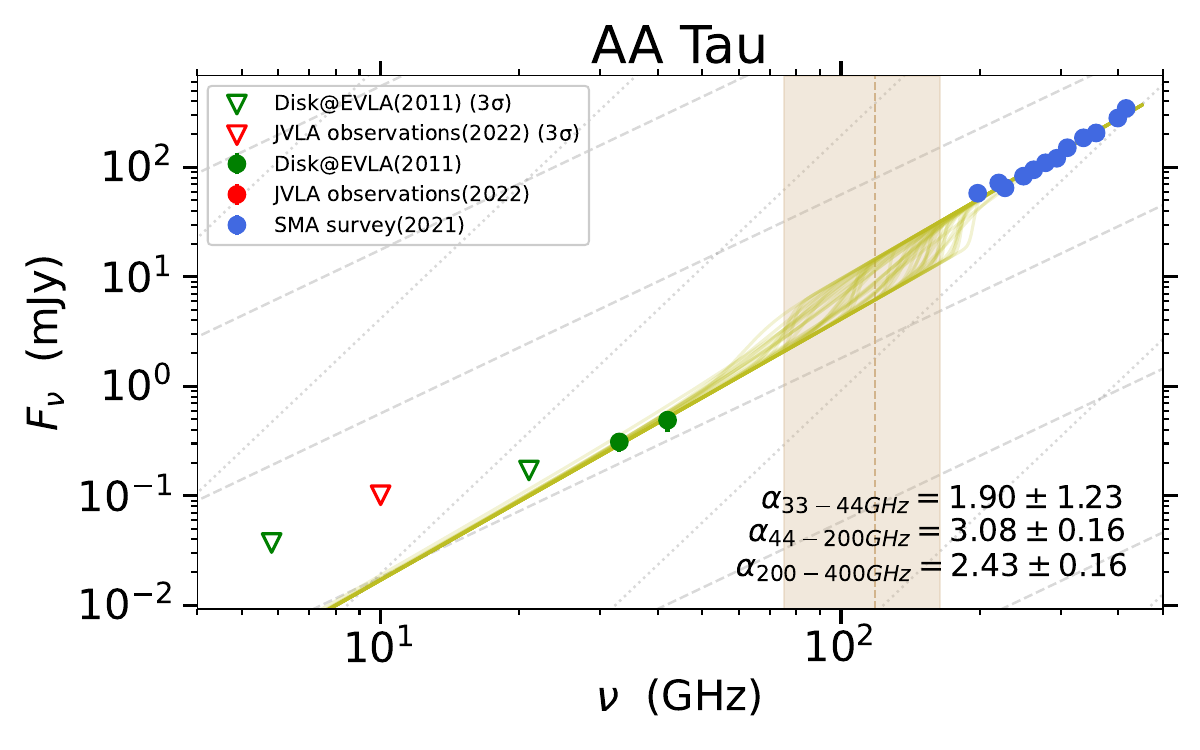}& 
        \hspace{-0.5cm}
        \includegraphics[width=6.5cm]{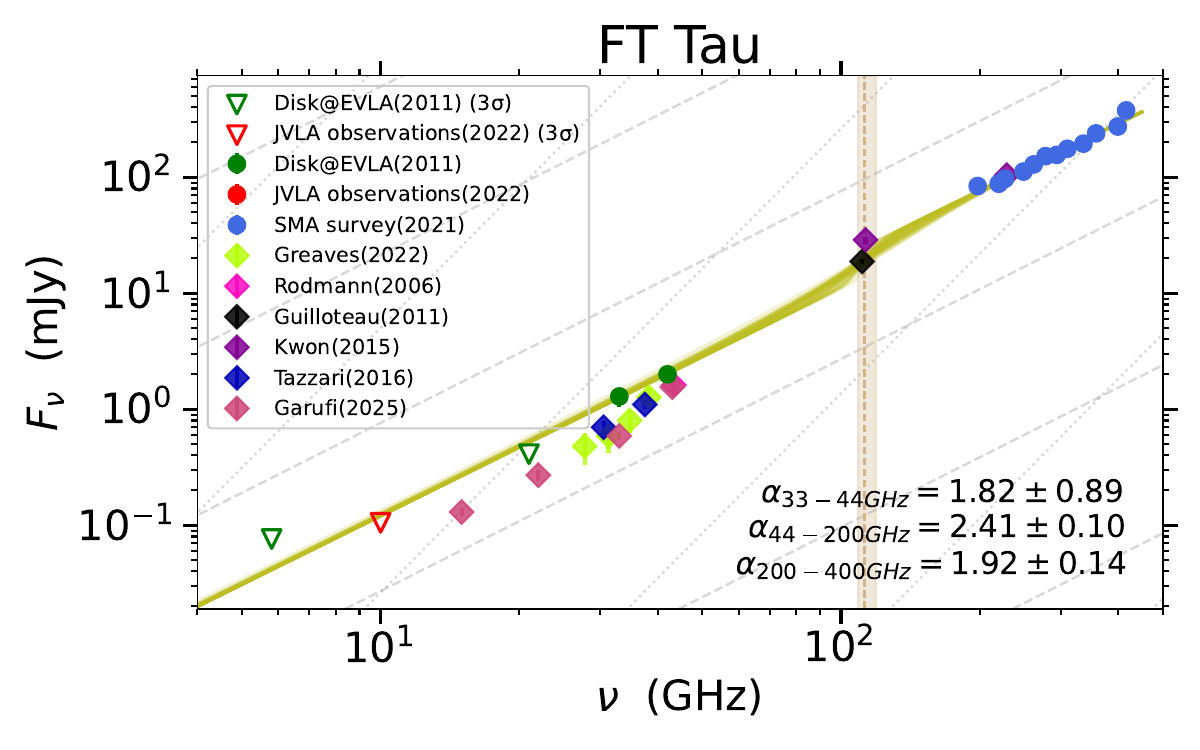}&
        \hspace{-0.5cm}
        \includegraphics[width=6.5cm]{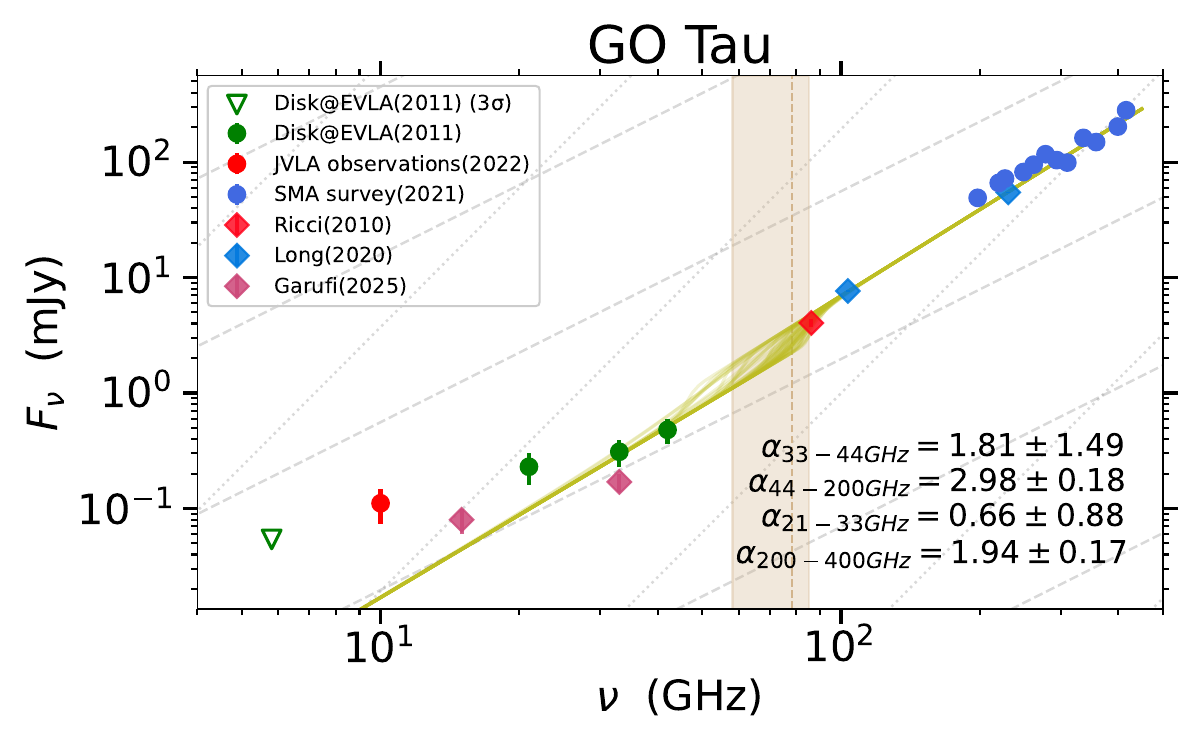}\\
        \includegraphics[width=6.5cm]{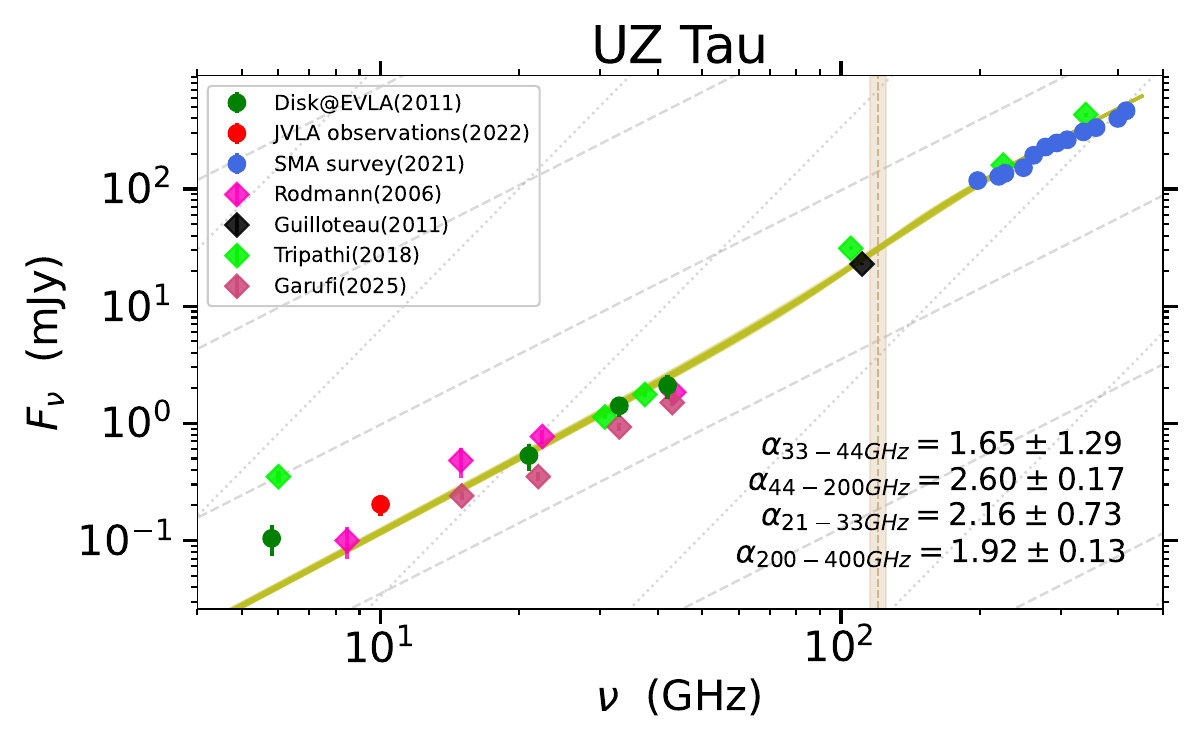}&
        \hspace{-0.5cm}
        \includegraphics[width=6.5cm]{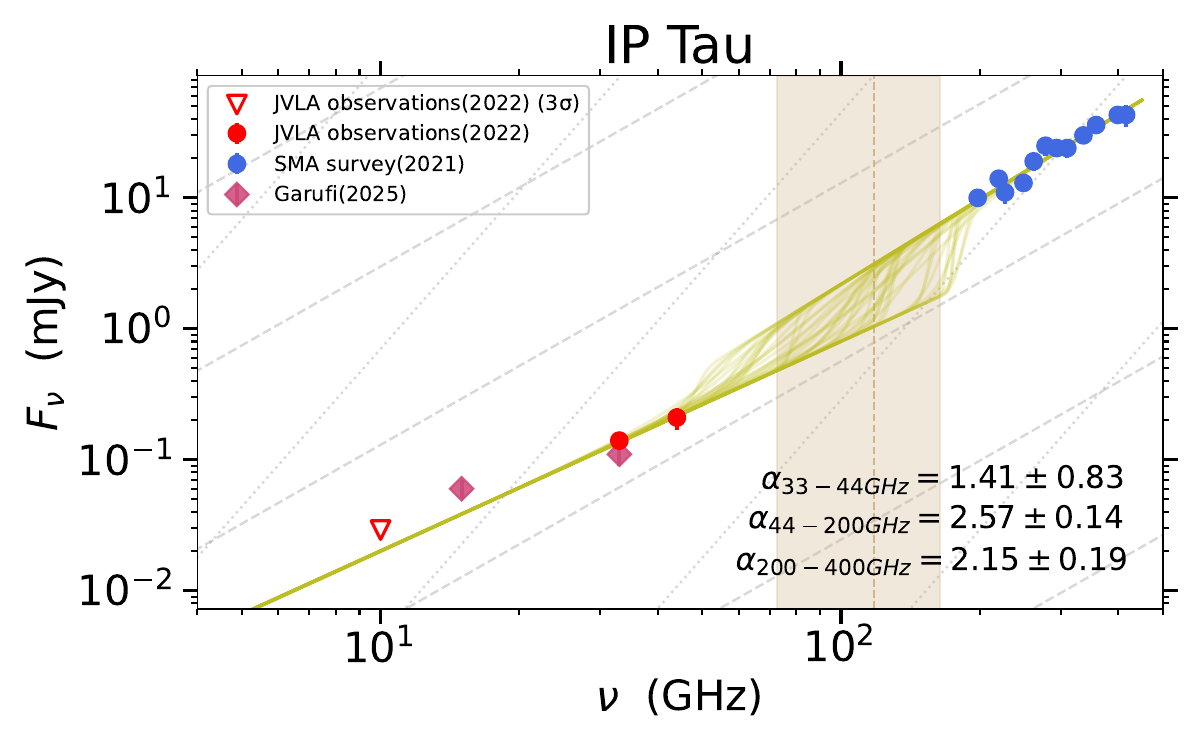}&
        \hspace{-0.5cm}
        \includegraphics[width=6.5cm]{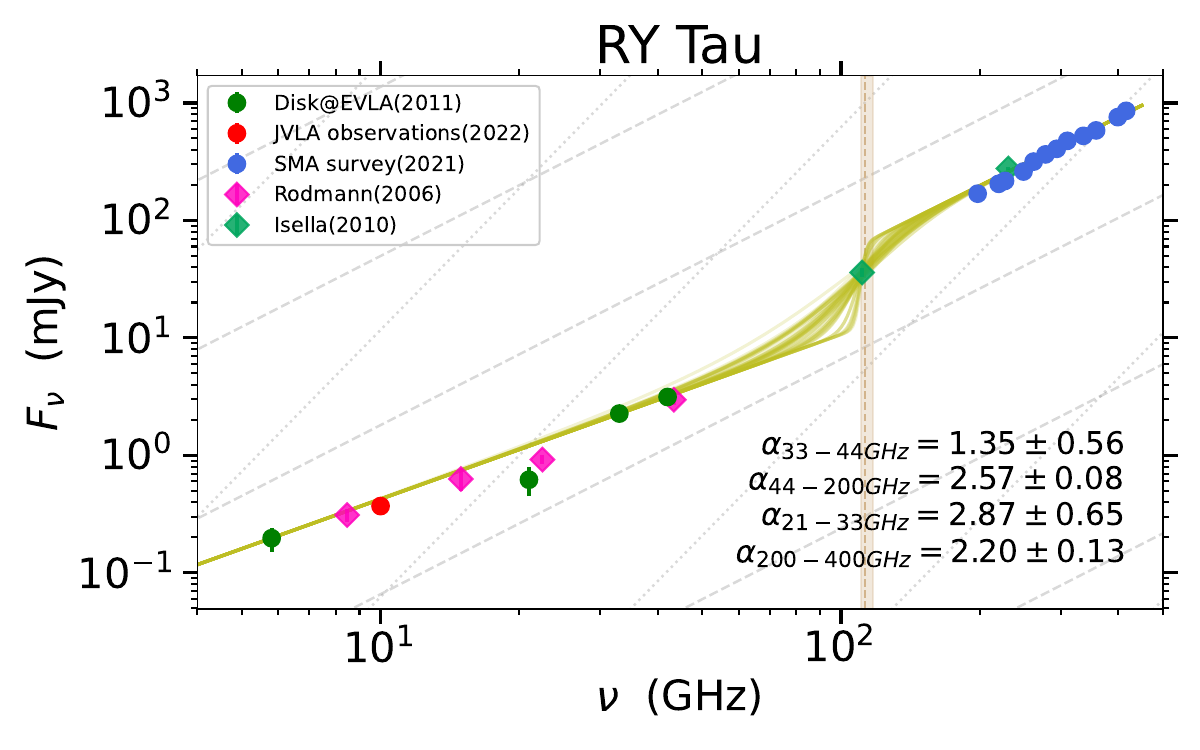}\\
        \includegraphics[width=6.5cm]{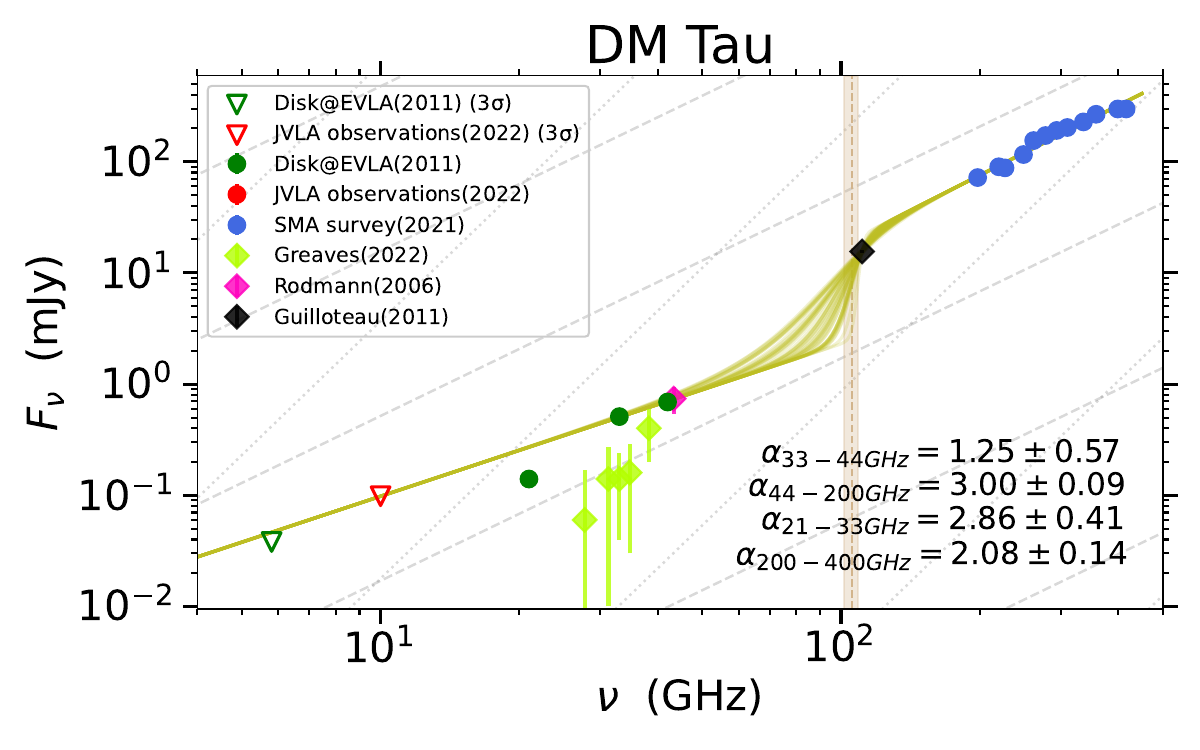}&
        \hspace{-0.5cm}
        \includegraphics[width=6.5cm]{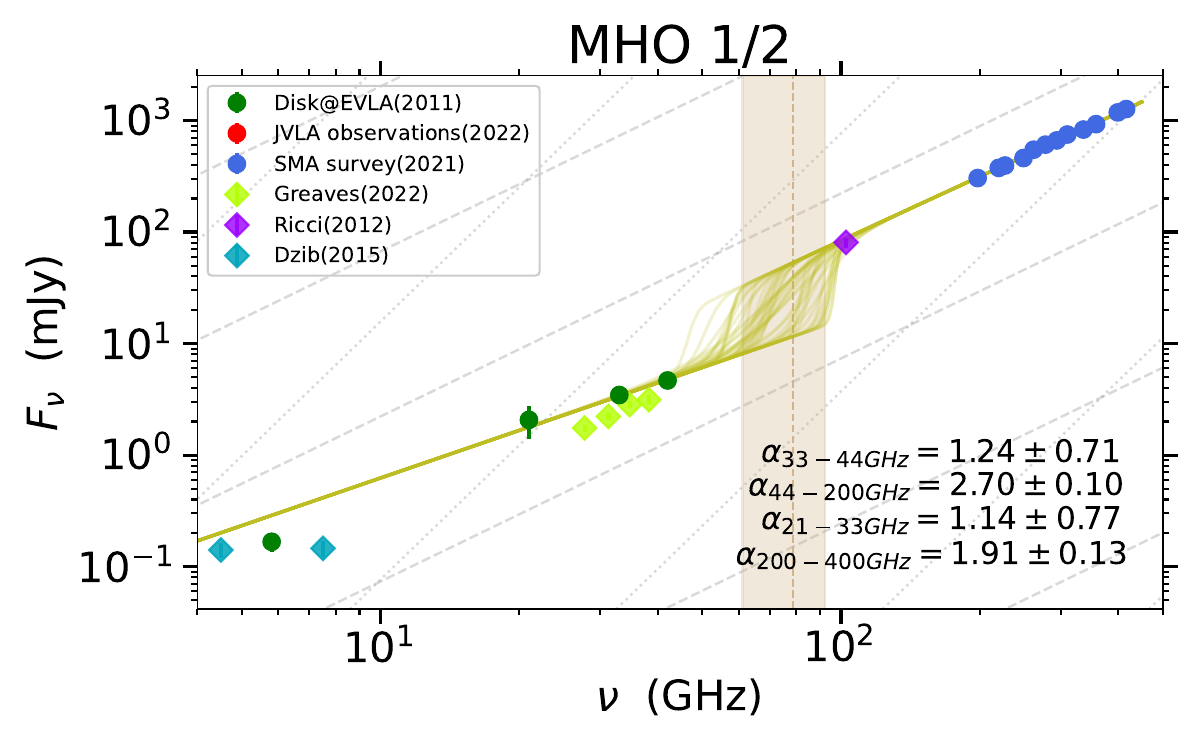}&
        \hspace{-0.5cm}
        \includegraphics[width=6.5cm]{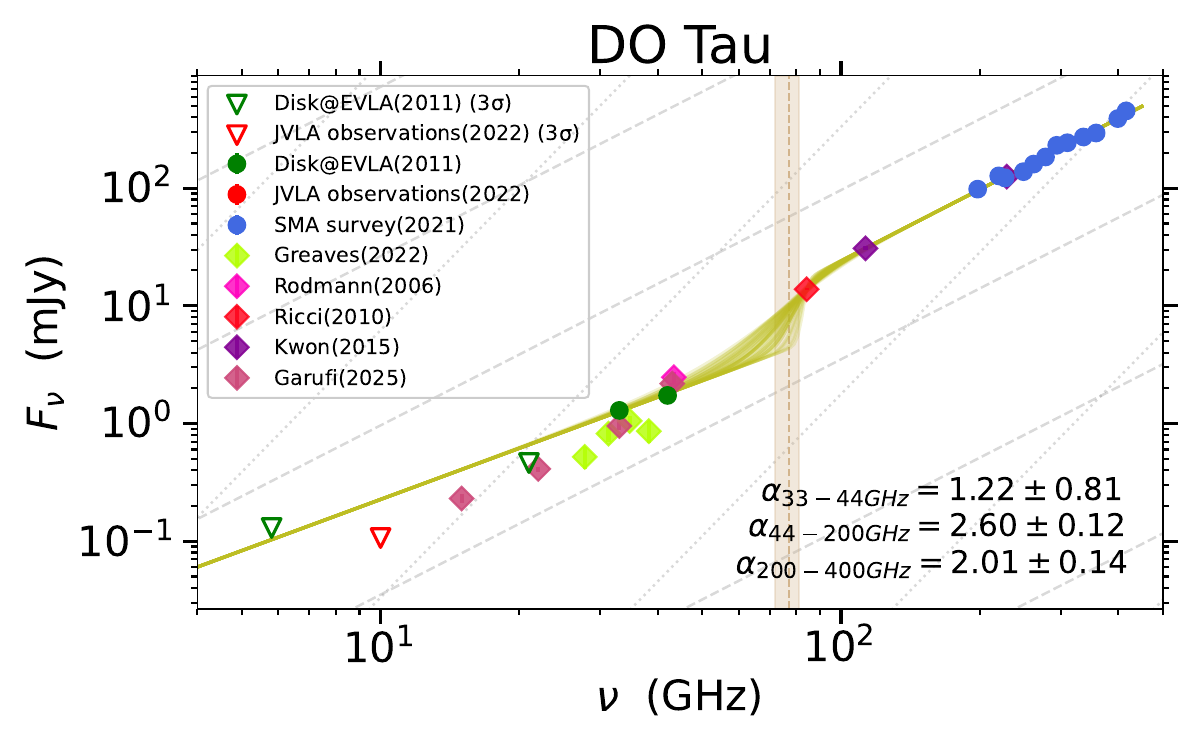}\\
        \includegraphics[width=6.5cm]{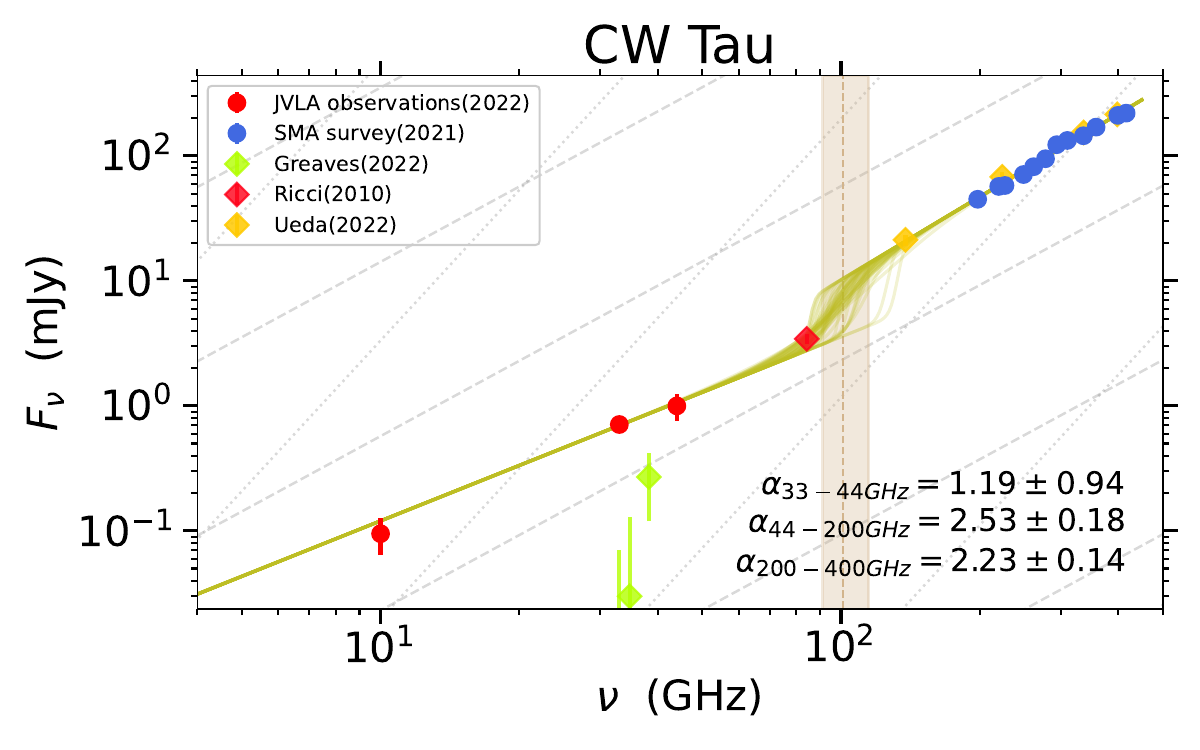}&
        \hspace{-0.5cm}
        \includegraphics[width=6.5cm]{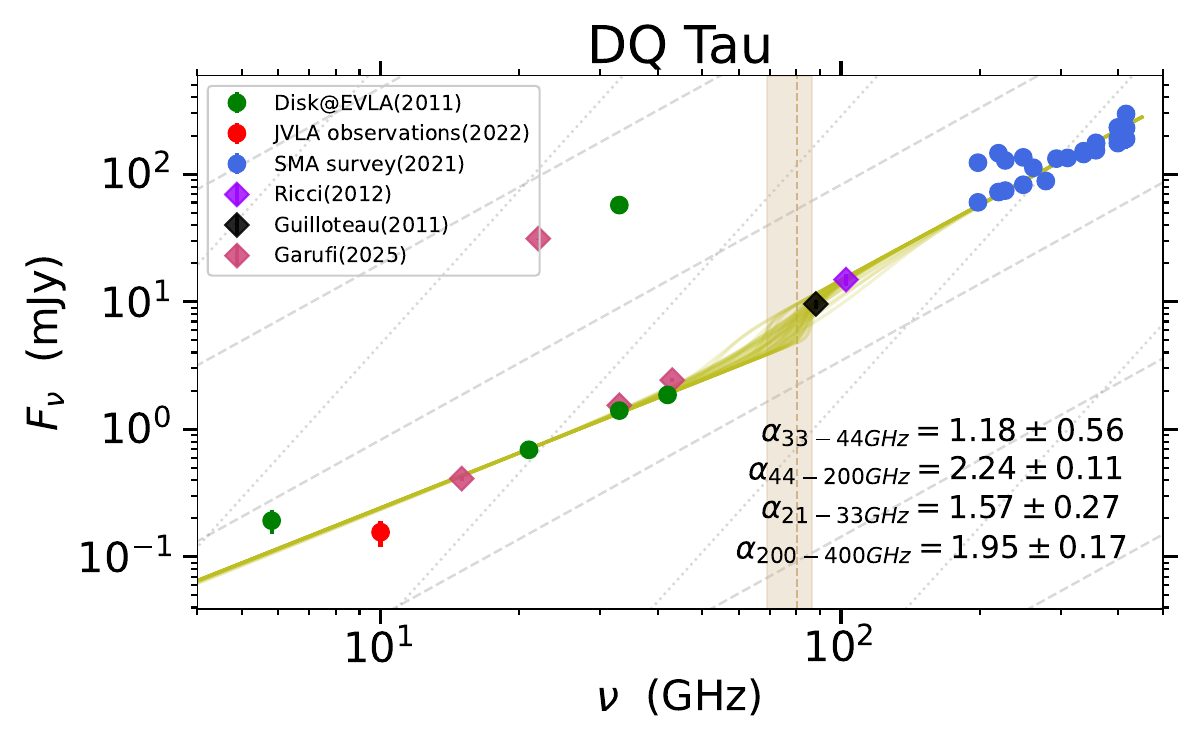}&
        \hspace{-0.5cm}
        \includegraphics[width=6.5cm]{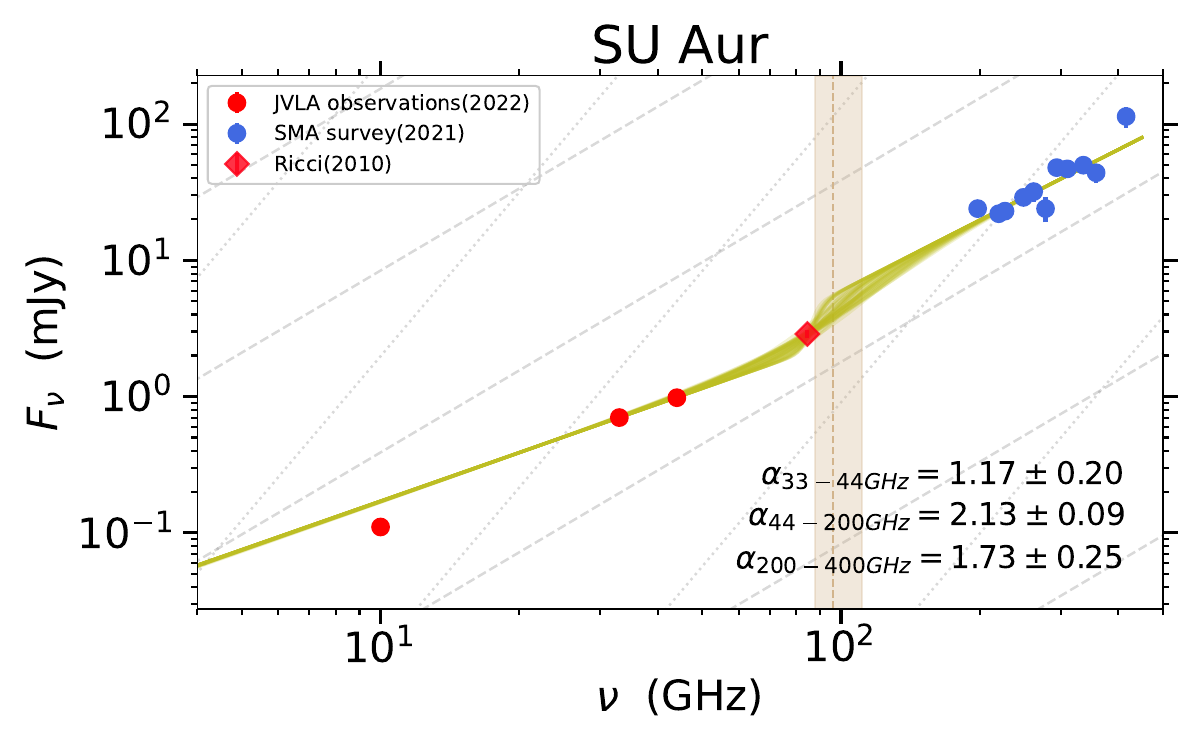}\\
        \end{tabular}
    \caption{Continuation of Figure \ref{fig:SED_0}.}
    \label{fig:SED_1}
\end{figure*}

\begin{figure*}
    \hspace{-1.8cm} 
        \begin{tabular}{ lll }
        \includegraphics[width=6.5cm]{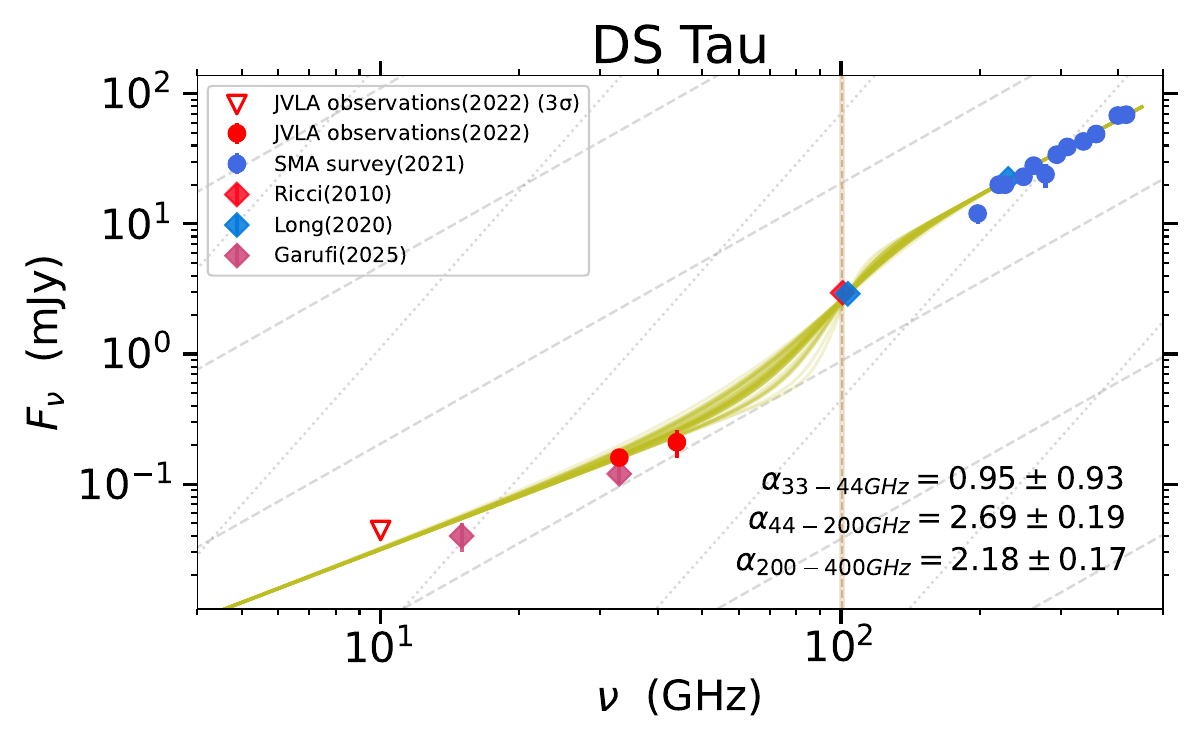}&
        \hspace{-0.5cm}
        \includegraphics[width=6.5cm]{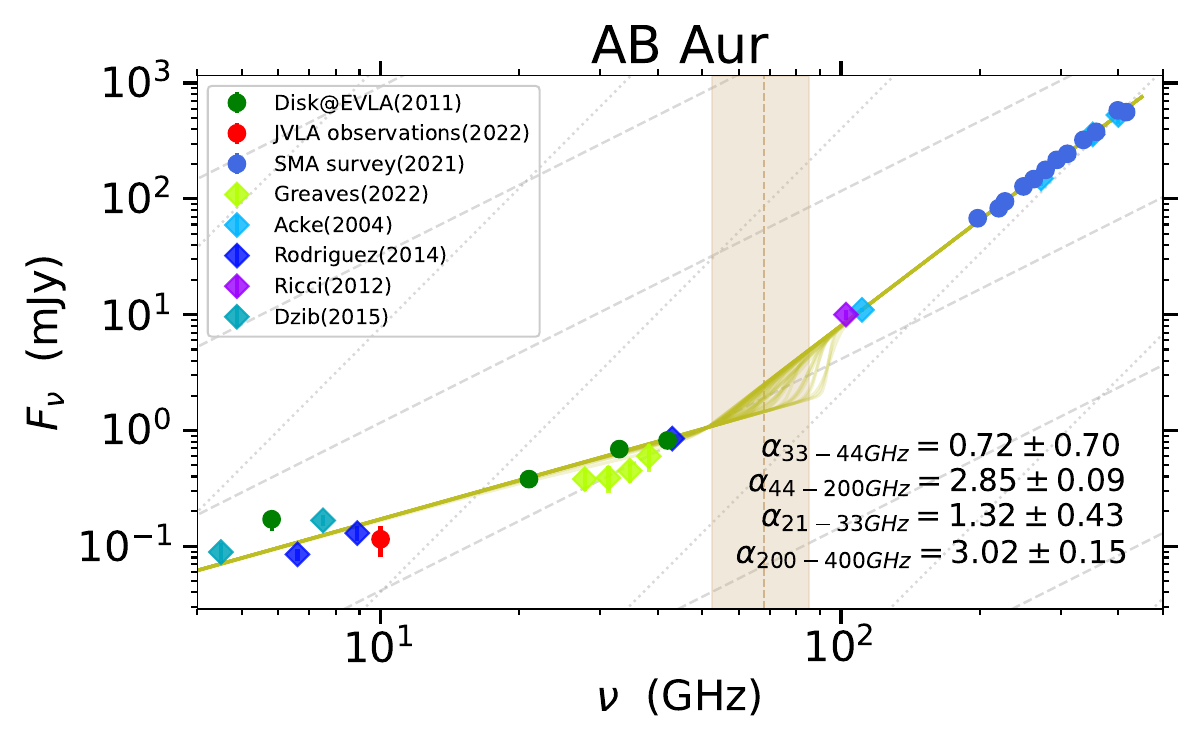}&
        \hspace{-0.5cm}
        \includegraphics[width=6.5cm]{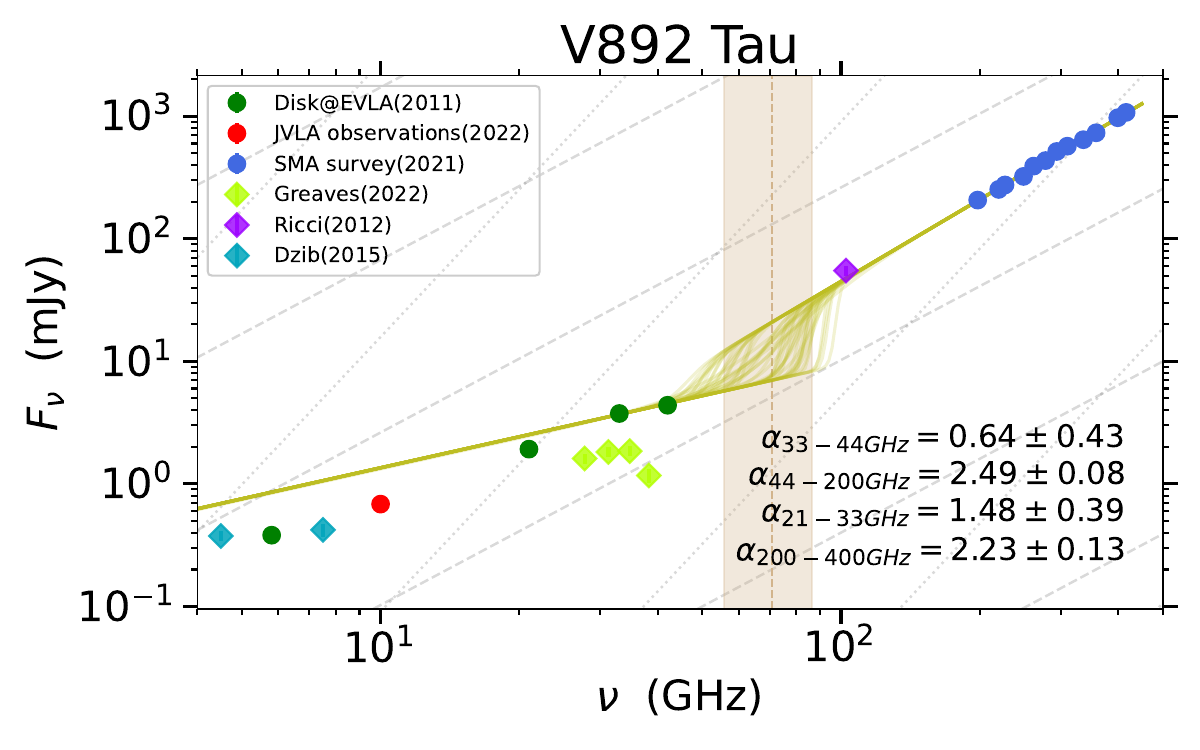}\\
        \includegraphics[width=6.5cm]{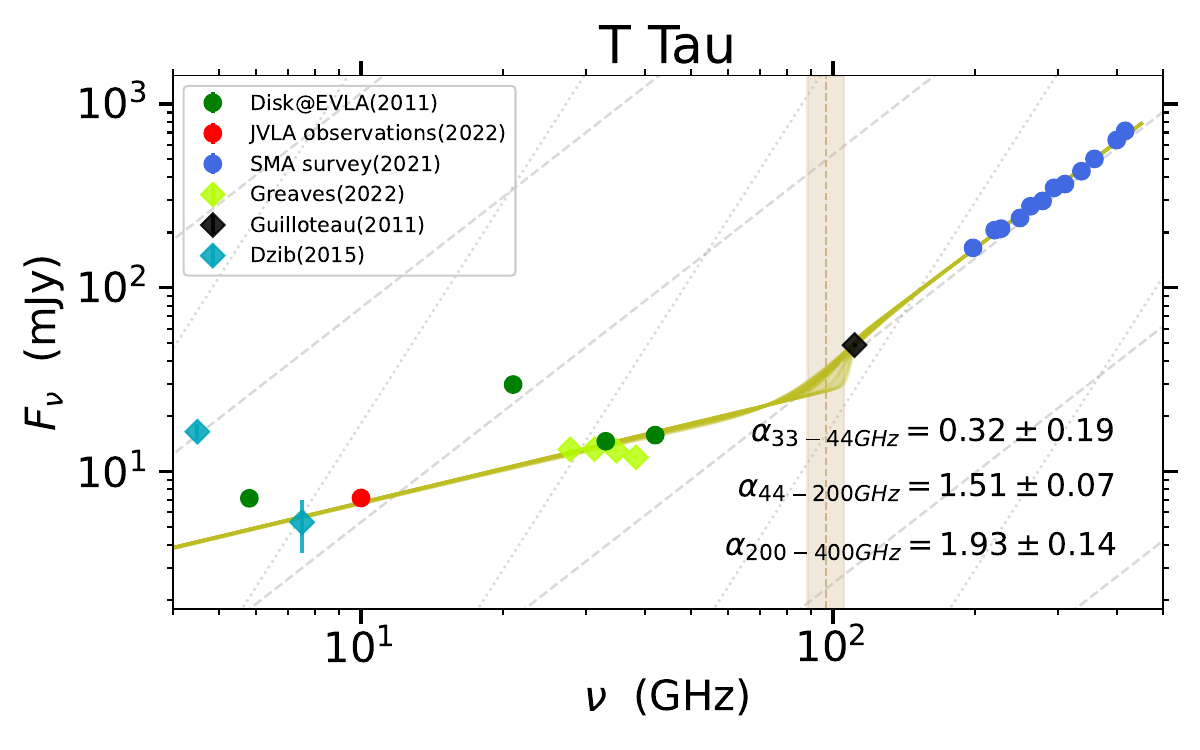}&
        \hspace{-0.5cm}
        \includegraphics[width=6.5cm]{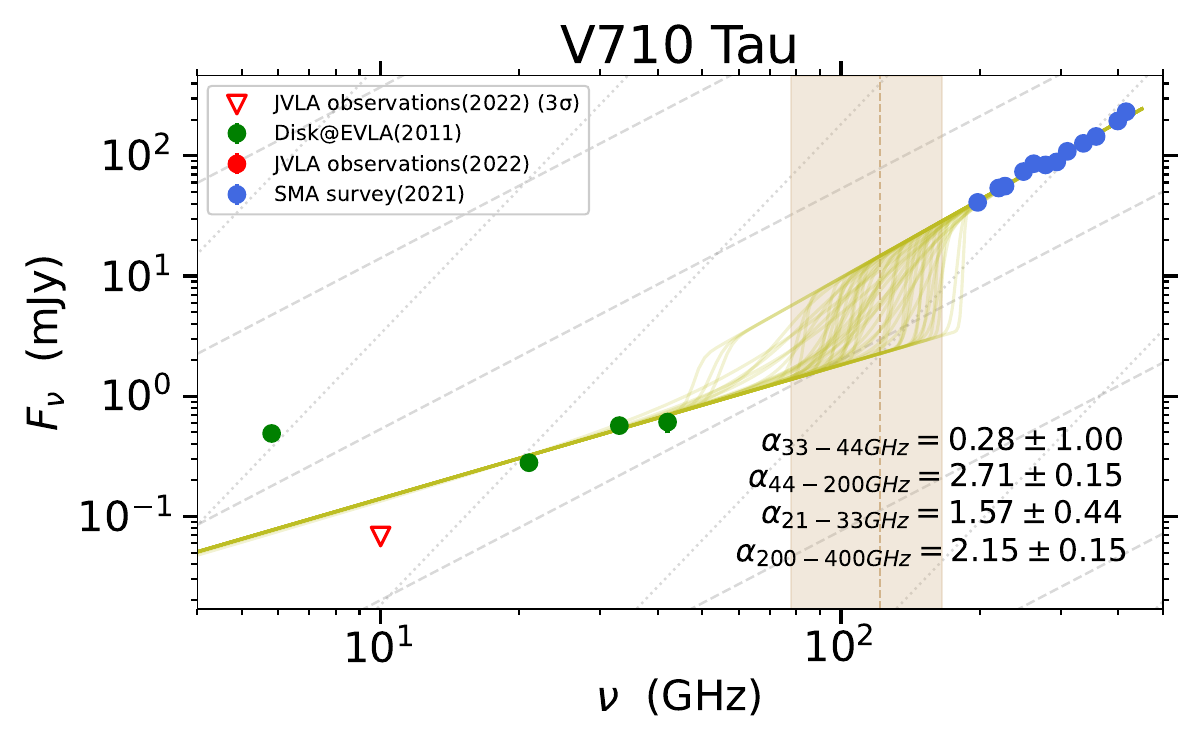}\\
        \end{tabular}
    \caption{Continuation of Figure \ref{fig:SED_0}.}
    \label{fig:SED_2}
\end{figure*}

For each target source, we estimated their flux densities at Ka and Q bands based on their $F_{\scriptsize \mbox{1.3 mm}}$ (\citealt{Chung2024ApJS..273...29C}) and assuming a spectral index of $\alpha$ = 3.5. 
We then accordingly allocated on-source integration time such that each target source can be detected at Ka and Q bands at $\gtrsim$5-$\sigma$ significances in 8 GHz aggregated bandwidth.
We allocated the X band on-source integration time such that the X band observations can achieve the same root-mean-square (RMS) noise levels as the Ka band observations. 
The RMS noises of our X band observations can discern the confusion of optically thin free-free emission, which has been typically assumed to have spectral indices $\alpha\sim$0.
This assumption is not necessarily realistic, which is discussed in Section \ref{sub:nonvariable}. 
For the observations on all sources and at all frequency bands, we adopted a minimum on-source time of 48 seconds in case of temporary technical issues during the observations. 

We observed the quasar, 3C84, at the beginning of each track of observations for passband calibrations.
We observed the quasar, 3C147, at the end of each track of observations for absolute flux calibrations.
For the Q, Ka, and X bands observations, we quoted the nominal absolute flux uncertainties of 10\%, 10\%, and 5\%, respectively. 
We selected the complex gain calibrators based on the flux densities, and the angular separations from our selected target sources. 
We adopted the officially recommended calibration cycle times for the C array configuration, which are 4 mins, 5 mins, and 10 mins at Q, Ka, and X bands, respectively. 

\subsection{Archival Disk@EVLA observations (AC982)}\label{sub:obsDiskatEVLA}

We retrieved the archival JVLA Q (41.0--43.0 GHz), Ka (31.7--33.6 GHz) , K (20.0--22.0 GHz) and C (4.8--6.8 GHz) band observations that were taken with the Disk@EVLA project (PI: C. Chandler).
These observations were taken in 2010, from July to December, in the C and D array configurations. 
The observational setup, observed sources, and adopted calibrators are listed in Table \ref{tab:EVLA_obs_summary}. 
The Ka band observations on 2010 September 10 (execution block ID: eb1922677) cannot be calibrated due to poor data quality; The Ka band observations on 2010 September 13 (scheduling block ID: sb1925247\_1) cannot be calibrated because the complex gain calibrators were detected at low signal-to-noise ratios (SNR).

\section{Data Reduction}\label{sec:reduction}

\subsection{Calibration}\label{sub:basic}
We manually calibrated our new JVLA observations and the archival data using the Common Astronomy Software Applications (CASA; version 6.4) software package (\citealt{CASA2022PASP..134k4501C}), following the standard calibration procedures. 
In the initial data inspection, we manually flagged the antennae and spectral windows that have poor response, and flagged the spectral channels that were subject to significant radio frequency interference (RFI). 
After applying antenna position correction, elevation-dependent gain curve, tropospheric opacity correction and requantizer gain, we bootstrapped delay and passband calibrations. 
We manually flagged the scans on the gain calibrators that were subject to poor phase coherence, as well as the associated scan on the science target sources, before performing complex gain calibrations.
We applied absolute flux scales to the complex gain solutions. 
Finally, we applied all the obtained calibration tables to the science targets. 

\begin{deluxetable*}{ccccccccccc}
\tabletypesize{\footnotesize}
\tablecolumns{11}
\tablewidth{0pt}
\tablecaption{ Flux density measurements \label{tab:flux_density}}
\tablehead{ 
 \colhead{Source} & \colhead{$F_{\mbox{\tiny 6.8 GHz}}$} & \colhead{$\Delta F_{\mbox{\tiny 6.8 GHz}}$} & \colhead{$F_{\mbox{\tiny 10.0 GHz}}$} & \colhead{$\Delta F_{\mbox{\tiny 10.0 GHz}}$} & \colhead{$F_{\mbox{\tiny 21.0 GHz}}$} & \colhead{$\Delta F_{\mbox{\tiny 21.0 GHz}}$} & \colhead{$F_{\mbox{\tiny 33.0 GHz}}$} & \colhead{$\Delta F_{\mbox{\tiny 33.0 GHz}}$} & \colhead{$F_{\mbox{\tiny 44.0 GHz}}$} & \colhead{$\Delta F_{\mbox{\tiny 44.0 GHz}}$} \\
\colhead{} & \colhead{(mJy)} & \colhead{(mJy)} & \colhead{(mJy)} & \colhead{(mJy)} & \colhead{(mJy)} & \colhead{(mJy)} & \colhead{(mJy)} & \colhead{(mJy)} & \colhead{(mJy)} & \colhead{(mJy)} }
\startdata 
AA Tau & $<0.037$ & $-$ & $<0.101$ & $-$ & $<0.17$ & $-$ & $0.31$ & $0.06$ & $0.49$ & $0.11$ \\
AB Aur & $0.171$ & $0.035$ & $0.115$ & $0.035$ & $0.38$ & $0.05$ & $0.69$ & $0.10$ & $0.82$ & $0.07$ \\
BP Tau & $-$ & $-$ & $<0.074$ & $-$ & $-$ & $-$ & $0.28$ & $0.04$ & $0.52$ & $0.11$ \\
CI Tau & $<0.035$ & $-$ & $<0.109$ & $-$ & $<0.16$ & $-$ & $0.53$ & $0.08$ & $0.91$ & $0.18$ \\
CW Tau & $-$ & $-$ & $0.095$ & $0.031$ & $-$ & $-$ & $0.71$ & $0.09$ & $1.00$ & $0.24$ \\
CY Tau & $<0.164$ & $-$ & $0.149$ & $0.050$ & $<0.30$ & $-$ & $0.88$ & $0.12$ & $1.68$ & $0.06$ \\
DE Tau & $-$ & $-$ & $<0.032$ & $-$ & $-$ & $-$ & $0.41$ & $0.04$ & $0.71$ & $0.16$ \\
DL Tau & $<0.082$ & $-$ & $<0.119$ & $-$ & $<0.34$ & $-$ & $0.93$ & $0.05$ & $2.54$ & $0.44$ \\
DM Tau & $<0.038$ & $-$ & $<0.098$ & $-$ & $0.14$ & $0.02$ & $0.51$ & $0.06$ & $0.69$ & $0.05$ \\
DN Tau & $<0.045$ & $-$ & $<0.102$ & $-$ & $0.31$ & $0.07$ & $0.55$ & $0.10$ & $1.17$ & $0.07$ \\
DO Tau & $<0.129$ & $-$ & $<0.106$ & $-$ & $<0.46$ & $-$ & $1.29$ & $0.15$ & $1.73$ & $0.27$ \\
DQ Tau & $0.192$ & $0.042$ & $0.156$ & $0.036$ & $0.69$ & $0.08$ & $1.40$\tablenotemark{a} & $0.05$\tablenotemark{a} & $1.86$ & $0.24$ \\
DR Tau & $0.061$ & $0.017$ & $<0.103$ & $-$ & $0.24$ & $0.02$ & $0.61$ & $0.07$ & $1.39$ & $0.06$ \\
DS Tau & $-$ & $-$ & $<0.044$ & $-$ & $-$ & $-$ & $0.16$ & $0.02$ & $0.21$ & $0.05$ \\
FT Tau & $<0.076$ & $-$ & $<0.105$ & $-$ & $<0.41$ & $-$ & $1.29$ & $0.24$ & $2.00$ & $0.21$ \\
GM Aur & $<0.062$ & $-$ & $<0.106$ & $-$ & $0.21$ & $0.02$ & $0.70$ & $0.03$ & $1.12$ & $0.33$ \\
GO Tau & $<0.054$ & $-$ & $0.111$ & $0.037$ & $0.23$ & $0.07$ & $0.31$ & $0.08$ & $0.48$ & $0.12$ \\
Haro 6-37 & $<0.221$ & $-$ & $<0.073$ & $-$ & $0.24$ & $0.04$ & $0.40$ & $0.08$ & $1.03$ & $0.09$ \\
Haro 6-39 & $-$ & $-$ & $0.064$ & $0.010$ & $-$ & $-$ & $0.16$ & $0.02$ & $<0.28$ & $-$ \\
IP Tau & $-$ & $-$ & $<0.029$ & $-$ & $-$ & $-$ & $0.14$ & $0.02$ & $0.21$ & $0.04$ \\
IQ Tau & $<0.066$ & $-$ & $<0.098$ & $-$ & $0.20$ & $0.05$ & $0.38$ & $0.05$ & $0.76$ & $0.17$ \\
LkCa 15 & $<0.025$ & $-$ & $<0.103$ & $-$ & $0.16$ & $0.04$ & $0.38$ & $0.06$ & $0.66$ & $0.15$ \\
MHO 1/2 & $0.167$ & $0.031$ & $-$ & $-$ & $2.07$ & $0.67$ & $3.46$ & $0.46$ & $4.67$ & $0.50$ \\
RY Tau & $0.197$ & $0.046$ & $0.370$ & $0.058$ & $0.62$ & $0.17$ & $2.27$ & $0.24$ & $3.14$ & $0.26$ \\
SU Aur & $-$ & $-$ & $0.110$ & $0.016$ & $-$ & $-$ & $0.70$ & $0.02$ & $0.98$ & $0.05$ \\
T Tau & $7.162$ & $0.053$ & $7.187$ & $0.086$ & $29.77$ & $0.39$ & $14.63$ & $0.37$ & $15.80$ & $0.59$ \\
UX Tau & $0.183$ & $0.024$ & $<0.077$ & $-$ & $0.56$ & $0.16$ & $0.79$ & $0.11$ & $1.32$ & $0.06$ \\
UY Aur & $0.110$ & $0.030$ & $0.150$ & $0.018$ & $0.54$ & $0.10$ & $1.03$ & $0.05$ & $2.41$ & $0.26$ \\
UZ Tau & $0.104$ & $0.031$ & $0.203$ & $0.041$ & $0.53$ & $0.14$ & $1.41$ & $0.28$ & $2.10$ & $0.50$ \\
V710 Tau & $0.490$ & $0.021$ & $<0.068$ & $-$ & $0.28$ & $0.04$ & $0.57$ & $0.08$ & $0.61$ & $0.12$ \\
V836 Tau & $-$ & $-$ & $<0.247$ & $-$ & $-$ & $-$ & $0.38$ & $0.04$ & $0.69$ & $0.11$ \\
V892 Tau & $0.381$ & $0.050$ & $0.683$ & $0.068$ & $1.92$ & $0.31$ & $3.75$ & $0.28$ & $4.38$ & $0.31$ \\
\enddata
\tablenotetext{a}{DQ~Tau shows varying Ka band flux densities between the JVLA observations on 2010 08 21 and 2010 11 01. The recorded $F_{\mbox{\tiny 33.0 GHz}} = 1.40 \pm 0.04$ mJy on the table is taken on 2010 11 01, and on 2010 08 21 the $F_{\mbox{\tiny 33.0 GHz}} = 57.24 \pm 1.63$ mJy. }
\tablecomments{ The $F$ and $\Delta F$ represent the measured flux density and the 1-$\sigma$ error of flux density which includes the absolute flux uncertainties. The upper limit indicates the 3-$\sigma$ detection limit for the flux density. The hyphen symbol implies no observations on the source at that frequency.}
\end{deluxetable*}

\subsection{Imaging}\label{sub:imaging}

We performed multi-frequency synthesis (mfs) imaging (\citealt{Rau2011A&A...532A..71R}) using the CASA 6.4 {\tt tclean} task. 
For the observations on each target at each frequency band, we produced one image using the aggregate bandwidth at that frequency band. 
We adopted natural weighting (Robust$=$2.0) to ensure the best continuum sensitivity.

In the X and C band observations, there are bright sources that are located outside of the $\sim$4$'$ and 7$'$ primary beam FWHM.
Imaging with the default primary beam gain cutoff (pblimit$=$0.2) failed to recover them, which led to imaging artifacts. 
To amend this issue, we performed widefield imaging for the X and C band data using a image size that is $4\times$ larger than the primary beam FWHM, by setting the keyword gridder$=$'widefield' and pblimit$=-$0.2.

Some target sources were observed multiple times at certain frequency bands.
To assess the data quality (e.g. residual phase error, rms noise level) of individual tracks of observations, we first imaged those tracks of observations separately and then compared the flux densities of the science target. 
For each target source, we omitted the tracks in which the target has $<$3-$\sigma$ detection, and omitted the tracks that high residual phase errors led to inconsistent flux measurements with the other tracks. 
Moreover, we identified the target sources which presented flux variability that is higher than 3 times the nominal 10\% flux density uncertainty, and left out the flux measurements in the flaring period. 
The flaring period is determined by the epoch when the flux density is $>$2 times larger than the average flux at other epochs. 
After verifying the consistency of flux densities, we jointly imaged the tracks at the same band to enhance the signal-to-noise ratio (SNR). 
The achieved synthesized beams and rms noise of the final images are summarized in Table \ref{tab:img_parameter}.

\begin{figure}
    \centering
    \includegraphics[width=8.5cm]{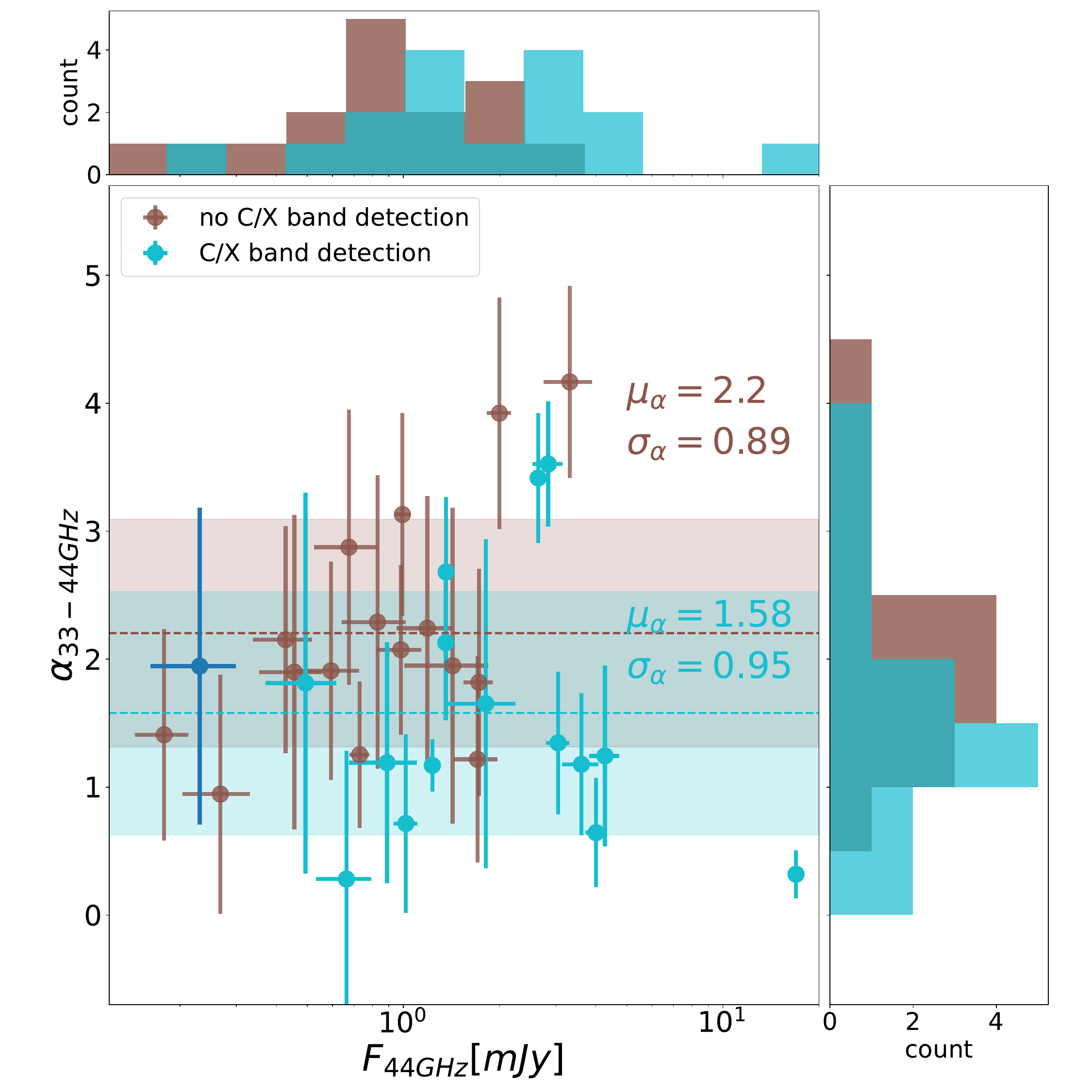}\\
    \caption{
    The Ka-Q band spectral index ($\alpha_{\mbox{\scriptsize 33-44 GHz}}$) versus the JVLA Q band (44 GHz) flux density ($F_{\mbox{\scriptsize 44 GHz}}$). 
    Cyan points in the scatter plot show the sources that were detected either in the C band (5.8 GHz; Section \ref{sub:obsDiskatEVLA}) observations of the Disk@EVLA project or in our new X band (10 GHz; Section \ref{sub:obs}) observations (i.e., radio detected sources).
    Blue point shows one of the radio detected sources, Haro~6-39, for which we adopted the 3-$\sigma$ upper limit of $F_{\mbox{\scriptsize 44 GHz}}$.
    The other sources are shown by brown points in the scatter plot (i.e., radio non-detected sources).
    Cyan/brown horizontal dashed lines and shaded areas in the scatter plot represent the means ($\mu_{\alpha}$) and standard deviations ($\sigma_{\alpha}$) of the $\alpha_{\mbox{\scriptsize 33-44 GHz}}$ of the radio detected/non-detected sources, respectively, which are also labeled in text.
    Histograms of $\alpha_{\mbox{\scriptsize 33-44 GHz}}$ and $F_{\mbox{\scriptsize 44 GHz}}$ for the radio detected and non-detected sources are shown in the right and top panels. 
    The $F_{\mbox{\scriptsize 44 GHz}}$ values have been rescaled to a distance of 140 pc, which makes them different from the directly measured flux densities listed in Table \ref{tab:flux_density}.
    }
    \label{fig:F44_alphaKaQ}
\end{figure}
 
\subsection{Measuring Flux Densities}\label{sub:flux}

We derived flux densities at each JVLA band based on (1) 2D Gaussian fitting, or (2) directly reading pixel values in the clean images. 
For images in which the target source was detected at $>$5-$\sigma$, we obtained the integrated flux density by performing 2D Gaussian fitting in a square region, using the CASA 6.4 {\tt imfit} task.
The square region is centered at the peak of the target source, and the side-length is $\sim$5 times the major axis of the synthesized beam.
For images where the source was detected at 3-$\sigma$ significance, we reported the peak intensity as the flux density. 
For non-detections, we recorded the 3-$\sigma$ detection limits instead. 
We note that the hierarchical quadruple system, UZ Tau, was marginally spatially resolved into two components at Q band but was not resolved at lower frequencies. 
Following the approach of the previous SMA survey (\citealt{Chung2024ApJS..273...29C}), we fit the UZ Tau system with one Gaussian regardless of the multiplicity. 
The binary system, MHO~1/2, was resolved at 44 GHz, marginally resolved at 33 and 21 GHz, but was not resolved at lower frequencies. 
At $\leq$33 GHz, we fit the MHO~1/2 system with one Gaussian; at 44 GHz, we reported the summed flux of the two components. 
The flux densities measurements are summarized in Table \ref{tab:flux_density}. 
For a few sources that the Ka and Q band flux density measurements made out of the Disk@EVLA data have been reported by previous studies (DR~Tau: \citealt{Tazzari2016A&A...588A..53T}; CY~Tau: \citealt{Perez2015ApJ...813...41P}; UZ~Tau: \citealt{Tripathi2018ApJ...861...64T}; AB~Aur: \citealt{Rodriguez2014ApJ...793L..21R}), our flux density measurements made out of the same Disk@EVLA data achieved good consistency with those previous studies. 
This justifies the robustness of our data processing (Section \ref{sec:reduction}).

\section{Results}\label{sec:result}

\subsection{Individual spectra}\label{sub:SEDresults}

Figure \ref{fig:SED_0}--\ref{fig:SED_2} compare the flux density measurements obtained from the JVLA observations detailed in Sections \ref{sub:obs} and \ref{sub:obsDiskatEVLA}, the recent SMA 200--400 GHz survey (\citealt{Chung2024ApJS..273...29C}), and some flux density measurements reported over the past 20 years. 
Spectral indices were evaluated based on the flux density measurements from two adjacent bands. 
The uncertainties of the spectral indices were estimated using standard error propagation. 
In the upper left of each panel in Figure \ref{fig:SED_0}--\ref{fig:SED_2}, we have labeled the inter-band spectral indices: the spectral indices between JVLA Ka and Q band ($\alpha_{\mbox{\scriptsize 33-44 GHz}}$), the spectral indices between JVLA K and Ka band ($\alpha_{\mbox{\scriptsize 21-33 GHz}}$), the spectral indices between JVLA Q band and SMA 200 GHz band ($\alpha_{\mbox{\scriptsize 44-200GHz}}$) , and the submillimeter spectral indices ($\alpha_{\mbox{\scriptsize 200-400GHz}}$) quoted from \citet{Chung2024ApJS..273...29C}.

\subsection{Spectra of variable sources}\label{sub:variable}

From the 32 disks sample, we found DQ~Tau, T~Tau, UZ~Tau, AB~Aur and UX~Tau show strong radio flux variability (at any epoch the measured flux density is $>$2 times larger or smaller than that at other epochs or the interpolated flux from the adjacent frequency bands). 
We also found some sources, DL~Tau, DR~Tau, CY~Tau, LkCa~15, GO~Tau and V710~Tau, show potential radio flux variability. 
The flux densities of DQ~Tau at $\gtrsim$200 GHz are varying with time, which were resolved in the previous observations of \citet{Salter2010A&A...521A..32S} and \citet{Chung2024ApJS..273...29C}.
The 33 GHz flux density of DQ~Tau also shows variability when comparing the two epochs of JVLA Ka band observations in August and November, 2010. 
Since the flux density of these two Ka band observations are different by two orders of magnitude, we did not image them jointly.
Instead, they were imaged and plotted separately in (Figure \ref{fig:SED_0}).
Another source, T~Tau, also exhibits significant radio flux variability, which is evident when comparing different epochs (August 20 and 23, 2010) of JVLA observations at frequencies below 50 GHz (see Figure \ref{fig:SED_2}).
We note that DQ~Tau and T~Tau are both multiple systems -- DQ Tau is a spectroscopic binary, and T Tau is a triple system.
UZ~Tau and AB~Aur also show radio flux variability at $<$10 GHz as the C band flux densities measured in July, 2010 and the X band flux densities measured in September, 2022 are $>$3-$\sigma$ different from those reported by \citet{Rodmann2006A&A...446..211R}, \citet{Rodriguez2014ApJ...793L..21R}, \citet{Dzib2015ApJ...801...91D} and \citet{Tripathi2018ApJ...861...64T}. 
In addition, DL~Tau, DR~Tau, CY~Tau, LkCa~15, UX~Tau, GO~Tau and V710~Tau, were not detected either in the C band observations in the Disk@EVLA observations in 2010 or in our new X band observations in 2022, while they were detected in other observations at similar frequencies  (e.g., \citealt{Dzib2015ApJ...801...91D,Zapata2017ApJ...834..138Z}). 
These sources may also exhibit radio variability at $<$10 GHz.
The millimeter and radio variabilities of these sources are noteworthy.
Previous JVLA monitoring observations (e.g., \citealt{Liu2014ApJ...780..155L,Dzib2015ApJ...801...91D,Coutens2019A&A...631A..58C}) indicated that significant radio flares are not common during the Class II stage. 
Our present new results reinforce that to diagnose the emission of grown dust at centimeter band with spatially unresolved spectra, it is more robust if one can simultaneously constrain the flux densities of dust thermal emission and free-free or synchrotron emission.

\begin{figure}
    \centering
    \includegraphics[width=8.5cm]{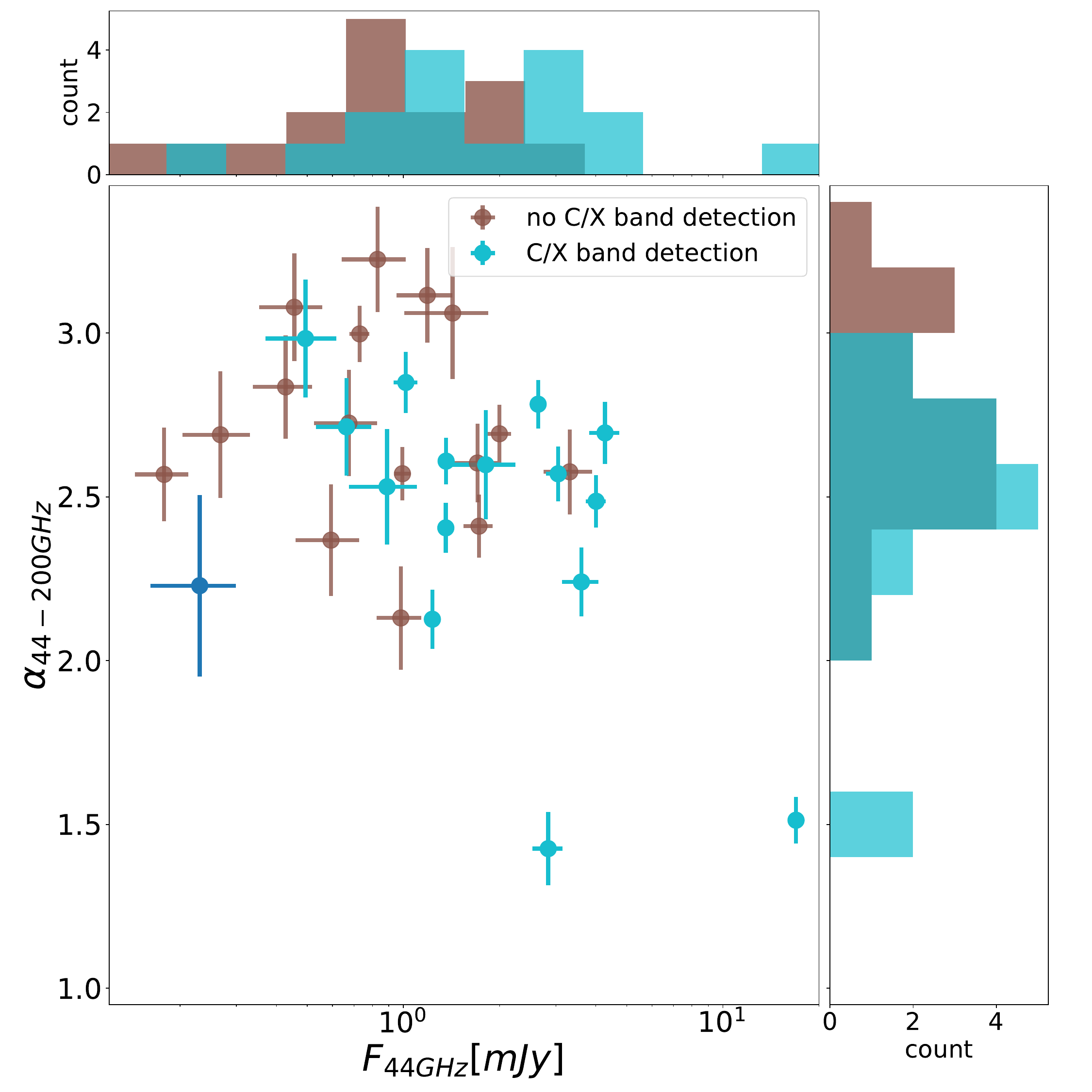}\\
    \caption{
    The Q band-200 GHz spectral index ($\alpha_{\mbox{\scriptsize 44-200GHz}}$) versus the JVLA Q band (44 GHz) flux density ($F_{\mbox{\scriptsize 44 GHz}}$). 
    Color codes and symbols are similar to those in Figure \ref{fig:F44_alphaKaQ}.
    Histograms of $\alpha_{\mbox{\scriptsize 44-200GHz}}$ and $F_{\mbox{\scriptsize 44 GHz}}$ for the radio detected and non-detected sources are shown in the right and top panels, respectively.
    The $F_{\mbox{\scriptsize 44 GHz}}$ values have been rescaled to a distance of 140 pc, which makes them different from the directly measured flux densities listed in Table \ref{tab:flux_density}.
    }
    \label{fig:F44_alphaQsubmm}
\end{figure}

\subsection{Spectra of sources not yet confirmed as millimeter/radio variables}\label{sub:nonvariable}

Apart from the variable sources noted above, the $>$10 GHz flux densities measured in different epochs of interferometric observations are in good agreement.
Previous single-dish observations made with the Green Bank Telescope (\citealt{Greaves2022MNRAS.513.3180G}) showed systematically lower flux densities in many, but not all cases (Figure \ref{fig:SED_0}--\ref{fig:SED_2}), which could be attributed to data calibration issues.

The Disk@EVLA observations we analyzed detected C band emission in the following 11 sources: UY~Aur, DR~Tau, UX~Tau, UZ Tau, RY~Tau, MHO~1/2, DQ~Tau, AB~Aur, V892~Tau, T~Tau, and V710~Tau.
This indicates that the K, Ka, and Q band flux densities measured from the Disk@EVLA data may have been influenced by free-free or synchrotron emission, as these observations were conducted within a short time frame of each other (Table \ref{tab:EVLA_obs_summary}).
The C band Disk@EVLA data we processed had high rms noise.
Therefore, the K, Ka, and Q band flux densities measured from the Disk@EVLA data might still be influenced by free-free or synchrotron emission, even if the sources were not detected at C band.
For instance, the 4 sources, DL~Tau,  CY~Tau, LkCa~15 and GO~Tau were not detected in the C band observations of the Disk@EVLA project, but were detected at similar frequencies in later observations reported by \citet{Dzib2015ApJ...801...91D} and \citet{Zapata2017ApJ...834..138Z}, and by our new X band observations.
The K, Ka, and Q band flux densities of these four sources measured from the Disk@EVLA data might also be influenced by free-free or synchrotron emission; another possibility is that the free-free and/or synchrotron emission was not prominent during the Disk@EVLA K, Ka, and Q bands observations but becomes stronger during the later JVLA observations.

Among the 8 sources that we performed new JVLA X, Ka, and Q band observations, we detected X band emission from 3 of them: Haro~6-39, CW~Tau, SU~Aur (Section \ref{sub:obs}).
As these X, Ka, and Q band observations were taken during a similar time period (Table \ref{tab:obs_summary}), the Ka and Q band flux densities of these three sources were likely partly contributed by free-free or synchrotron emission.
Furthermore, recent JVLA X and Ku band observations indicate significant free-free emission from DM~Tau (\citealt{Terada2023ApJ...953..147T,Liu2024A&A...685A..18L}). 
We also observed time variations of free-free or synchrotron emission in some of the sources, which can be seen when comparing our observations with those from previous studies (Figure \ref{fig:SED_0}--\ref{fig:SED_2}).
For sources that present strong free-free emission, the spectral index $\alpha_{\mbox{\scriptsize 21-33 GHz}}$ may be smaller than $\alpha_{\mbox{\scriptsize 33-44 GHz}}$.
On the other hand, in cases where $\alpha_{\mbox{\scriptsize 21-33 GHz}}$ is higher than $\alpha_{\mbox{\scriptsize 33-44 GHz}}$, all of the three emission mechanisms, free-free, synchrotron, or dust emission may contribute significantly. 

\begin{figure}
    \centering
    \includegraphics[width=8.5cm]{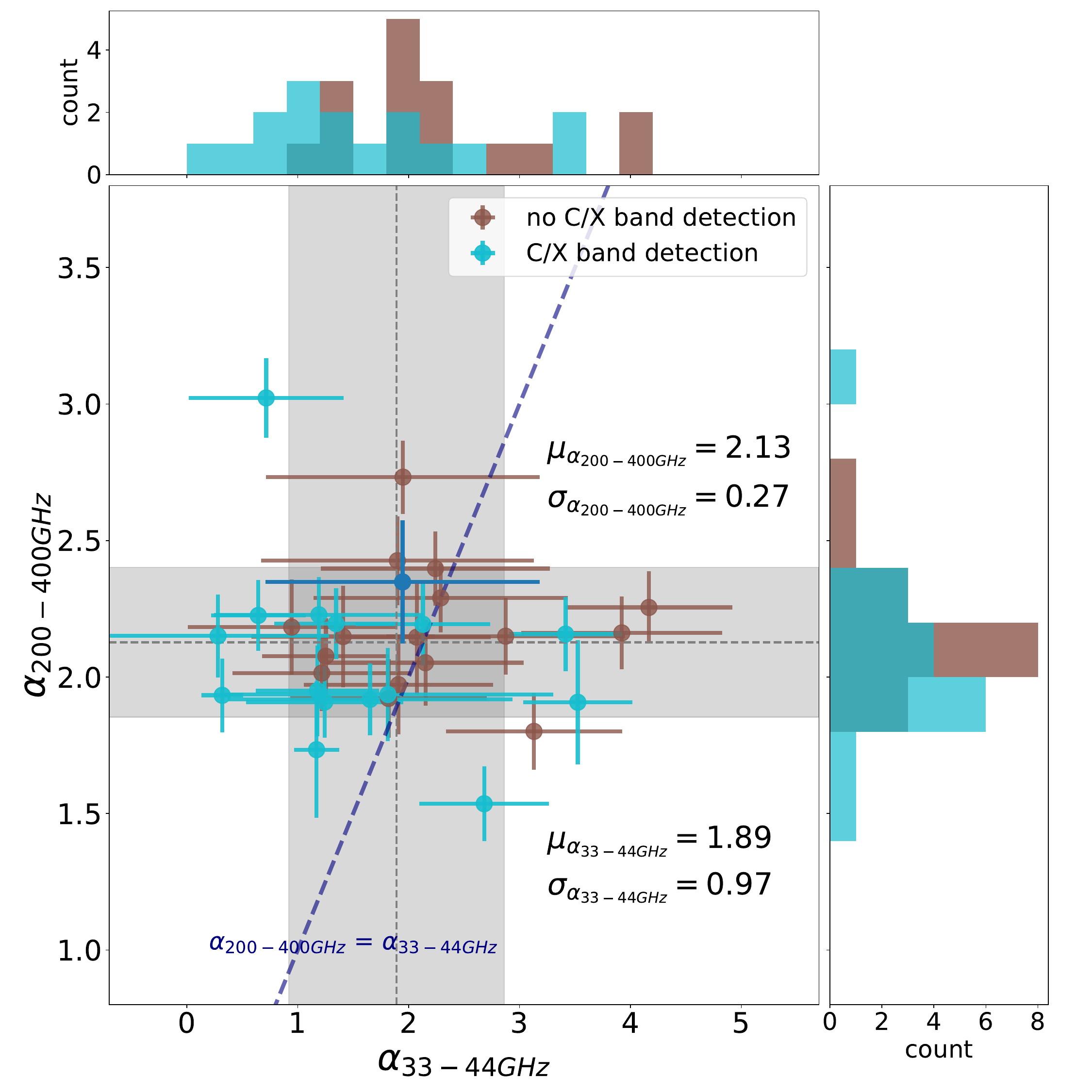}\\
    \caption{
    The Ka-Q band spectral index ($\alpha_{\mbox{\scriptsize 33-44 GHz}}$) versus the submillimeter spectral index ($\alpha_{\mbox{\scriptsize 200-400GHz}}$) quoted from \citet{Chung2024ApJS..273...29C}.  
    Color codes and symbols are similar to those in Figure \ref{fig:F44_alphaKaQ}.
    Gray vertical and horizontal dashed lines (and shaded areas) in the scatter plot represent the means ($\mu_{\alpha}$) (and standard deviations ($\sigma_{\alpha}$)) of the $\alpha_{\mbox{\scriptsize 33-44 GHz}}$ and $\alpha_{\mbox{\scriptsize 200-400GHz}}$ of all the sources, respectively, which are also labeled in text.
    The dark blue dashed line shows where the two spectral indices are equal. 
    Histograms of $\alpha_{\mbox{\scriptsize 200-400GHz}}$ and $\alpha_{\mbox{\scriptsize 33-44 GHz}}$ for the radio detected and non-detected sources are shown in the right and top panels. }
    \label{fig:alphaKaQ_alphasubmm}
\end{figure}

With our data in hand, it is not possible to quantitatively assess the fractional contribution of free-free emission at Ka and Q bands using a uniform approach. 
We are subject to this issue since our experimental design assumed that free-free emission is mostly optically thin at $\gtrsim$6 GHz frequencies, which had been a common approach.
If it is indeed this case, we can extrapolate the flux densities of free-free emission from C or X band to higher frequencies based on an assumption of spectral indices $\alpha\sim$0 (\citealt{Reynolds1986ApJ...304..713R}).  
However, from none of the observed sources we found evidence that this assumption is true (Figure \ref{fig:SED_0}--\ref{fig:SED_2}).
Instead, from the sources that the free-free emission was relatively well constrained by the present observations (e.g. LkCa~15, RY~Tau, DQ~Tau, AB~Aur, V892~Tau), we found that the spectral indices are considerably higher than 0, indicating that free-free emission may be partly optically thick. 
In these cases, extrapolating flux densities of free-free emission from C and X bands to higher frequencies based on an optically thin assumption will lead to significant underestimates. 
The observations we utilized only made measurements at one of the C (6 GHz) and X (10 GHz) bands during a similar time period, not both.
Therefore, we did not obtain good constraints on the optical depth of free-free emission.
The extrapolation to higher frequency then becomes largely arbitrary.
Nevertheless, it is still possible to qualitatively discuss the fractional contribution of free-free emission at 30--50 GHz based on the distribution of the spectral indices $\alpha_{\mbox{\scriptsize 33-44 GHz}}$ (more below).

\subsection{Spectral variations and transition regions across 50–200 GHz}\label{sub:trans_region}

We observed a wide range of spectral indices $\alpha_{\mbox{\scriptsize 33-44 GHz}}$, varying from 0.3 to 4.2 (see Figure \ref{fig:F44_alphaKaQ}), with most values clustered around $\sim$2.0, but showing some prominent outliers. 
This indicates a broader scatter compared to the more narrowly concentrated distribution of the 200--400 GHz spectral index, $\alpha_{\mbox{\scriptsize 200-400GHz}}\sim$2.0, reported in the SMA survey (\citealt{Chung2024ApJS..273...29C}).
In most cases, the flux densities at frequencies below 50 GHz are lower than the extrapolation from 200--400 GHz based on a constant spectral index assumption (see Figure \ref{fig:SED_0}--\ref{fig:SED_2}).
This suggests that there are transitional regions in the spectra between 50--200 GHz frequencies where the flux densities rapidly increase with frequency. 
In these transition regions, the spectral indices may be around 3--4, although they are not spectrally resolved in all observed cases (more below).
The observed $\alpha_{\mbox{\scriptsize 44-200GHz}}$ of most sources lies between 2.0 and 3.5 (Figure \ref{fig:F44_alphaQsubmm}), which is likely a consequence of averaging the high spectral indices in the transition regions and the lower spectral indices at frequencies higher and lower than those of the transition regions.

Incorporating the previous flux density measurements taken between 50 GHz and 200 GHz frequencies helps in gaining a better understanding of the spectral profiles in the transition regions (50--200 GHz). 
In some cases, such as BP~Tau (Figure \ref{fig:SED_0}), we observed that the spectral profile, resolved at 200--400 GHz smoothly transitions to a sharp turning point at $\sim$100 GHz.
In this case, there is another turning point at $\gtrsim$50 GHz where the spectral index rapidly increases with frequency. 
The frequency ranges of the transition regions (or the frequencies of the turning points) change from source to source (Appendix \ref{appendix:SED_fitting}). 
For example (Figure \ref{fig:SED_1}), in the spectra of Haro~6-37 and UY~Aur, we resolved relatively steep transitions at $\sim$35--50 GHz; the transition regions are close to 90--110 GHz in the cases of BP~Tau and CI~Tau; the transition regions are close to 150 GHz in the cases of CY~Tau and DE~Tau, etc.
In the cases of IQ~Tau and AA~Tau, the frequencies of the transition regions are loosely constrained to be in the range of 50--200 GHz, due to the lack of observations at 90--150 GHz.
For all of the selected sources, the frequencies of the transition regions can be better determined by performing Northern Extended Millimetre Array (NOEMA) Band 1 (70.384---119.872 GHz) and Band 2 (127.000--182.872 GHz) observations and/or ALMA Band 3 (84--116 GHz), 4 (125--163 GHz), and 5 (163--211 GHz) observations.
 

\subsection{Statistical results}\label{sub:statisticalresults}

Figure \ref{fig:F44_alphaKaQ} and \ref{fig:F44_alphaQsubmm} show the $\alpha_{\mbox{\scriptsize 33-44 GHz}}$ and $\alpha_{\mbox{\scriptsize 44-200GHz}}$ of all 32 sources in the selected sample, which are plotted against the Q band flux density.
For Haro~6-39, which was not detected in Q band, we adopted the 3-$\sigma$ upper limit as the Q band flux density in statistical analysis.
To assess whether or not free-free and/or synchrotron emission was prominent during the Ka and Q band observations (Sections \ref{sub:obs}, \ref{sub:obsDiskatEVLA}), we distinguished the 16 sources detected in the C band observations of the Disk@EVLA project or in our X band observations as "radio detected sources," while the remaining samples were referred to as "radio non-detected sources".

\begin{figure}
    \centering
    \includegraphics[width=8.5cm]{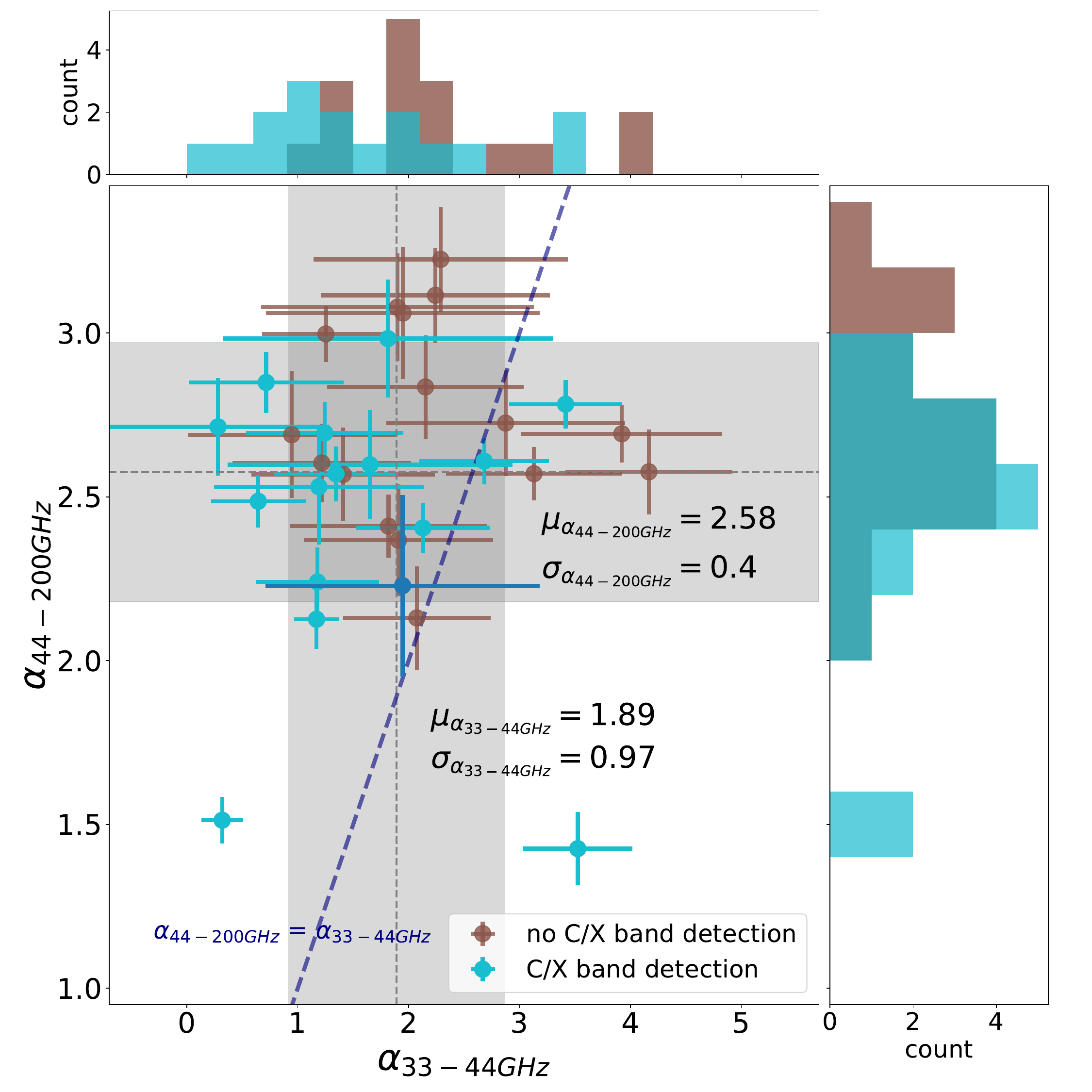}\\
    \caption{
    The Ka-Q band spectral index ($\alpha_{\mbox{\scriptsize 33-44 GHz}}$) versus the Q band-200 GHz spectral index ($\alpha_{\mbox{\scriptsize 44-200GHz}}$).
    Color codes and symbols are similar to those in Figure \ref{fig:F44_alphaKaQ}.
    Gray vertical and horizontal dashed lines (and shaded areas) in the scatter plot represent the means ($\mu_{\alpha}$) (and standard deviations ($\sigma_{\alpha}$)) of the $\alpha_{\mbox{\scriptsize 33-44 GHz}}$ and $\alpha_{\mbox{\scriptsize 44-200GHz}}$ of all the sources, respectively, which are also labeled in text.
    The dark blue dashed line shows where the two spectral indices are equal. 
    Histograms of $\alpha_{\mbox{\scriptsize 44-200GHz}}$ and $\alpha_{\mbox{\scriptsize 33-44 GHz}}$ for the radio detected and non-detected sources are shown in the right and top panels. }
    \label{fig:alphaQsubmm_alphaKaQ}
\end{figure}

We observed that a significant portion of the radio non-detected sources have $\alpha_{\mbox{\scriptsize 33-44 GHz}}$ values consistent with 2.0 (Figure \ref{fig:F44_alphaKaQ}). 
On the other hand, the radio detected sources show systematically lower $\alpha_{\mbox{\scriptsize 33-44 GHz}}$ values. 
Moreover, sources with low $\alpha_{\mbox{\scriptsize 33-44 GHz}}$ values tend to be bright Q band emitters. 
As a result, the radio detected sources tend to occupy the lower right part of Figure \ref{fig:F44_alphaKaQ}, while the radio non-detected sources tend to occupy the upper left part of that figure.
However, we note that both sub-samples exhibit a similarly large scatter.

The $\alpha_{\mbox{\scriptsize 44-200GHz}}$ values are distributed in a narrower range of 2.0--3.5 than that of $\alpha_{\mbox{\scriptsize 33-44 GHz}}$ (Figure \ref{fig:F44_alphaQsubmm}). 
Most radio detected and non-detected sources have comparable $\alpha_{\mbox{\scriptsize 44-200 GHz}}$.
Exceptions are the variable sources, T~Tau, which has $\alpha_{\mbox{\scriptsize 33-44 GHz}}=$1.51$\pm$0.07 (Figure \ref{fig:SED_2}); and UY~Aur, which has $\alpha_{\mbox{\scriptsize 33-44 GHz}}=$1.43$\pm$0.11 (Figure \ref{fig:SED_0}).

\begin{figure}
    \centering
    \includegraphics[width=8.5cm]{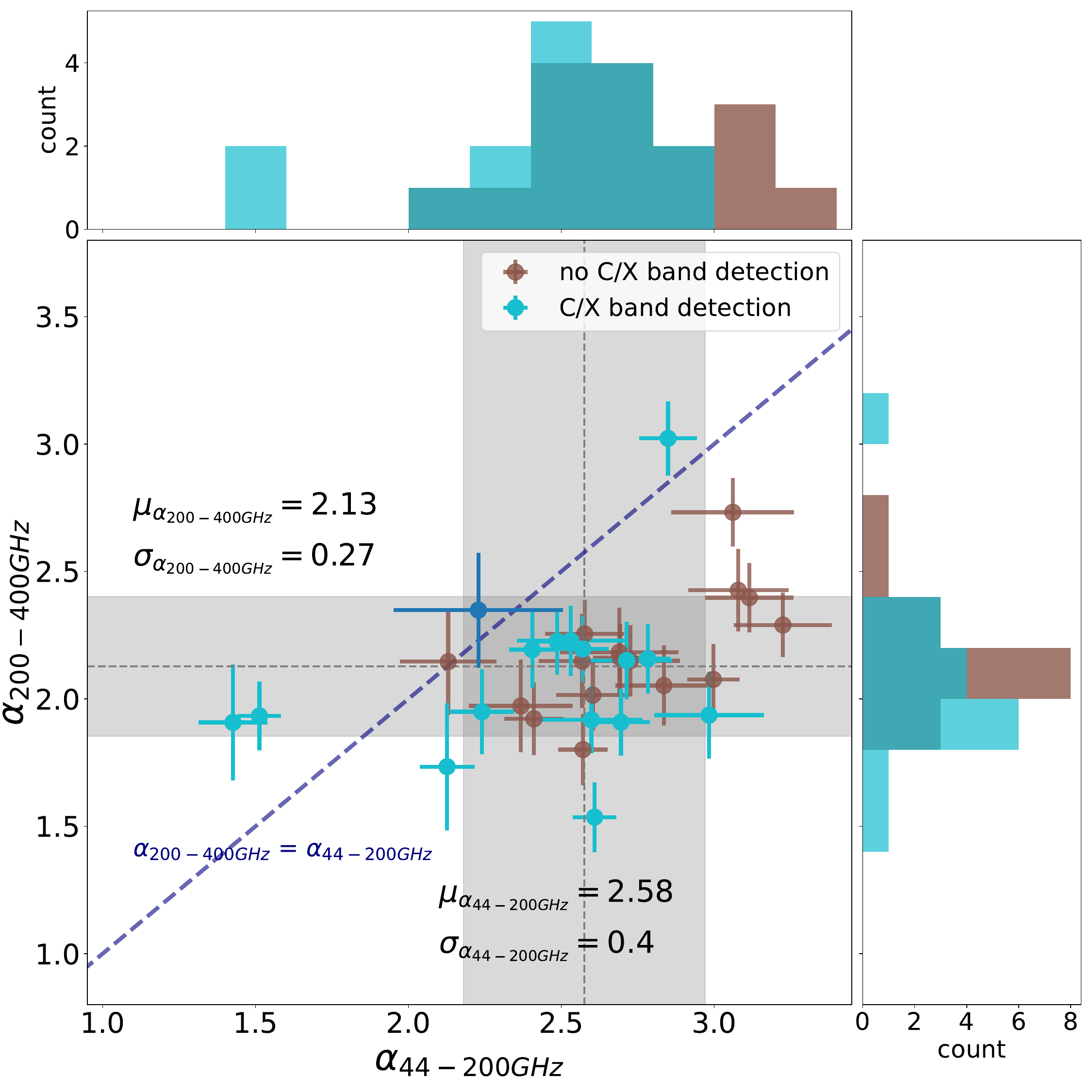}\\
    \caption{
    The Q band-200 GHz spectral index ($\alpha_{\mbox{\scriptsize 44-200GHz}}$) versus the submillimeter spectral index ($\alpha_{\mbox{\scriptsize 200-400GHz}}$) quoted from \citet{Chung2024ApJS..273...29C}.  
    Color codes and symbols are similar to those in Figure \ref{fig:F44_alphaKaQ}.
    Gray vertical and horizontal dashed lines (and shaded areas) in the scatter plot represent the means ($\mu_{\alpha}$) (and standard deviations ($\sigma_{\alpha}$)) of the $\alpha_{\mbox{\scriptsize 44-200GHz}}$ and $\alpha_{\mbox{\scriptsize 200-400GHz}}$ of all the sources, respectively, which are also labeled in text.
    The dark blue dashed line shows where the two spectral indices are equal. 
    Histograms of $\alpha_{\mbox{\scriptsize 200-400GHz}}$ and $\alpha_{\mbox{\scriptsize 44-200GHz}}$ for the radio detected and non-detected sources are shown in the right and top panels. }
    \label{fig:alphaQsubmm_alphasubmm}
\end{figure}

To examine the change in spectral index from submillimeter to millimeter wavelengths, we compare $\alpha_{\mbox{\scriptsize 33-44 GHz}}$ with $\alpha_{\mbox{\scriptsize 200-400GHz}}$, $\alpha_{\mbox{\scriptsize 44-200GHz}}$ with $\alpha_{\mbox{\scriptsize 33-44 GHz}}$, and $\alpha_{\mbox{\scriptsize 44-200GHz}}$ with $\alpha_{\mbox{\scriptsize 200-400GHz}}$ in Figure \ref{fig:alphaKaQ_alphasubmm}, Figure \ref{fig:alphaQsubmm_alphaKaQ}, and Figure \ref{fig:alphaQsubmm_alphasubmm}, respectively.
In Figure \ref{fig:alphaKaQ_alphasubmm}, the $\alpha_{\mbox{\scriptsize 200-400GHz}}$ are concentrated to a narrow range of 2.1$\pm$0.3, which is consistent with the measurement for a larger sample reported by \citet{Chung2024ApJS..273...29C}. 
We do not observe any significant differences in the $\alpha_{\mbox{\scriptsize 200-400GHz}}$ values between the radio-detected and non-detected sources, although this may be partly due to our limited sample size.
In general, at frequencies $>$200 GHz, free-free and synchrotron emission are likely negligible compared to dust emission, although there are exceptions (e.g., DQ~Tau, \citealt{Salter2010A&A...521A..32S,Chung2024ApJS..273...29C}).
In Figure \ref{fig:alphaQsubmm_alphasubmm}, most of the sources, especially the radio non-detected sources, show higher $\alpha_{\mbox{\scriptsize 44-200GHz}}$ than $\alpha_{\mbox{\scriptsize 200-400GHz}}$.
In most of the sources, $\alpha_{\mbox{\scriptsize 44-200GHz}}$ is also higher than $\alpha_{\mbox{\scriptsize 33-44 GHz}}$ (Figure \ref{fig:alphaQsubmm_alphaKaQ}). 
As discussed in Section \ref{sub:trans_region}, this indicates that in a spectrum, there is a transition region between 50--200 GHz where the spectral index is high (approximately 3--4).
In over 50\% of our sample, the value of $\alpha_{\mbox{\scriptsize 33-44 GHz}}$ is lower than $\alpha_{\mbox{\scriptsize 200-400GHz}}$, which is contrary to the trend expected from a single emission component that has a dust origin (cf. \citealt{Hildebrand1983QJRAS..24..267H,Liu2019ApJ...877L..22L}, and references therein; for more discussion see the following section).



\section{Discussion}\label{sec:discussions}

\subsection{Characterizing the 20--400 GHz broadband spectra of the Class II disks
}

In our present study, we expect that dust emission dominates at $>$200 GHz frequencies. 
At $<$100 GHz, it appears that both dust emission and free-free/synchrotron emission may be prominent, which are not clearly distinguished from the present observations. 
In most of our observed cases, this hinders fitting with physical functions.
For example, the source CI Tau (in Figure \ref{fig:SED_0}) has 2 constraints and 3 upper limits at $<$50 GHz frequency, and some constraints at $>$90 GHz. 
At frequencies below 50 GHz, it is possible to fit the data using a dust emission spectrum or a combination of dust and free-free emission, while the resulting parameters are too degenerate to yield any meaningful physical interpretation. 
Although the fitting functions themselves are physically motivated, the parameters derived from these fits are not. 
The spectral index is the only concrete information in this frequency range.

In addition, when extrapolating the fitting from frequencies below 50 GHz to above 90 GHz, there is an arbitrary choice of whether (i) the dominant emitter of <50 GHz emission should be obscured by the dust emission source that dominates the $>$200 GHz spectrum, or (ii) we should simply coadd the emission from the emission sources, assuming no mutual obscuration. 
These arbitrary choices make it difficult to extract further information beyond the spectral index and the transition frequency. 
The sources, V836\,Tau (Figure \ref{fig:SED_0}), GM\,Aur, Haro\,6-39, AA\,Tau, DE\,Tau, FT\,Tau (Figure \ref{fig:SED_1}), and many others, are subject to the same problem as CI Tau. 
BP Tau may allow fitting a physical function at $>$90 GHz, while the lower frequency can only be characterized by a power-law with no physical meaning.

For most sources, the observed spectra at $\gtrsim$20 GHz can be characterize by two power laws connected by a sigmoid function (Figure \ref{fig:SED_0}--\ref{fig:SED_2}; Table \ref{tab:fit_parameter}), which are detailed in Appendix \ref{appendix:SED_fitting}. 
The fittings in Appendix \ref{appendix:SED_fitting} characterized the spectral indices at high and low frequencies, the frequency of the transition regions, and the steepness of the transition. 
The spectral profiles at $<$20 GHz are more complicated, which may be due to the contribution of multiple emission mechanisms (e.g., free-free, synchrotron, and dust) and their potential variability. 
The transition region is necessary because the intersection of the two power laws does not naturally meet in the middle. 

To characterize the transition region, we chose to use a sigmoid function in this paper to systematically derive the transitional frequency without explicitly deriving the spectral index in the transition region. 
The spectral index in this range likely lacks physical meaning due to the mixed contributions of different emission mechanisms. 
While alternative approaches, such as piecewise power-law fits, could also characterize this region, the sigmoid function provides a practical and systematic framework for estimating the transition frequency. 
We note that piecewise power-laws fit (as well as other similar approaches) are not necessarily more physical than connecting two power-laws with a sigmoid function, in the sense that the slopes of the piecewise power-laws are step functions. 
On the contrary, the functional form of the sigmoid function allows a relatively smooth transition (e.g., the cases of DL\,Tau and LkCa\,15 in Figure \ref{fig:SED_0}, etc). 

It is important to note that the sigmoid function is not intended to imply a physically meaningful model for the transition but rather provides a practical and systematic framework for estimating the transition frequency. 
We also note that due to the lack of data points in the transition region, the slope of the transition cannot be reliably constrained, as shown in Figures \ref{fig:SED_0}–\ref{fig:SED_2} and Table \ref{tab:fit_parameter}. 
Our conclusions, particularly the differences in emission components between high and low frequencies, remain unaffected by the choice of transition model.

When the data are more complete, we can do a fairly sophisticated fitting of dust thermal emission (e.g., DM\,Tau, \citealt{Liu2024A&A...685A..18L}). 
In general, the data presented in this study provide useful constraints on certain physical parameters, such as the spectral index at 20--50 GHz, at 200--400 GHz, and the transition frequency. 
However, deriving physical insights in a uniform and systematic way across all sources remains challenging. 
We aim to address these complexities in follow-up case studies of individual sources as additional data become available. 

\subsection{Interpretation of emission components in Class II disk spectra and their implications}
 
In general, the 4--400 GHz broadband spectra of the Class II protoplanetary disks in the Taurus-Auriga region require at least two distinct emission components for interpretation.
The first is the dust emission in the bulk of the disk that dominates the emission at $>$200 GHz frequencies.
Other components (e.g., high-density dust crescents or clumps, free-free emission, synchrotron emission, and emission from spinning nanometer-sized dust, c.f. \citealt{Hildebrand1983QJRAS..24..267H,Liu2019ApJ...877L..22L,Reynolds1986ApJ...304..713R,Anglada1998AJ....116.2953A,Gudel2002ARA&A..40..217G,Hoang2018ApJ...862..116H}) dominate at lower frequencies.

In the case of radio non-detected sources, flux densities in the 30--50 GHz range with $\alpha_{\mbox{\scriptsize 33-44 GHz}}\sim$2 indicate either dust emission, optically thick free-free emission, or a combination of both.
For the radio detected sources, the consistently lower $\alpha_{\mbox{\scriptsize 33-44 GHz}}$ values (Figure \ref{fig:F44_alphaKaQ}) suggest that the 30--50 GHz flux densities are partly due to optically thin free-free emission (\citealt{Reynolds1986ApJ...304..713R,Anglada1998AJ....116.2953A}) or synchrotron emission (\citealt{Gudel2002ARA&A..40..217G}). 
Dust thermal emission could be significant in this frequency range.

For dust thermal emission to have $\alpha_{\mbox{\scriptsize 33-44 GHz}}\sim$2, the dust emission source needs to have either very high optical depth or very large $a_{\rm max}$ (\citealt{Hildebrand1983QJRAS..24..267H}).
Besides, high optical depths could be achieved by concentrating a lot of dust mass, another possibility to locally enhance the dust optical depth may be dust grain growth that yields millimeter-centimeter grain sizes. 
In this case, the dust absorption opacities can be maximized at 33--44 GHz.
Otherwise, including some specific form of carbonaceous dust (e.g. graphite or certain types of amorphous dust) in the dust composition (i.e. segregation of dust composition) may also significantly enhance the dust absorption opacity, although it is not yet clear how much of these materials can be included (see the discussion in \citealt{Draine2006ApJ...636.1114D}).

\begin{deluxetable*}{c|ccc|cccc}
\tabletypesize{\footnotesize}
\tablecolumns{8}
\tablewidth{0pt}
\tablecaption{ Observed spectral indices and best fit parameters in spectrum fitting \label{tab:fit_parameter}  }
\tablehead{ 
 \colhead{Source} & \colhead{$\alpha_{\mbox{\tiny 33-44 GHz}}$} & \colhead{$\alpha_{\mbox{\tiny 200-400GHz}}$} & \colhead{$\alpha_{\mbox{\tiny 44-200GHz}}$} & \colhead{$\alpha_{\mbox{\tiny mm}}$} & \colhead{$\alpha_{\mbox{\tiny submm}}$} &  \colhead{$\omega$} & \colhead{${b/\omega}$} \\
 \colhead{} & \colhead{} & \colhead{} & \colhead{} & \colhead{} & \colhead{} & \colhead{} & \colhead{(GHz)} }
\startdata
AA Tau & $1.9 \pm 1.23$ & $2.43 \pm 0.16$ & $3.08 \pm 0.16$ & $2.38 \pm ^{1.1} _{1.09} $ & $2.43 \pm ^{0.17} _{0.17} $ & $0.23 \pm ^{0.049} _{-0.153} $ & $118.52 \pm ^{45.22} _{43.51} $ \\
AB Aur & $0.72 \pm 0.7$ & $3.02 \pm 0.15$ & $2.85 \pm 0.09$ & $1.11 \pm ^{0.62} _{0.42} $ & $3.02 \pm ^{0.13} _{0.14} $ & $0.349 \pm ^{0.127} _{-0.086} $ & $67.91 \pm ^{17.17} _{15.47} $ \\
BP Tau & $2.15 \pm 0.89$ & $2.05 \pm 0.16$ & $2.84 \pm 0.16$ & $2.37 \pm ^{0.82} _{0.82} $ & $2.04 \pm ^{0.16} _{0.15} $ & $0.274 \pm ^{0.027} _{-0.07} $ & $90.59 \pm ^{2.08} _{1.26} $ \\
CI Tau & $2.24 \pm 1.03$ & $2.4 \pm 0.14$ & $3.12 \pm 0.14$ & $2.39 \pm ^{0.94} _{0.97} $ & $2.4 \pm ^{0.14} _{0.13} $ & $0.066 \pm ^{-0.131} _{-0.054} $ & $111.29 \pm ^{2.83} _{2.13} $ \\
CW Tau & $1.19 \pm 0.94$ & $2.23 \pm 0.14$ & $2.53 \pm 0.18$ & $1.47 \pm ^{0.78} _{0.63} $ & $2.2 \pm ^{0.11} _{0.12} $ & $0.169 \pm ^{-0.047} _{-0.096} $ & $100.83 \pm ^{13.88} _{9.69} $ \\
CY Tau & $2.68 \pm 0.58$ & $1.54 \pm 0.14$ & $2.61 \pm 0.07$ & $2.71 \pm ^{0.59} _{0.56} $ & $1.54 \pm ^{0.13} _{0.13} $ & $0.289 \pm ^{0.097} _{-0.118} $ & $165.15 \pm ^{24.75} _{27.53} $ \\
DE Tau & $1.91 \pm 0.85$ & $1.97 \pm 0.18$ & $2.37 \pm 0.17$ & $2.05 \pm ^{0.78} _{0.8} $ & $1.98 \pm ^{0.18} _{0.17} $ & $0.227 \pm ^{0.046} _{-0.153} $ & $150.95 \pm ^{31.24} _{32.64} $ \\
DL Tau & $4.17 \pm 0.75$ & $2.25 \pm 0.13$ & $2.58 \pm 0.13$ & $1.23 \pm ^{0.1} _{0.09} $ & $2.19 \pm ^{0.1} _{0.11} $ & $0.059 \pm ^{-0.01} _{-0.009} $ & $81.44 \pm ^{1.99} _{1.86} $ \\
DM Tau & $1.25 \pm 0.57$ & $2.08 \pm 0.14$ & $3.0 \pm 0.09$ & $1.37 \pm ^{0.58} _{0.48} $ & $2.09 \pm ^{0.13} _{0.14} $ & $0.118 \pm ^{-0.107} _{-0.063} $ & $105.4 \pm ^{3.42} _{3.84} $ \\
DN Tau & $3.13 \pm 0.79$ & $1.8 \pm 0.14$ & $2.57 \pm 0.08$ & $1.67 \pm ^{0.76} _{0.64} $ & $1.8 \pm ^{0.13} _{0.13} $ & $0.072 \pm ^{-0.146} _{-0.073} $ & $75.6 \pm ^{36.81} _{23.51} $ \\
DO Tau & $1.22 \pm 0.81$ & $2.01 \pm 0.14$ & $2.6 \pm 0.12$ & $1.44 \pm ^{0.66} _{0.59} $ & $2.01 \pm ^{0.13} _{0.13} $ & $0.136 \pm ^{-0.094} _{-0.068} $ & $76.97 \pm ^{4.22} _{5.06} $ \\
DQ Tau & $1.18 \pm 0.56$ & $1.95 \pm 0.17$ & $2.24 \pm 0.11$ & $1.43 \pm ^{0.57} _{0.54} $ & $1.95 \pm ^{0.16} _{0.17} $ & $0.12 \pm ^{-0.112} _{-0.081} $ & $80.18 \pm ^{6.15} _{11.25} $ \\
DR Tau & $3.42 \pm 0.51$ & $2.16 \pm 0.14$ & $2.78 \pm 0.07$ & $2.17 \pm ^{0.29} _{0.33} $ & $2.17 \pm ^{0.14} _{0.13} $ & $0.03 \pm ^{-0.008} _{-0.007} $ & $110.67 \pm ^{4.73} _{4.1} $ \\
DS Tau & $0.95 \pm 0.93$ & $2.18 \pm 0.17$ & $2.69 \pm 0.19$ & $1.35 \pm ^{0.74} _{0.55} $ & $1.91 \pm ^{0.12} _{0.13} $ & $0.06 \pm ^{-0.042} _{-0.028} $ & $100.66 \pm ^{0.91} _{1.06} $ \\
FT Tau & $1.82 \pm 0.89$ & $1.92 \pm 0.14$ & $2.41 \pm 0.1$ & $1.96 \pm ^{0.87} _{0.76} $ & $1.93 \pm ^{0.14} _{0.13} $ & $0.056 \pm ^{-0.132} _{-0.053} $ & $112.45 \pm ^{6.75} _{3.59} $ \\
GM Aur & $1.95 \pm 1.23$ & $2.73 \pm 0.13$ & $3.06 \pm 0.2$ & $2.56 \pm ^{1.08} _{1.13} $ & $2.76 \pm ^{0.11} _{0.12} $ & $1.677 \pm ^{1.091} _{0.611} $ & $82.23 \pm ^{20.69} _{24.79} $ \\
GO Tau & $1.81 \pm 1.49$ & $1.94 \pm 0.17$ & $2.98 \pm 0.18$ & $2.37 \pm ^{1.31} _{1.09} $ & $2.48 \pm ^{0.12} _{0.14} $ & $0.282 \pm ^{0.064} _{-0.084} $ & $78.23 \pm ^{7.06} _{20.13} $ \\
Haro 6-37 & $3.92 \pm 0.91$ & $2.16 \pm 0.13$ & $2.69 \pm 0.09$ & $1.33 \pm ^{0.53} _{0.45} $ & $2.17 \pm ^{0.13} _{0.14} $ & $0.168 \pm ^{-0.064} _{-0.138} $ & $49.09 \pm ^{17.59} _{4.53} $ \\
Haro 6-39 & $1.95 \pm 1.24$ & $2.35 \pm 0.23$ & $2.23 \pm 0.28$ & $0.82 \pm ^{0.18} _{0.18} $ & $2.35 \pm ^{0.21} _{0.22} $ & $0.195 \pm ^{0.016} _{-0.155} $ & $111.6 \pm ^{53.34} _{51.25} $ \\
IP Tau & $1.41 \pm 0.83$ & $2.15 \pm 0.19$ & $2.57 \pm 0.14$ & $1.6 \pm ^{0.8} _{0.67} $ & $2.15 \pm ^{0.17} _{0.18} $ & $0.211 \pm ^{0.036} _{-0.161} $ & $117.97 \pm ^{45.87} _{45.46} $ \\
IQ Tau & $2.87 \pm 1.08$ & $2.15 \pm 0.14$ & $2.73 \pm 0.16$ & $1.93 \pm ^{1.12} _{0.75} $ & $2.15 \pm ^{0.13} _{0.13} $ & $0.23 \pm ^{0.051} _{-0.145} $ & $118.54 \pm ^{45.24} _{46.13} $ \\
LkCa 15 & $2.29 \pm 1.15$ & $2.29 \pm 0.13$ & $3.22 \pm 0.16$ & $2.53 \pm ^{1.05} _{1.04} $ & $2.05 \pm ^{0.09} _{0.09} $ & $0.049 \pm ^{-0.025} _{-0.017} $ & $121.78 \pm ^{3.15} _{2.87} $ \\
MHO 1 & $1.24 \pm 0.71$ & $1.91 \pm 0.13$ & $2.41 \pm 0.1$ & $1.42 \pm ^{0.66} _{0.55} $ & $1.9 \pm ^{0.13} _{0.12} $ & $0.342 \pm ^{0.133} _{-0.089} $ & $78.55 \pm ^{13.87} _{17.45} $ \\
RY Tau & $1.35 \pm 0.56$ & $2.2 \pm 0.13$ & $2.57 \pm 0.08$ & $1.4 \pm ^{0.54} _{0.49} $ & $1.94 \pm ^{0.1} _{0.1} $ & $0.077 \pm ^{-0.128} _{-0.055} $ & $112.58 \pm ^{4.49} _{2.06} $ \\
SU Aur & $1.17 \pm 0.2$ & $1.73 \pm 0.25$ & $2.13 \pm 0.09$ & $1.18 \pm ^{0.21} _{0.22} $ & $1.72 \pm ^{0.22} _{0.23} $ & $0.069 \pm ^{-0.135} _{-0.062} $ & $95.92 \pm ^{15.35} _{8.14} $ \\
T Tau & $0.32 \pm 0.19$ & $1.93 \pm 0.14$ & $1.51 \pm 0.07$ & $0.62 \pm ^{0.12} _{0.08} $ & $1.93 \pm ^{0.14} _{0.13} $ & $0.112 \pm ^{-0.11} _{-0.057} $ & $96.76 \pm ^{8.63} _{8.45} $ \\
UX Tau & $2.13 \pm 0.61$ & $2.19 \pm 0.15$ & $2.41 \pm 0.08$ & $2.15 \pm ^{0.69} _{0.58} $ & $2.19 \pm ^{0.16} _{0.14} $ & $0.076 \pm ^{-0.157} _{-0.078} $ & $88.94 \pm ^{9.09} _{16.79} $ \\
UY Aur & $3.52 \pm 0.49$ & $1.91 \pm 0.23$ & $1.43 \pm 0.11$ & $1.46 \pm ^{0.47} _{0.42} $ & $1.48 \pm ^{0.07} _{0.06} $ & $0.782 \pm ^{0.13} _{0.122} $ & $38.52 \pm ^{1.33} _{1.27} $ \\
UZ Tau & $1.65 \pm 1.29$ & $1.92 \pm 0.13$ & $2.6 \pm 0.17$ & $2.08 \pm ^{1.13} _{0.94} $ & $2.05 \pm ^{0.09} _{0.09} $ & $0.026 \pm ^{-0.016} _{-0.01} $ & $120.19 \pm ^{5.14} _{4.59} $ \\
V710 Tau & $0.28 \pm 1.0$ & $2.15 \pm 0.15$ & $2.71 \pm 0.15$ & $1.11 \pm ^{0.72} _{0.43} $ & $2.15 \pm ^{0.15} _{0.15} $ & $0.24 \pm ^{0.057} _{-0.144} $ & $121.24 \pm ^{44.14} _{43.42} $ \\
V836 Tau & $2.07 \pm 0.66$ & $2.15 \pm 0.21$ & $2.13 \pm 0.16$ & $2.14 \pm ^{0.04} _{0.04} $ & $1.83 \pm ^{0.12} _{0.11} $ & $0.375 \pm ^{0.147} _{-0.072} $ & $72.64 \pm ^{15.6} _{15.46} $ \\
V892 Tau & $0.64 \pm 0.43$ & $2.23 \pm 0.13$ & $2.49 \pm 0.08$ & $0.84 \pm ^{0.34} _{0.24} $ & $2.22 \pm ^{0.13} _{0.13} $ & $0.399 \pm ^{0.15} _{-0.062} $ & $70.75 \pm ^{15.85} _{15.04} $ \\
\enddata
\tablecomments{ \footnotesize
$\alpha_{\mbox{\tiny 33-44 GHz}}$: spectral index between JVLA Ka band and Q band. $\alpha_{\mbox{\tiny 200-400GHz}}$: submillimeter spectral index quoted from \citet{Chung2024ApJS..273...29C}. $\alpha_{\mbox{\tiny 44-200GHz}}$: spectral index between JVLA Q band and SMA 200 GHz band. $\alpha_{\mbox{\tiny mm}}$: spectral index of the submillimeter power law, $P_{\mbox{\scriptsize submm}}(\nu)$. $\alpha_{\mbox{\tiny submm}}$:spectral index of the millimeter power law, $P_{\mbox{\scriptsize mm}}(\nu)$. $\omega$: steepness of transition in the sigmoid function. $\sigma_{\omega, b}(\nu)$, ${b/\omega}$: mean frequency of the transition region in $\sigma_{\omega, b}(\nu)$. }
\end{deluxetable*}

Further complexity arises across the snow line, a transition boundary where dust composition and spectral index can change rapidly, as noted by \citet{Testi2014prpl.conf..339T}. 
Such transitions might be relevant for understanding changes in spectral index, which could exhibit distinct features tied to compositional changes across this line (e.g. \citealt{Liu2021ApJ...923..270L,Houge2024MNRAS.527.9668H}).

Our current understanding of the 4--400 GHz spectra (Section \ref{sub:trans_region}) and the statistical results (Section \ref{sub:statisticalresults}) suggests that the dust processing scenario described in Section \ref{sec:intro} may be applicable in a good fraction of Class II protoplanetary disks.
In the bulk of the disk, the emission is optically thick at 200--400 GHz, resulting in $\alpha_{\rm 200-400GHz}\sim$2.0. 
In certain disks that the observed spectra have allowed for sophisticated SED fittings (e.g.,  HD\,163296, DM\,Tau; \citealt{Guidi2022A&A...664A.137G,Liu2024A&A...685A..18L}), or in the disks where linearly polarized (sub)millimeter dust polarization have been detected (e.g., HL\,Tau, HD\,163296, IM\,Lup, HD\,142527, etc; \citealt{Kataoka2016ApJ...820...54K,Stephens2023Natur.623..705S,Dent2019MNRAS.482L..29D,Hull2018ApJ...860...82H,Kataoka2016ApJ...831L..12K}), the 200--400 GHz emission in the bulk of these disks may be dominated by optically thick thermal emission of small dust grains (e.g., $a_{\rm max}\sim$50--150 $\mu$m).
In the remaining disks, the 200--400 GHz emission in the bulk of the disks is consistent with being dominated by the optically thick thermal emission of small dust grains (e.g., $a_{\rm max}\lesssim$0.1 mm), although the optically thick thermal emission of grown dust grains (e.g., $a_{\rm max}\gtrsim$3 mm) is also consistent with the observations (c.g. \citealt{Chung2024ApJS..273...29C}).
In addition, in many disks, there are embedded substructures (regions of grown dust) with high $\Sigma_{\mbox{\scriptsize dust}}$ and/or high $a_{\mbox{\scriptsize max}}$ values. 
These dust substructures have small linear and angular dimensions, and therefore contribute only a small fraction of the 200-400 GHz flux densities.
As the bulk of the disk becomes optically thin at around 50--200 GHz, its flux density drops rapidly at low frequencies.
Consequently, at $\lesssim$50 GHz, the flux densities of dust substructures become more prominent compared to the emission in the bulk of the disks. 
Such dust substructures may be distinguished from the free-free or synchrotron emission sources in high angular resolution imaging observations.
The dust substructures may be important sites for forming planetesimals and rocky planets.

\section{Conclusion}\label{sec:conclusion}
We have performed the JVLA X (8--12 GHz), Ka (29--37 GHz), and Q (40--48 GHz) observations towards 8 selected Class II protoplanetary disks in the Taurus-Auriga region.
Comparing with the C (4.8--6.8 GHz), K (20.0--22.0 GHz), Ka (31.7--33.6 GHz), and Q (41.0--43.0 GHz) band observations taken from the Disk@EVLA project (PI: C. Chandler) in the C and D array configurations, the previous SMA 200--400 GHz survey (\citealt{Chung2024ApJS..273...29C}), and various flux density measurement published over the last 20 years, we have compiled the 4--400 GHz spectra for 32 Class II protoplanetary disks in the Taurus-Auriga region. 
At submillimeter band, this sample represents the top 33 percentile of the brighest Class II disks observed in this region.
We found that in general, the spectra at $>$20 GHz frequency can be described with a piecewise function: a power law with spectral index $\sim$2 in the $>$200 GHz region, a power law with spectral index in the range of 0.3--4.2 in the 20--50 GHz region, and a transition region in between these two power laws which can be characterized by a sigmoid function. 
This implies that the flux densities at $>$200 GHz and $<$50 GHz are dominated by distinct emission components.
At $>$200 GHz, the emission is likely dominated by the optically thick dust thermal emission in the bulk of the disks. 
In some sources that were not detected in the C and/or X band observations, the embedded, high-density (grown, e.g., $a_{\scriptsize max}\gtrsim$1 mm) dust substructures may contribute to a significant fraction of the flux densities at 30--50 GHz, and thus there 30--50 GHz spectral indices are mostly consistent with 2.0.
Among the selected samples, examples with previously spatially resolved high-density grown dust substructures include DM~Tau (\citealt{Liu2024A&A...685A..18L}) and LkCa~15 (\citealt{Isella2014ApJ...788..129I}).
However, at 30--50 GHz, free-free and/or synchrotron emission may be significant. 
In fact, the 30--50 GHz spectral indices of the bright Q band emitters in our sample tend to be systematically lower than 2.0, meaning that free-free emission and/or synchrotron are what made some sources bright at Q band.
Based on these results, we hypothesize that the embedded, high-density grown dust substructures are often found in resolved Class II protoplanetary disks. 
Investigating these dust substructures may be a key to understanding the formation of kilometer-sized planetesimals and rocky planets. 
To robustly investigate the properties of the embedded dust structures, one need to carry out high angular resolution observations at $\lesssim$50 GHz band to separate them from the free-free and/or synchrotron emissions. 
Otherwise, in the analyses of the spatially unresolved spectra, one need to simultaneously constrain the spectral profiles of the dust emission, free-free emission, and synchrotron emission. 
Based on the present observations, we caution that, in general, free-free emission is not necessarily optically thin at $>$10 GHz frequencies.
To correctly assess the dust properties in the (grown) dust substructures, it is important to constrain the optical depths of free-free emission with the radio spectra that are well sampled in the frequency domain. 


\begin{acknowledgments}
We thank the anonymous referee for useful suggestions. 
We thank Dr. Sean Andrews for useful discussion. 
The National Radio Astronomy Observatory is a facility of the National Science Foundation operated under cooperative agreement by Associated Universities, Inc.
The Submillimeter Array is a joint project between the Smithsonian Astrophysical Observatory and the Academia Sinica Institute of Astronomy and Astrophysics and is funded by the Smithsonian Institution and the Academia Sinica.
We recognize that Maunakea is a culturally important site for the indigenous Hawaiian people; we are privileged to study the cosmos from its summit.
C.Y.C. and H.B.L. are supported by the National Science and Technology Council (NSTC) of Taiwan (Grant Nos. 111-2112-M-110-022-MY3, 113-2112-M-110-022-MY3).
\end{acknowledgments}

%

\vspace{5mm}
\facilities{JVLA, ALMA, SMA}


\software{
          astropy \citep{astropy2022ApJ...935..167A},  
          Numpy \citep{VanDerWalt2011}, 
          CASA \citep[][]{CASA2022PASP..134k4501C},
          emcee \citep{Foreman-Mackey2013PASP},
          }



\bibliography{main}{}
\bibliographystyle{aasjournal}

\appendix

\section{Spectrum characterization}\label{appendix:SED_fitting}

When the observations do not sufficiently constrain physical functions, or when the target sources are not yet well understood to the level that can enable physical modeling, we can characterize the observations using functions that are not directly linked to physics. 
For example, when analyzing a spectrum with only 2 or 3 independent measurements in the frequency domain, we need to make the choice of
\begin{enumerate}
    \item fitting a power-law (which does not necessarily have a physical meaning) with a spectral index and report it, or
    \item fitting the spectrum of a presumed dust emission source, and report the free-parameters such as dust column density, dust size (or beta), temperature, and solid angle, etc.
\end{enumerate}
In many cases (e.g., when we cannot resolve degeneracy), only the first option is possible. 
For example, when observing $\alpha=$3.2 in a spatially unresolved source with only 200 GHz and 300 GHz measurements, it is not clear whether the source is optically thick or thin, and thus it is hard to extract useful physical parameters.

In light of this, in this work, we adopted an approach similar to that of the aforementioned option 1, to systematically characterize the spectra observed at $>$20 GHz.
The emission in this frequency range is not entirely dominated by free-free or synchrotron emission and thus is less subject to time variability. 
The main purpose is to quantitatively, jointly constrain the $\alpha$ at high and low frequencies, and a transitional frequency. 
Specifically, we fit piecewise functions using the following strategy:

Since emission at $>$200--400 GHz and at $<$50 GHz likely have distinct origins (e.g., the bulk of disks and localized high-density dust substructures, etc), we first separately fit the spectral profile at $\lesssim$50 GHz and at 200--400 GHz by two power-law functions, which are expressed as:
\begin{equation}
P_{\mbox{\scriptsize mm}}(\nu) = F_{\mbox{\scriptsize 44 GHz}} \left(\frac{\nu}{\mbox{44 GHz}}\right)^{\alpha_{\mbox{\tiny mm}}},
\end{equation}
\begin{equation}
P_{\mbox{\scriptsize submm}}(\nu) = F_{\mbox{\scriptsize 230 GHz}} \left(\frac{\nu}{\mbox{230 GHz}}\right)^{\alpha_{\mbox{\tiny submm}}},
\end{equation}
where $\nu$ is frequency in GHz units, $F_{\mbox{\scriptsize 44 GHz}}$ and $F_{\mbox{\scriptsize 230 GHz}}$ are flux densities at the 44 GHz and 230 GHz reference frequencies, and $\alpha_{\mbox{\scriptsize mm}}$ and $\alpha_{\mbox{\scriptsize submm}}$ are the power law indices in the frequency ranges $\sim$20--50 GHz and at 200--400 GHz, respectively. 
The power-law indices provide indications to the properties of the dominant emission sources at these two frequency ranges. 

We then use the sigmoid function to characterize the smooth transition between the two power laws in the 50--200 GHz frequency range
\begin{equation}
\sigma_{\omega, b}(\nu) = \frac{1}{1+e^{-\omega \nu + b}} = \frac{1}{1+e^{-\omega (\nu - b/\omega)}}
\label{eqn:sigmoid}
\end{equation}
where $\omega$ is the parameter that determines the steepness of the transition, and $b$ is the parameter that shifts the transition region in the frequency domain.
When the transition is steeper, the value of $\omega$ is larger. 
The mean frequency of the transition region is given by $b/\omega$, which indicates the frequency that the dust emission in the bulk of the dusty disk is becoming optically thin.

Using this strategy, each 20--400 GHz spectrum is effectively described by
\begin{equation}
F(\nu) = \left(1 - \sigma_{\omega, b}(\nu)\right) P_{\mbox{\scriptsize mm}}(\nu) + \sigma_{\omega, b}(\nu) P_{\mbox{\scriptsize submm}}(\nu),
\label{eqn:specmodel}
\end{equation}
where the six free parameters to fit are $F_{\mbox{\scriptsize 44 GHz}}$, $F_{\mbox{\scriptsize 230 GHz}}$, $\alpha_{\mbox{\scriptsize mm}}$, $\alpha_{\mbox{\scriptsize submm}}$, $b/\omega$ and $\omega$.

When fitting $P_{\mbox{\scriptsize submm}}(\nu)$, we utilized the data taken from the  SMA 200--400 GHz survey (\citealt{Chung2024ApJS..273...29C}). 
To determine $P_{\mbox{\scriptsize mm}}(\nu)$, we focused on the Q, Ka, and K band data that were either taken from our JVLA observations (Section \ref{sub:obs}) or from the Disk@EVLA observations (Section \ref{sub:obsDiskatEVLA}).
Since our data base is not uniform, we classified the sources into the following cases and treated the fitting to each case slightly differently:
(1) For the few sources that the high $\alpha_{\mbox{\tiny 33-44 GHz}}$ values are indicative of optically thin dust emission, if their $\alpha_{\mbox{\tiny 33-44 GHz}}$ is higher than $\alpha_{\mbox{\tiny 21-33 GHz}}$ (Haro~6-37, DR~Tau, DN~Tau and IQ~Tau), we fit $P_{\mbox{\scriptsize mm}}(\nu)$ to the K and Ka band data taken from the Disk@EVLA observations, 
(2) For the sources that missed Q or K band detection (Haro~6-39 and DL~Tau) we treated them as special cases and addressed their fitting individually,
(3) the fittings of the two sources UY~Aur and V836~Tau are treated individually as exceptional cases, since their spectral indices increase with frequency in the 50--200 GHz and $\sim$100--200 GHz frequency ranges, respectively,
(4) for the rest of the sources, which are the majority, we fit $P_{\mbox{\scriptsize mm}}(\nu)$ to the Q and Ka band data taken from the Disk$@$EVLA or our new JVLA observations.

For many sources that the low ($\lesssim$50 GHz) and high ($>$200 GHz) frequency range spectral indices are similar (e.g., $\sim$2), one may wonder if the {\it vertical offset} in the spectra that requires the sigmoid function is a flux calibration problem. 
This is not very likely, as the vertical offset in the spectra will appear large if the flux densities are presented in linear scale instead of log scale. 
Taking the observations on DN~Tau as an example (Figure \ref{fig:SED_0}), to fit the data with a single smooth power law would require that the combined radio and/or the millimeter systematic flux errors are roughly a factor of 6.
This is unrealistically huge as compared to the typically considered 3\%--5\% errors in the radio band and $\sim$10\% errors in the millimeter bands.
Moreover, if it was a flux scale issue, it should affect all the targets in a similar way. 
However, there is a wide variety of vertical offsets between the low and high frequency regimes, again much larger than one could expect from flux scale concerns.

The individually treated cases are discussed as follows.
\paragraph{Haro~6-39} It was not observed in the Disk@EVLA project and was not detected at Q band in our new JVLA observations. We did not find other useful flux density measurements at Q band. As a tentative compromise, we fit $P_{\mbox{\scriptsize mm}}(\nu)$ to the Ka and X band data taken from our new JVLA observations. 
\paragraph{DL~Tau} In this source, the Ka and Q band emission with a high $\alpha_{\mbox{\tiny 33-44 GHz}}$ value may trace the optically thin tail of the small dust emission in the bulk of the disk. This source was not detected at K band in the Disk@EVLA observations, and was not detected in our new JVLA X band observations. For the sake of 
using the same functional form in the fitting, we fit  $P_{\mbox{\scriptsize mm}}(\nu)$ to the Ka band data taken from the Disk@EVLA observations and the 7.5 GHz flux density measurement quoted from \citet{Dzib2015ApJ...801...91D}. In this case, the frequency of the transition region ($b/\omega$) is constrained by the fitting, while the $P_{\mbox{\scriptsize mm}}(\nu)$ does not have a physical meaning. 
\paragraph{UY~Aur} Its spectral index at 200--400 GHz is $\alpha_{\mbox{\scriptsize 200-400GHz}}=$1.91$\pm$0.23. Its spectral index becomes smaller than 1.91 in a region between $\sim$50 GHz and 200 GHz (Figure \ref{fig:SED_0}). The spectral index at 33--44 GHz is relatively high ($\alpha_{\mbox{\scriptsize 33-44 GHz}}=$4.04$\pm$0.55). The most probable explanation for these spectra features is that the emission at 50--400 GHz and at higher frequency trace the optically thick dust emission in the bulk of the disk, and the $\sim$30--50 GHz emission trace the optically thin tail of the dust emission from the bulk of the disk. The low spectral index at $\sim$50--400 GHz is the most likely explained by the effect of dust self-scattering with $a_{\mbox{\scriptsize maz}}\lesssim$1 mm (\citealt{Liu2019ApJ...877L..22L}), while the $\alpha_{\mbox{\scriptsize 33-44 GHz}}$ value is also consistent with $a_{\mbox{\scriptsize maz}}<$1 mm.
Our intention is to describe the optically thick dust emission in the bulk of the disk with the power law function $P_{\mbox{\scriptsize submm}}(\nu)$.
Therefore, for UY~Aur, we fit $P_{\mbox{\scriptsize submm}}(\nu)$ to both the SMA 200--400 GHz data (\citealt{Chung2024ApJS..273...29C}) and the Q band data taken from the Disk@EVLA survey. 
\paragraph{V836~Tau} Comparing the 102.5 GHz flux density reported by \citet{Ricci2012A&A...540A...6R} with the SMA 200--400 GHz measurements (\citealt{Chung2024ApJS..273...29C}) indicates that its spectral index is $<$2.0 at 100--200 GHz, which signifies optically thick dust emission in the bulk of the disk with $a_{\mbox{\scriptsize max}}\lesssim$500 $\mu$m. For a reason similar to that in the case of UY~Aur, we fit $P_{\mbox{\scriptsize submm}}(\nu)$ to the SMA 200--400 GHz measurements (\citealt{Chung2024ApJS..273...29C}) and the 102.5 GHz measurement (\citealt{Ricci2012A&A...540A...6R}).

When fitting the 4--400 GHz spectral profile with Equation \ref{eqn:specmodel}, we included some flux densities at $\sim$100 GHz quoted from previous work (\citealt{Pietu2006A&A...460L..43P,Isella2010ApJ...714.1746I,Ricci2010A&A...512A..15R,Ricci2012A&A...540A...6R,Guilloteau2011A&A...529A.105G,Pinilla2014A&A...564A..51P,Rodriguez2014ApJ...793L..21R,Dzib2015ApJ...801...91D,Perez2015ApJ...813...41P,Tripathi2018ApJ...861...64T,Huang2020ApJ...891...48H,Long2020ApJ...898...36L,Ueda2022ApJ...930...56U}). 
These data points help constrain the steepness ($\omega$) and the frequencies of the turning points ($b/\omega$) in the sigmoid function (Equation \ref{eqn:sigmoid}).
For Haro~6-37, DR~Tau, DN~Tau, IQ~Tau and DL~Tau, which we did not used the Q band data in the fittings of $P_{\mbox{\scriptsize mm}}(\nu)$, we also included the Q band data when fitting $F(\nu)$ in the last step. 

We fit the six free parameters in the piecewise functions using the Markov chain Monte Carlo (MCMC) method, employing the Python package {\tt emcee}.
When fitting $P_{\mbox{\scriptsize mm}}(\nu)$ and $P_{\mbox{\scriptsize submm}}(\nu)$, we assumed uniform prior functions within the ranges of 0.0--0.2 Jy, 0.0--2.0 Jy, 1.0--5.0 and 1.0--5.0, for $F_{\mbox{\scriptsize 44 GHz}}$, $F_{\mbox{\scriptsize 230 GHz}}$, $\alpha_{\mbox{\scriptsize mm}}$ and $\alpha_{\mbox{\scriptsize submm}}$, respectively.
The initial values of $F_{\mbox{\scriptsize 44 GHz}}$, $F_{\mbox{\scriptsize 230 GHz}}$, $\alpha_{\mbox{\scriptsize mm}}$ and $\alpha_{\mbox{\scriptsize submm}}$ were set to be 0.02, 0.2, 4.0, 2.0, respectively. 
We extracted the best fit parameters from the results of the power law fittings and assumed them as fixed parameters when fitting $F(\nu)$. 
After fixing the two power law functions, $P_{\mbox{\scriptsize mm}}(\nu)$ and $P_{\mbox{\scriptsize submm}}(\nu)$, there are two remaining free parameters, $b/\omega$ and ln($\omega$). 
The prior function of ln($\omega$), is uniform in the range of $-$3.9--0. 
The upper limit of $\omega$ defines the steepest transition which is close to the step function, and the lower limit defines the smoothest transition. 
We set the prior function of $b/\omega$ to be uniform within $f_{\mbox{\scriptsize mm, max}}-f_{\mbox{\scriptsize submm, min}}$, where $f_{\mbox{\scriptsize mm, max}}$ is the maximum frequency of the data used in $P_{\mbox{\scriptsize mm}}(\nu)$ fitting (e.g. 44 GHz), and $f_{\mbox{\scriptsize submm, min}}$ is the minimum frequency of the data used in $P_{\mbox{\scriptsize submm}}(\nu)$ fitting (e.g. 200 GHz). 
The range were chosen to restrict $b/\omega$ within $f_{\mbox{\scriptsize mm, max}}-f_{\mbox{\scriptsize submm, min}}$. 
The initial values of ln($\omega$) and $b/\omega$ were set to be $-$3.0 and 100.0, respectively. 
The best fit parameters are summarizedin Table \ref{tab:fit_parameter}.

In some sources, the best fit $P_{\mbox{\scriptsize mm}}(\nu)$ functions present consistency with the measurements at $<$10 GHz frequencies (e.g., RY~Tau, CW~Tau, AB~Aur).
Many sources (e.g., LkCa~15, UX~Tau, GO~Tau, UZ~Tau) at $<$10 GHz frequencies, which may be attributed to free-free and/or synchrotrion emission.

In some cases, our best fits appear like step functions, which is largely a presentation effect. 
This is because we defined the best fit as the 50\% percentile of our MCMC sample, which happened to be the ones that have relatively high $\omega$ value (e.g., the case of V710\,Tau in Figure \ref{fig:SED_2}). 
As can be seen in Figure \ref{fig:SED_2} and Table \ref{tab:fit_parameter}, the same SED can be equally well fit with a lower $\omega$ value (i.e., smoother transition).
For most of the sources, the sigmoid functions have modest steepness with $\omega\sim$0.5--0.8 (e.g., BP~Tau, CI~Tau, etc), although the value of $\omega$ are subject to large error bars in many cases (e.g., IP~Tau, Haro~6-39, V710~Tau).
The three sources, DL~Tau, DR~Tau, and UZ~Tau, may have relatively small $\omega$ values ($<$0.1). 
For sources that present relatively steep transitions, the bulk of the disks may have relatively uniform dust column densities and maximum grain sizes. 
Conversely, DL~Tau, DR~Tau, and UZ~Tau may have non-uniform distributions of dust column densities and maximum grain sizes in the bulk of their disks, such that different surface regions transits from optically thick to optically thin at different frequencies. 
Characterizing or categorizing the steepness of the transition therefore provides useful information about dust processing in the protoplanetary disks.



\end{CJK}
\end{document}